\documentclass[nofootinbib,a4paper,11pt]{article}
\usepackage{jheppub}
\usepackage[T1]{fontenc}
\usepackage[utf8]{inputenc}
\usepackage[compat=1.1.0]{tikz-feynman}
\usepackage{float}
\usepackage{slashed}
\usepackage{multicol,multirow}
\usepackage{amssymb}
\usepackage{amsmath}
\usepackage{subcaption}
\usepackage{array}
\usepackage{color}
\usepackage{hyperref}
\hypersetup{colorlinks=true}
\usepackage{longtable}
\usepackage{pifont}
\usepackage{graphics,graphicx,epsfig,ulem,booktabs}
\makeatletter
\newcommand\footnoteref[1]{\protected@xdef\@thefnmark{\ref{#1}}\@footnotemark}
\makeatother

\newcolumntype{C}[1]{>{\centering\let\newline\\\arraybackslash\hspace{0pt}}m{#1}}

\graphicspath{{images/}}

\definecolor{green}{rgb}{0.1,0.5,0.0}

\title{Revisiting Inclusive Decay Widths of Charmed Mesons}

\author[a]{Daniel King,}
\author[b]{Alexander Lenz,}
\author[b]{Maria Laura Piscopo,}
\author[c]{Thomas Rauh,}
\author[b]{Aleksey V. Rusov,}
\author[a]{Christos Vlahos}

\affiliation[a]{IPPP, Department of Physics, University of Durham, DH1 3LE, UK}
\affiliation[b]{Physik Department, Universit\"{a}t Siegen, Walter-Flex-Str. 3, 57068 Siegen, Germany}
\affiliation[c]{Albert Einstein Center for Fundamental Physics, Institute for Theoretical Physics, University of Bern, Sidlerstrasse 5, CH-3012 Bern, Switzerland}

\emailAdd{daniel.j.king@durham.ac.uk}
\emailAdd{alexander.lenz@uni-siegen.de}
\emailAdd{maria.piscopo@uni-siegen.de}
\emailAdd{rauh@itp.unibe.ch}
\emailAdd{rusov@physik.uni-siegen.de}
\emailAdd{christos.vlahos@durham.ac.uk}

\abstract{Determining for the first time the Darwin operator contribution for the non-leptonic charm-quark decays 
and using new non-perturbative results for the matrix elements of $\Delta C=0$ four-quark operators, including 
eye-contractions, we present a comprehensive study of the lifetimes of charmed mesons and inclusive semileptonic  decay rates as well as the ratios,
within the framework of the Heavy Quark Expansion (HQE).
We find good agreement with experiment for the ratio $\tau(D^+)/\tau(D^0)$, for the total $D_s^+$-meson decay rate, for the semileptonic rates of all three mesons 
$D^0$, $D^+$ and $D_s^+$, and for the semileptonic ratio $\Gamma_{sl}^{D^+}/\Gamma_{sl}^{D^0}$.  
The total decay rates of the $D^0$ and $D^+$ mesons are underestimated in our HQE  approach and we suspect that this is 
due to missing  higher-order QCD corrections to the free charm quark decay and the Pauli interference contribution. 
For the $SU(3)_F$ breaking ratios $\tau (D_s^+) / \tau (D^0) $ and $\Gamma_{sl}^{D_s^+}/\Gamma_{sl}^{D^0} $
our predictions lie closer to one than experiment. This might originate from  the poor knowledge of the non-perturbative parameters
$\mu_G^2$, $\mu_\pi^2$ and $\rho_D^3$ in the $D^0$ and $D_s^+$ systems.
These parameters could be determined by experimental studies of the moments of inclusive semileptonic $D$ meson decays.
}

\begin{document}

\hfill SI-HEP-2021-23, SFB-257-P3H-21-058, IPPP/21/22
\maketitle
\flushbottom

\clearpage

\section{Introduction}
Lifetimes of charm mesons are determined
experimentally very precisely 
\cite{Zyla:2020zbs} \footnote{New results from Belle
II have recently been made public
\cite{Belle-II:2021cxx}: $\tau(D^0)= 410.5 \pm 1.1 \pm
0.8$~fs, \, $\tau (D^+) = 1030.4 \pm 4.7 \pm 3.1$~fs.
} 
and show a pattern which is clearly less monotonous
than in the $b$-sector, with values spreading over a
rather large range.
Moreover, also inclusive semileptonic branching
fractions have been measured~\cite{Zyla:2020zbs}, and
recently an update for the $D_s^+$-meson has been
released by the BESIII Collaboration
\cite{Ablikim:2021qvs}. A summary of the current
experimental status is presented in
Table~\ref{tab:exp-data}.
\begin{table}[h]
\centering
\renewcommand{\arraystretch}{1.6}
    \begin{tabular}{|c||C{3cm}|C{3cm}|C{3cm}|}
    \hline
         & $D^0$ & $D^+$ & $D_s^+$ 
         \\
         \hline
         \hline
    $\tau \, [{\rm ps}]$ & $0.4101(15)$ & $1.040(7)$ & $0.504 (4)$
    \\
    \hline
     $\Gamma \, [{\rm ps}^{-1}]$ & $2.44(1)$ & $0.96(1)$ & $1.98 (2)$
    \\
    \hline
    $\tau (D_q)/\tau (D^0)$ & $1$ & $2.54(2)$ & $1.20 (1)$
    \\
    \hline
    \hline
     ${\rm Br}(D_q \to X e^+ \nu_e)  [\%]$ & 
     $6.49(11)$  & $16.07(30)$ & $6.30(16)$
     \\
    \hline
    $\displaystyle\frac{\Gamma (D_q \to X e^+ \nu_e)}{\Gamma (D^0 \to X e^+ \nu_e)}$ & $1$ & $0.977(26)$ & $0.790 (26)$    \\
    \hline
    \end{tabular}
    \caption{Status of the experimental determinations of the lifetime and the semileptonic branching fractions of the lightest charmed mesons ($D_q \in \left\{ D^0, D^+, D_s^+ \right\}$). All values are taken from the 
    PDG~\cite{Zyla:2020zbs} apart from the semileptonic 
    $D_s^+$-meson decays which were recently measured by the BESIII Collaboration~\cite{Ablikim:2021qvs}. }
    \label{tab:exp-data}
\end{table}
While in the bottom sector, the approximation that the
meson decay can be described in terms of the free $b$-quark decay
is experimentally well accommodated, 
for the charm system this is poorly justified. A
systematic way to study  this assumption is provided
by the heavy quark expansion (HQE)  - see
Refs.~\cite{Khoze:1983yp,Shifman:1986mx} for 
early references or Ref.~\cite{Lenz:2015dra} for a
recent review, according to which  the inclusive decay
width of a meson containing a heavy charm quark can be
written as 
\begin{equation}
\Gamma(D) = 
\Gamma_3  +
\Gamma_5 \frac{\langle {\cal O}_5 \rangle}{m_c^2} + 
\Gamma_6 \frac{\langle {\cal O}_6 \rangle}{m_c^3} + ...  
 + 16 \pi^2 
\left( 
  \tilde{\Gamma}_6 \frac{\langle \tilde{\mathcal{O}}_6 \rangle}{m_c^3} 
+ \tilde{\Gamma}_7 \frac{\langle \tilde{\mathcal{O}}_7 \rangle}{m_c^4} + ... 
\right),
\label{eq:HQE}
\end{equation}
with the matrix element of the $\Delta C = 0$
operators given by  $\langle {\cal O}_Y \rangle =
\langle D | {\cal O}_Y|D  \rangle/(2 m_D)$. Their
numerical size is expected to be of the order of the
hadronic scale  $\Lambda_{\rm QCD} \leq 1$ GeV, but the
actual value  must be determined with a
non-perturbative calculation.  Note that in
Eq.~\eqref{eq:HQE} quantities labelled by a tilde
refer to the contribution of four-quark operators,
while those without a tilde correspond to two-quark
operators, c.f.\ Fig.~\ref{fig:HQE}. The Wilson
coefficients $\Gamma_i$ in Eq.~\eqref{eq:HQE} can be
computed perturbatively and admit the following
expansion in the strong coupling $\alpha_s$, i.e.
\begin{equation}
\Gamma_i = 
\Gamma_i^{(0)} + 
 \frac{\alpha_s (m_c)}{ 4 \pi} \Gamma_i^{(1)} +
 \left[ \frac{\alpha_s (m_c)}{ 4 \pi} \right]^2 \Gamma_i^{(2)} + ... \,.
\label{eq:Wilson}
\end{equation}
In the present work we will try to shed further light
into the question, whether the expansion parameters
$\alpha_s (m_c)$ and $\Lambda_{\rm QCD} / m_c$ are small
enough in order to ensure a meaningful convergence of
the HQE.
The Particle Data Group \cite{Zyla:2020zbs} quotes,
for the pole and $\overline{\rm MS}$ mass of the charm
quark, the values
\begin{equation}
m_c^{\rm Pole} = (1.67  \pm 0.07) \; {\rm GeV}
\; ,
\hspace{1cm}
\overline{m}_c(\overline{m}_c) = (1.27  \pm 0.02) \; {\rm GeV},
\label{eq:Mass}
\end{equation}
while the dependence of the strong coupling on both
the charm scale and the loop order (obtained using the
RunDec package \cite{Herren:2017osy}) is shown in
Table~\ref{tab:alphas}. In our numerical analysis we use 
the 5-loop running of the strong coupling. 
\begin{table}[h]
\centering
\renewcommand{\arraystretch}{1.4}
\begin{tabular}{|c||c|c|c|}
\hline
$\alpha_s(m_c)$ &
$m_c = 1.67 \, {\rm GeV}$ & 
$m_c = 1.48 \, {\rm GeV}$ &
$m_c = 1.27 \, {\rm GeV}$
\\
\hline
\hline
\mbox{2-loop}     &  
$0.322$ &
$0.346$ &
$0.373$
\\
\hline
\mbox{5-loop}     &  
$0.329$ & 
$0.356$ &
$0.387$ 
\\
\hline
\end{tabular} 
\caption{ Numerical values of the strong coupling
$\alpha_s$ evaluated at different scales and loop
order, obtained using the RunDec
package~\cite{Herren:2017osy}.}
\label{tab:alphas}
\end{table}
While the determination of the $\overline{\rm MS}$
mass is theoretically well founded, that of the pole
mass seems to be affected by a potential breakdown of
perturbation theory. On the other side, the pole mass
is the natural expansion parameter of the HQE. The relation between the two
mass schemes, up to third order 
in the strong coupling, reads \cite{Chetyrkin:1999qi,Chetyrkin:1999ys,Melnikov:2000qh}
\begin{eqnarray}
  m_c^{\rm Pole}  & = &
   \overline{m}_c(\overline{m}_c) 
   \left[ 1 
   + \frac43 \frac{\alpha_s(  \overline{m}_c)}{\pi} 
   + 10.43
  \left( \frac{\alpha_s(  \overline{m}_c)}{\pi} \right)^2 
   + 116.5
  \left( \frac{\alpha_s(  \overline{m}_c)}{\pi} \right)^3
  \right] 
  \nonumber
  \\
  & = &
  \overline{m}_c(\overline{m}_c) 
   \left[ 1 
   + 0.1642
   + 0.1582 
   + 0.2176 
  \right]       \, ,
   \label{eq:MSpole}
\end{eqnarray}
where we have used the
5-loop result for the strong coupling 
at the scale 1.27 GeV. Due to the fact that $\Gamma_3$
depends on the fifth power of the charm pole mass, see
Section~\ref{SubSec:Lead}, one obtains quite different
results according to how higher orders in
Eq.~(\ref{eq:MSpole}) are treated. 
Specifically, by truncating the expansion in
Eq.~\eqref{eq:MSpole} at first order in $\alpha_s$, from 
$\overline{m}_c(\overline{m}_c)
= 1.27$ GeV, we obtain for the pole mass the value 
$m_c^{\rm Pole} = 1.479 $ GeV, which leads respectively to
\begin{eqnarray}
\left( m_c^{\rm Pole} \right)^5 
& = &
  \overline{m}_c(\overline{m}_c)^5 
  \left[ 1 +0.1642 \right]^5
= 2.14 \, \overline{m}_c(\overline{m}_c)^5,
\label{eq:prefactor1}
\end{eqnarray}
taking the fifth power of $m_c^{\rm Pole}$, and
\begin{eqnarray}
 \left( m_c^{\rm Pole} \right)^5  &  \approx & 
  \overline{m}_c (\overline{m}_c)^5 
  \left[ 1 + 5 \cdot 0.1642 \right]
  = 1.82 \, \overline{m}_c(\overline{m}_c)^5 
   \, ,
   \label{eq:MSpole5}
\end{eqnarray}
further expanding up to the first 
order in $\alpha_s$. The result in Eq.~(\ref{eq:MSpole5}) 
is about 15 $\%$ smaller than the one in 
Eq.~(\ref{eq:prefactor1}).
Instead, by including also all the higher order terms given in 
Eq.~(\ref{eq:MSpole}), we get
\begin{eqnarray}
\left( m_c^{\rm Pole} \right)^5 
& = &
  \overline{m}_c(\overline{m}_c)^5 
  \left[  1 
  + 0.1642
   + 0.1582
   + 0.2176
   \right]^5
   =  8.66 \,
  \overline{m}_c(\overline{m}_c)^5 
   \, ,
   \label{eq:MSpole5b}
\end{eqnarray}
which is roughly a factor four larger than
the value in Eq.~(\ref{eq:prefactor1}).

In the following, we will thus consider four different quark mass schemes:
\begin{enumerate}
\item Use Eq.~(\ref{eq:MSpole}) to first order in  
      $\alpha_s$, since this is the order to which most of
      the Wilson coefficients are known. In this case we fix
      $m_c^{\rm Pole} = 1.48$ GeV and express everything 
      in terms of the pole mass.
      A further possibility would be to consider the
      expansion in Eq.~(\ref{eq:MSpole}) to be an 
      asymptotic one, whose smallest correction appears at
      order $\alpha_s^2$, which is where we stop the
      expansion. In this case we get the pole mass value
      from PDG, $m_c^{\rm Pole} = 1.67 $ GeV. We did 
      numerical tests for this large value of the charm quark mass and the results for decay rates are roughly $70 - 90 \%$ larger than the values obtained using $m_c^{\rm Pole} = 1.48$~GeV. Since we expect
      this enhancement to be compensated by missing NNLO
      corrections to the non-leptonic decay rates,
      we will not separately present results for $m_c^{\rm Pole} = 1.67 $ GeV.
       
\item Express the $c$-quark mass in terms of the 
     $\overline{\rm MS}$ mass \cite{Bardeen:1978yd}, 
     \begin{equation}
     m_c^{\rm Pole} = \overline m_c(\overline m_c) 
     \left[ 1 + \frac{4}{3} \frac{\alpha_s(\overline m_c) }{\pi} \right]\,,
     \label{eq:c-quark-mass-Pole-to-MS}
     \end{equation}
     taking $\overline{m}_c(\overline{m}_c) = 1.27$ GeV~\cite{Zyla:2020zbs}, 
     and expand consistently up to order $\alpha_s$. Because
     of the dependence on the fifth power of the charm-quark
     mass, in this case $\Gamma_3$ is affected by a large
     correction~$5 \times (4/3)(\alpha_s /\pi)$.
     
\item Express the $c$-quark mass in terms of the kinetic mass \cite{Bigi:1994ga,Bigi:1996si}.  The kinetic scheme has been introduced in order to obtain a short distance definition of the heavy quark mass which allows a faster convergence of the perturbative series and is still valid at small scales $\mu \sim 1$ GeV. The relation between the kinetic scheme and the $\overline{\rm MS}$ and Pole schemes can be found, up to N$^3$LO corrections, in Ref.~\cite{Fael:2020njb}. At order $\alpha_s$ one has
\begin{equation}
      m_c^{\rm Pole}  = m_c^{\rm Kin}
      \left[ 1 + \frac{4 \alpha_s}{3 \pi} 
      \left( 
      \frac43 \frac{\mu^{\rm cut}}{ m_c^{\rm Kin} }
      +
      \frac12 \left(\frac{\mu^{\rm cut}}{ m_c^{\rm Kin}}\right)^2
      \right)\right] \, ,
      \label{eq:Pole-Kin-scheme}
      \end{equation}
where $\mu^{\rm cut}$ is the Wilsonian cutoff 
separating the perturbative and non-perturbative regimes.
Using $\overline m_c(\overline m_c)$ as an input, the authors of Ref.~\cite{Fael:2020njb} obtain
\begin{eqnarray}
m_c^{\rm Kin} (1 {\rm GeV}) & = & 1.128 \, {\rm GeV} \quad {\rm (N }^3{\rm LO)} \,, 
\\
m_c^{\rm Kin} (1 {\rm GeV}) & = & 1.206 \,  {\rm GeV} \quad  {\rm (NLO)} \,.
\end{eqnarray}
Comparing with Eq.~\eqref{eq:c-quark-mass-Pole-to-MS} it follows that the kinetic scheme might
be preferred to the $\overline{\rm MS}$ scheme if the term in the round brackets of Eq.~(\ref{eq:Pole-Kin-scheme}) would give a suppression factor. For $\mu^{\rm cut} = 1 \, {\rm GeV}$ and $m_c^{\rm Kin}  = 1.2 $ GeV, this is not the case, while using lower values i.e.\ $\mu^{\rm cut} <1$~GeV, the convergence of the series could be improved, however this would bring in an additional uncertainty due to the closeness to
 the non-perturbative scale $ \Lambda_{\rm QCD} $.
 In our numerical analysis we will investigate the kinetic scheme with $\mu^{\rm cut} = 0.5$~GeV. 
 From Ref.~\cite{Fael:2020njb} we take the following value
 \begin{equation}
  m_c^{\rm kin} (0.5 \, {\rm GeV}) = 1.363 \, {\rm GeV}  \,,
 \end{equation}
 obtained for consistency at NLO in $\alpha_s$ and using 
 as an input $\overline{m}_c (\overline{m}_c)$.   
 
\item  
In addition, we will consider the $1S$-mass scheme defined as
 \cite{Hoang:1998ng,Hoang:1998hm, Hoang:1999zc}:
\begin{equation}
m_c^{\rm Pole} = m_c^{1S}
\left(1 + \frac{(C_F \, \alpha_s)^2}{8} \right) \,,
\label{eq:mc-1S}
\end{equation}
where $C_F = 4/3$, and the $1S$ mass $m_c^{1S} \approx 1.44$ GeV
is obtained using the conversion from the $\overline{\rm MS}$-scheme 
(implemented in the RunDec package~\cite{Herren:2017osy}) at one-loop level. 
Note that the correction within the $1S$ scheme in fact starts at order~$\alpha_s^2$ 
which however is still considered to be a NLO (not NNLO) effect \cite{Hoang:1998ng}.\footnote{Similarly, another possibility would be to study the potential subtracted mass \cite{Beneke:1998rk}.}

\end{enumerate}
The above arguments clearly indicate the importance of
including higher order perturbative QCD corrections to the
decay rates.

With this work we present a study of
the total decay rate of the $D^0$, $D^+$ and $D_s^+$ mesons,
of their lifetime ratios $\tau (D^+) /\tau (D^0) $ and $\tau
(D_s^+) /\tau (D^0) $ and of the semileptonic branching
fractions~${\rm Br}(D_q\to X e^+ \nu_e)$ using 
state-of-the-art expressions for the $\Delta C =
0$ Wilson coefficients and for the non-perturbative
parameters. $\Gamma_3$ is known at NLO-QCD \cite{Hokim:1983yt,Altarelli:1991dx,Voloshin:1994sn,Bagan:1994zd,Bagan:1995yf,Lenz:1997aa,Lenz:1998qp,Krinner:2013cja}
for non-leptonic decays. NNLO-QCD \cite{Czarnecki:1997hc,Czarnecki:1998kt,vanRitbergen:1999gs,Melnikov:2008qs,Pak:2008cp,Pak:2008qt,Dowling:2008ap,Bonciani:2008wf,Biswas:2009rb,Brucherseifer:2013cu} and  NNNLO-QCD \cite{Fael:2020tow,Czakon:2021ybq} corrections have been
computed for semileptonic decays, while for non-leptonic
decays NNLO corrections have been determined in the massless
case and in full QCD  (i.e.\ no effective Hamiltonian was
used) in Ref.~\cite{Czarnecki:2005vr}. $\Gamma_5$ was determined at LO-QCD for both semileptonic and non-leptonic decays 
\cite{Bigi:1992su,Blok:1992hw,Blok:1992he,Bigi:1992ne}. For the semileptonic modes even 
NLO-QCD corrections are available \cite{Alberti:2013kxa,Mannel:2014xza,Mannel:2015jka}.
In the $b$-system,
$\Gamma_6$~was first computed at LO-QCD in Ref.~\cite{Gremm:1996df} 
and recently the NLO-QCD corrections were determined in Ref.~\cite{Mannel:2019qel}, 
both for the semileptonic case only. Very recently $\Gamma_6$ has been determined also for non-leptonic decays  \cite{Lenz:2020oce,Mannel:2020fts,Moreno:2020rmk}
and the coefficient was found to be large. 
For semileptonic $D$-meson decays, 
$\Gamma_6$ was determined in Ref.~\cite{Gambino:2010jz}, 
see also the recent Ref.~\cite{Fael:2019umf}, while the corresponding results for the non-leptonic charm modes are presented for the first time in this work.
$\tilde{\Gamma}_6$ is known at NLO-QCD for lifetimes of $B$-meson
\cite{Beneke:2002rj,Franco:2002fc}
and of $D$-meson \cite{Lenz:2013aua}, while $\tilde \Gamma_7$ and $\tilde \Gamma_8$ have been estimated in LO-QCD in Refs.~\cite{Gabbiani:2003pq, Gabbiani:2004tp}.
 
On the non-perturbative side, at dimension-five, the matrix element 
of the chromomagnetic operator can be determined from spectroscopy, while for the kinetic operator there exist several 
heavy quark effective theory (HQET) determinations with lattice simulations \cite{Bazavov:2018omf,Gambino:2017vkx,Aoki:2003jf,Kronfeld:2000gk,Gimenez:1996av} and using sum rules  \cite{Ball:1993xv,Bigi:1994ga,Neubert:1996wm}.
The matrix elements of the four-quark operators $\langle \tilde{\mathcal{O}}_6 \rangle$ have been computed
using  HQET sum rules \cite{Kirk:2017juj}. 
Violations of  $SU(3)_F$ and so far undetermined eye-contractions could yield visible effects and a
calculation of these corrections with HQET sum rules -
following Ref.~\cite{King:2019lal} - has been performed in
Ref.~\cite{King:2020}.
Corresponding lattice results for the matrix elements of the four-quark operators would be highly desirable.
We emphasise that the matrix element of the dimension-six
Darwin operator,  $\langle {\cal O}_6 \rangle $,
can be expressed in terms of the above Bag parameters by
taking into account the equation of motion for the gluon field strength tensor. 

The paper is organised as follows. 
In Section~\ref{Sec:total}, after briefly introducing the
effective Hamiltonian describing the $c$-quark decays, we
analyse in detail the structure of the HQE, discussing each
of the short-distance contributions in Eq.~\eqref{eq:HQE}.
In Section~\ref{sec:HQE-NP-parameters}, we describe how the
corresponding non-perturbative parameters are determined. 
Numerical results for the total $D$-meson decay widths,
their ratios, as well as for the semileptonic branching
fractions, are presented in Section~\ref{sec:Results}.
Finally, we conclude in Section~\ref{sec:conclusion} with an
outlook on how to further improve the theoretical
predictions in the charm sector.
The numerical input used in the analysis are collected in Appendix~\ref{Appendix-A}, 
the complete expressions for the coefficients of the Darwin operator for non-leptonic  
$c$-quark decays are presented in Appendix~\ref{Appendix-B}, while in Appendix~\ref{Appendix-C} we show the parametrisation of the matrix elements of the four-quark operators.

\section{The Total Decay Rate}
\label{Sec:total}
\subsection{Effective Hamiltonian and HQE}
\label{SubSec:Heff}
The non-leptonic decay of a charm quark $c \to q_1 \bar{q}_2 u$ ($q_i = u,d,s$) is governed by the effective $\Delta C = 1$ Hamiltonian
(see e.g. Ref.~\cite{Buchalla:1995vs})
\begin{eqnarray}
  {\cal H}_{\rm eff}^{\rm NL} & = & 
  \frac{G_F}{\sqrt{2}} 
  \left[
   \sum \limits_{q_{1,2}=d,s} \lambda_{q_1 q_2} 
  \left[C_1 (\mu_1) \, Q_1^{q_1q_2} + C_2 (\mu_1) \, Q_2^{q_1q_2} \right]
   - \lambda_b \sum \limits_{j=3}^6  C_j (\mu_1) Q_j 
   \right] + {\rm h.c.},
   \label{eq:Heff-NL}
\end{eqnarray}
where $\lambda_{q_1q_2} = V_{cq_1}^* V_{uq_2} $
and $\lambda_{b} = V_{cb}^* V_{ub} $ are the CKM factors, 
$C_i (\mu_1)$ denote the Wilson coefficients at the renormalisation scale $\mu_1 \sim m_c$, 
$Q_{1,2}^{q_1q_2}$ are tree-level $\Delta C = 1$ 
operators
\footnote{In our notation, $Q_1^{q_1q_2}$ is the colour-singlet operator.}
\begin{eqnarray}
Q_1^{q_1 q_2} & = &  
\left(\bar{q}_1^i \gamma_\rho (1- \gamma_5) c^i \right)
\, \left(\bar{u}^j \gamma^\rho (1- \gamma_5) q_2^j \right),
\label{eq:Q1}
\\
Q_2^{q_1 q_2} & = &  
\left(\bar{q}_1^i \gamma_\rho (1- \gamma_5) c^j \right)
\, \left(\bar{u}^j \gamma^\rho (1- \gamma_5) q_2^i \right),
\label{eq:Q2}
\end{eqnarray}
while $Q_j, \, j = 3...6 $ are penguin operators, which can only arise in the singly Cabibbo 
suppressed decays $c \to s \bar{s} u$  and $c \to d \bar{d} u$ or in further suppressed pure penguin decays like 
$c \to u \bar{u} u$. 
Values of the Wilson coefficients at different scales are shown in Table~\ref{tab:WCs} both at NLO-QCD and LO-QCD. 
\begin{table}[th]
\renewcommand{\arraystretch}{1.3}
\centering
   \begin{tabular}{|C{1.8cm}||C{1.8cm}|C{1.8cm}|C{1.8cm}|C{1.8cm}|C{1.8cm}|C{1.8cm}|}
   \hline
     $\mu_1  [{\rm GeV}]$  & 
     $1$     &  
     $1.27 $ &  
     $1.36 $ &  
     $1.44 $ &   
     $1.48 $ &
     $3 $ 
     \\
    \hline \hline
     $C_1 (\mu_1) $  & 
     $1.25$ \qquad $(1.34)$ & 
     $1.20$ \qquad $(1.27)$ & 
     $1.19$ \qquad $(1.26)$ & 
     $1.18$ \qquad $(1.25)$ & 
     $1.18$ \qquad $(1.24)$ & 
     $1.10$ \qquad $(1.15)$ 
     \\ 
     \hline
     $C_2 (\mu_1) $  & 
     $-0.48$ \qquad $(-0.62)$ & 
     $-0.39$ \qquad $(-0.50)$ & 
     $-0.40$ \qquad $(-0.53)$ & 
     $-0.37$ \qquad $(-0.49)$ & 
     $-0.37$ \qquad $(-0.48)$ & 
     $-0.24$ \qquad $(-0.32)$ 
     \\ 
     \hline
     $C_3 (\mu_1) $  & 
     $0.03$ \qquad $(0.02)$ &
     $0.02$ \qquad $(0.01)$ & 
     $0.02$ \qquad $(0.01)$ & 
     $0.01$ \qquad $(0.01)$ & 
     $0.01$ \qquad $(0.01)$ & 
     $0.00$ \qquad $(0.00)$ 
     \\
     \hline
     $C_4 (\mu_1) $  & 
     $-0.06$ \qquad $(-0.04)$ &
     $-0.05$ \qquad $(-0.03)$ &
     $-0.04$ \qquad $(-0.03)$ &
     $-0.04$ \qquad $(-0.02)$ &
     $-0.04$ \qquad $(-0.02)$ &
     $-0.01$ \qquad $(-0.01)$ 
     \\
     \hline
     $C_5 (\mu_1) $  & 
     $0.01$ \qquad $(0.01)$ &
     $0.01$ \qquad $(0.01)$ & 
     $0.01$ \qquad $(0.01)$ & 
     $0.01$ \qquad $(0.01)$ & 
     $0.01$ \qquad $(0.01)$ & 
     $0.00$ \qquad $(0.00)$ 
     \\
     \hline
     $C_6 (\mu_1) $  & 
     $-0.08$ \qquad $(-0.05)$ &
     $-0.05$ \qquad $(-0.03)$ &
     $-0.05$ \qquad $(-0.03)$ &
     $-0.04$ \qquad $(-0.03)$ &
     $-0.04$ \qquad $(-0.03)$ &
     $-0.01$ \qquad $(-0.01)$ 
     \\
     \hline 
    \end{tabular} 
    \caption{Comparison of
    the Wilson coefficients 
    at NLO-QCD (LO-QCD) for different values of $\mu_1$.}
    \label{tab:WCs}
\end{table}
\\
We see that the Wilson coefficients of the penguin operators are very small, additionally their contributions are also strongly suppressed by the CKM factor $\lambda_b \ll \lambda_{q_1 q_2}$. Therefore, in our analysis, we neglect the effect of the penguin operators, given the current
limited theoretical accuracy in the charm sector.

\begin{figure}[ht]
\centering
\includegraphics[scale=0.37]{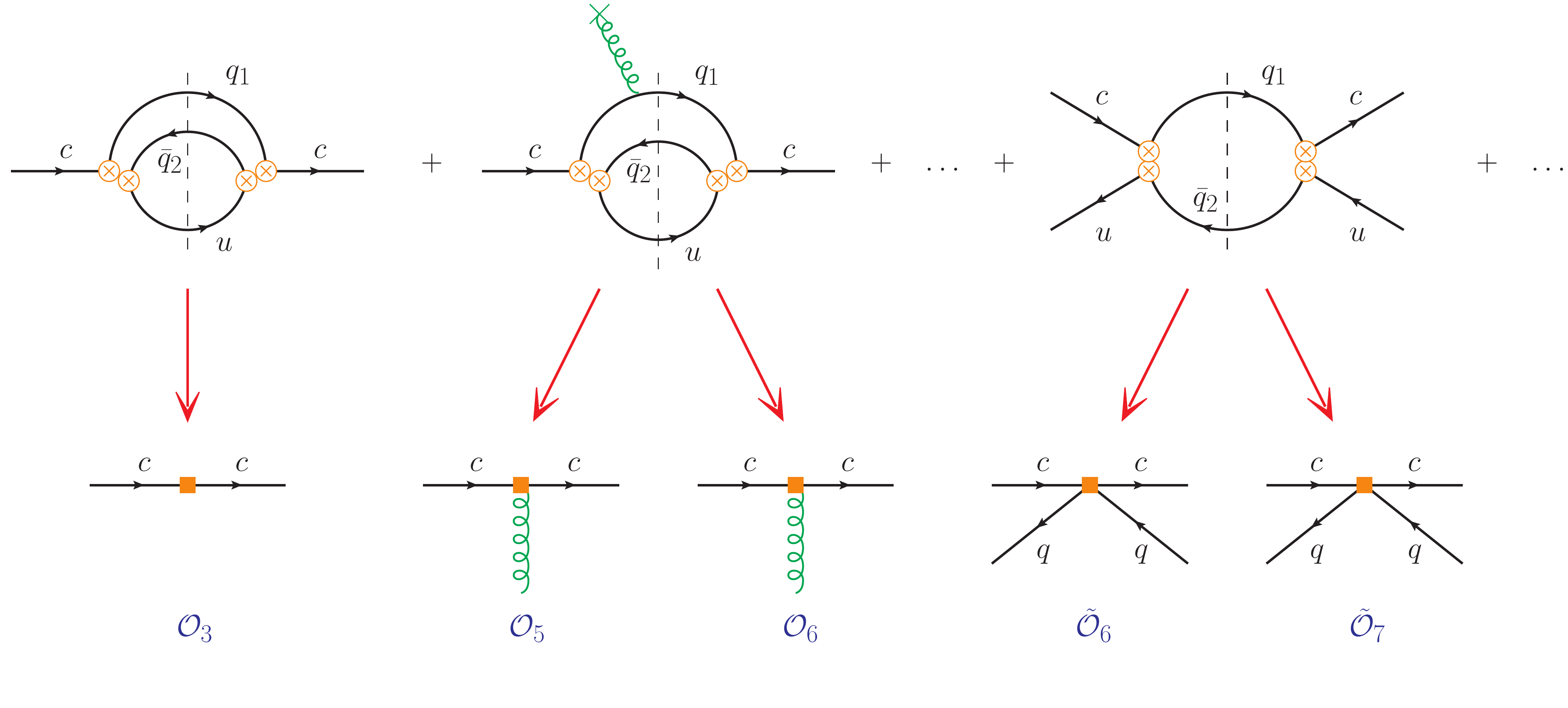}
\caption{The diagrams describing contributions to the HQE in
Eq.~(\ref{eq:HQE}).  The crossed circles denote the
$ \Delta  C =1$
operators $Q_i$  of the effective Hamiltonian while the squares
denote the local $ \Delta C = 0$ operators ${\cal O}_i$ and
$\tilde{{\cal O}_i}$. 
The two-loop and the phase space enhanced
one-loop diagrams correspond respectively to the two-quark operators
${\cal O}_i$ and to the four-quark operators $\tilde{{\cal O}_i}$ 
in the HQE.}
\label{fig:HQE}
\end{figure}

The complete effective Hamiltonian  describing all possible $c$-quark decays is a sum of non-leptonic, semileptonic and radiative contributions, namely
\begin{eqnarray}
  {\cal H}_{\rm eff} & = &  {\cal H}_{\rm eff}^{\rm NL} + {\cal H}_{\rm eff}^{\rm SL} + {\cal H}_{\rm eff}^{\rm rare} \, ,
\end{eqnarray}
where ${\cal H_{\rm eff}^{\rm NL}}$ is given in Eq.~\eqref{eq:Heff-NL},
\begin{equation}
{\cal H}_{\rm eff}^{\rm SL} 
= 
\frac{G_F}{\sqrt 2} \sum_{q = d, s} \sum_{\ell = e, \mu}
V_{cq}^* \, Q^{q \ell} + {\rm h.c.}\,,
\label{eq:Heff-SL}
\end{equation}
with the semileptonic operator 
$Q^{q \ell} =\left(\bar{q} \gamma^\mu (1- \gamma_5) c \right)
\left(\bar{\nu}_\ell \gamma_\mu (1 - \gamma_5) \ell \right)$, while 
${\cal H}_{\rm eff}^{\rm rare}$ describes decays like
$D \to \pi \ell^+ \ell^-$, whose branching fraction is much smaller than those corresponding to
tree-level transitions. Hence, in the following  we neglect
rare decays  and we do not show an explicit expression for ${\cal H}_{\rm eff}^{\rm rare}$. \\
The total decay width of a heavy $D$ meson with mass~$m_{D}$ and four-momentum $p_{D}^\mu$ can be written~as 
\begin{equation}
\Gamma ({D})  = 
\frac{1}{2 m_{{D}}} \sum_{X}  \int \limits_{\rm PS} (2 \pi)^4  \delta^{(4)}(p_{D}- p_X) \, \,
|\langle X(p_X)| {\cal H}_{\rm eff} | {D}(p_{D}) \rangle |^2,
\label{eq:Gamma-D}
\end{equation}
where PS~denotes the phase space integration and we have summed over all possible final states $X$ into which the $D$ meson can decay. Eq.~(\ref{eq:Gamma-D}) can be related, via the optical theorem, to the discontinuity of the forward scattering matrix element of the time ordered product of the double insertion of the effective Hamiltonian, i.e.
\begin{equation}
\Gamma (D) =    \frac{1}{2 m_D} {\rm Im}
\langle D | {\cal T}| D \rangle \, ,
\label{eq:Gamma_opt_th}
\end{equation}
with the transition operator
\begin{equation}
{\cal T}  =  
i \int d^4x 
\,  T \left\{ {\cal H} _{\rm eff} (x) \, ,
 {\cal H} _{\rm eff} (0)  \right\} \, .
 \label{eq:optical_theorem}
\end{equation}
In the framework of the HQE, the four-momentum of the decaying $c$-quark 
is parametrised in terms of "large" and "small" components as
\begin{equation}
p_c^\mu = m_c \, v^\mu + k^\mu,
\label{eq:c-quark-momentum}
\end{equation}
where $v^\mu = p^\mu/ m_D$ denotes the four-velocity of the $D$-meson and $k^\mu \to i D^\mu$, where
 $D^\mu$ is the covariant derivative with respect to the background gluon field, accounting for the residual
interaction of the $c$-quark with the light degrees of freedom, i.e.\ soft gluons and quarks.
At the same time, the heavy charm quark field is redefined as  
\begin{equation}
c (x) = e^{ - i m_c v \cdot x} c_v (x)\,,  
\label{eq:phase-redef}    
\end{equation}
to remove the large fraction of the $c$-quark momentum.
Using Eqs.~\eqref{eq:c-quark-momentum} and \eqref{eq:phase-redef}, $\Gamma(D)$ in Eq.~(\ref{eq:Gamma_opt_th}) can be expanded in the small quantity 
$D^\mu/m_c \sim \Lambda_{\rm QCD}/m_c$, leading to the series in Eq.~(\ref{eq:HQE}), 
for more details, see e.g. Ref.~\cite{Dassinger:2006md}
or Ref.~\cite{Lenz:2020oce} for a more recent treatment. The result is 
schematically shown in Fig.~\ref{fig:HQE}. The first diagram on the top line of Fig.~\ref{fig:HQE}, corresponding to the limit $m_c \to \infty$, 
represents the decay of a free $c$-quark, while
power corrections due to the interaction of the heavy quark with soft gluons and quarks are described respectively by the second and third diagrams on the top line of Fig.~\ref{fig:HQE}. Finally, before discussing the individual terms in Eq.~(\ref{eq:HQE}) separately, it is worth emphasizing that 
the field $c_v$ is related to the effective heavy quark field $h_v$, introduced in the framework of the  HQET (see e.g. Ref.~\cite{Neubert:1993mb}), by
\begin{equation}
c_v (x) = h_v (x) + \frac{i \slashed D_\perp}{2 m_c} h_v (x) 
+ {\cal O}\left(\frac{1}{m_c^2}\right),
\label{eq:cv-hv-relation}
\end{equation}
where $D_\perp^\mu = D^\mu - (v \cdot D) \, v^\mu$.
\subsection{Dimension-three Contribution}
\label{SubSec:Lead}
\begin{figure}[t]
    \centering
    \includegraphics[scale=1.0]{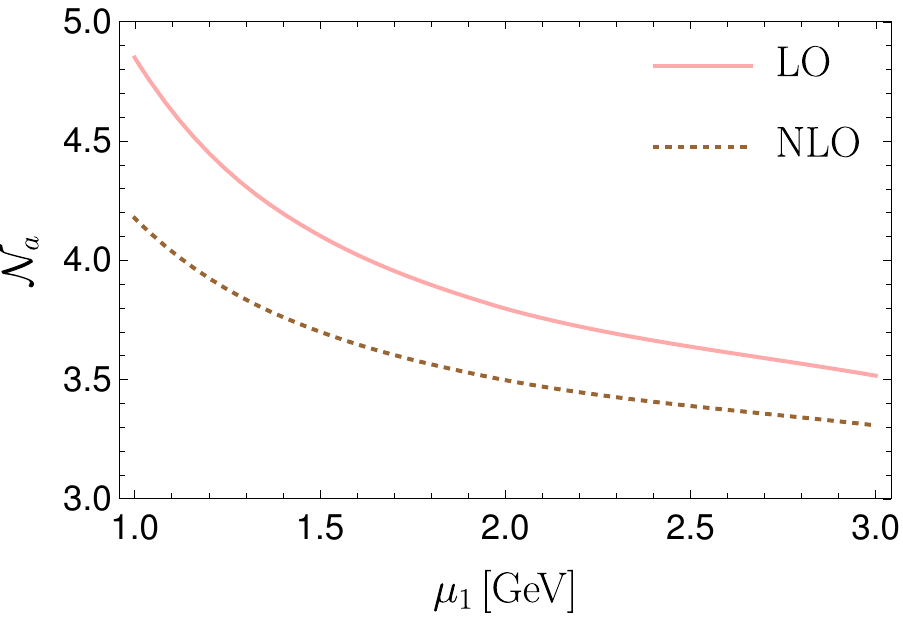}
    \caption{Scale dependence of the Wilson coefficient combination 
    ${\cal N}_a = 3 \, C_1^2 + 3 \, C_2^2 + 2 \, C_1 C_2$.}
    \label{fig:Na}
\end{figure}

The leading term in Eq.~(\ref{eq:HQE}), $\Gamma_3^{(0)}$, can be schematically
written as
\begin{eqnarray}
    \Gamma_3^{(0)} & = & \Gamma_0 \, c_3 =
    \Gamma_0 
    \left[
    f\left(z_s, z_e, z_{\nu_e} \right) + 
    f\left(z_s, z_\mu, z_{\nu_\mu} \right) 
    + |V_{ud}|^2 {\cal N}_a \,  
    f\left(z_s, z_u, z_d \right) 
    + \ldots \right]\,,
    \label{eq:Gamma_3}
\end{eqnarray}
where we define 
\begin{equation}
\Gamma_0 = \frac{G_F^2 m_c^5}{192 \pi^3} |V_{cs}|^2   \,,  
\label{eq:Gamma_0}
\end{equation}
and we introduce the dimensionless mass parameter $z_q = m_q^2/m_c^2$.
Note than we neglect the neutrino as well as the electron, the up and down quarks masses, i.e. $z_\nu = z_e = z_u = z_d = 0$,
while  $z_s \not =  0 \not =   z_\mu $. 
The first two terms in $c_3$ in Eq.~(\ref{eq:Gamma_3}) 
correspond to the semileptonic modes
$c \to s e^+ \nu_e$ and $c \to s \mu^+ \nu_\mu$, 
while the third term to the Cabibbo favoured 
decay $c \to s u \bar{d}$. 
The ellipsis stand for CKM suppressed contributions.
The dependence on the $\Delta C = 1$ Wilson coefficients
is absorbed in the combination 
${\cal N}_a = 3 \, C_1^2 + 3 \, C_2^2 + 2 \, C_1 C_2$.
The behaviour of ${\cal N}_a$ as function of the renormalisation scale,
both at LO- and NLO-QCD, is shown in Table~\ref{tab:Na} 
and in Fig.~\ref{fig:Na}, indicating a visible shift 
from LO to NLO and a moderate reduction of the scale 
uncertainty in the NLO result.
Else there are no cancellations
in ${\cal N}_a $ that might lead to numerical instability.
The phase-space function $f(a,b,c)$ in Eq.~\eqref{eq:Gamma_3} describes 
the effect of the final state masses. 
In the case of one massive particle, it reduces to the well-known expression
\begin{eqnarray}
  f(z,0,0)  =  1 - 8 z + 8 z^3 - z^4 - 12 z^2 \ln z \, ,
  & \quad &
  f \left(z_s, 0, 0\right) \approx  1 - 0.03 \, ,
\end{eqnarray}
which shows that the contribution due to the finite $s$-quark mass is small.
The analytic expression of $f(a,b,c)$ for two different masses in the final state, can be found e.g. in the Appendix of Ref.~\cite{Mannel:2017jfk}.
\begin{table}[t]
\centering
\renewcommand{\arraystretch}{1.25}
   \begin{tabular}{|c||C{1.3cm}|C{1.3cm}|C{1.3cm}|C{1.3cm}|C{1.3cm}|C{1.3cm}|}
   \hline
   $\mu_1$ [GeV] & 1 & 1.27 & 1.36 & 1.44 & 1.48 & 3 
   \\
   \hline 
   \hline
   ${\cal N}_a ({\rm LO})$ &
   4.85 & 
   4.35 &
   4.23 &
   4.15 &
   4.12 &
   3.52 
   \\
   ${\cal N}_a ({\rm NLO})$ &  
   4.18 &
   3.86 &
   3.79 &
   3.74 &
   3.71 &
   3.31 
   \\
   \hline
  \end{tabular} 
  \caption{Comparison of ${\cal N}_a$  at LO- and NLO-QCD, for different values of the renormalisation scale~$\mu_1$. }
  \label{tab:Na}
\end{table}
\\
By including also NLO-QCD corrections,  $\Gamma_3$
can be schematically presented as 
\begin{equation}
    \Gamma_3 = \Gamma_0 \left[
    3 \, C_1^2    \, {\cal C}_{3,11} 
 +  2 \, C_1 C_2  \, {\cal C}_{3,12} 
 +  3 \, C_2^2    \, {\cal C}_{3,22} 
 +             {\cal C}_{3, \rm SL} 
    \right]  \, ,
\end{equation}
where a summation over all the modes is implicitly assumed.
At NLO, the expressions for ${\cal C}_{3,11}$, ${\cal C}_{3,22}$
and ${\cal C}_{3,\rm SL}$
are taken from Ref.~\cite{Hokim:1983yt}, where the computation was done 
for three different final state masses, hence we can easily use these results for all $c$-quark decay modes.
For the coefficient ${\cal C}_{3,12}$ we use Ref.~\cite{Bagan:1994zd} for 
the $c \to s \bar d u$, $c \to d \bar s u$ and $c \to d \bar d u$ decay channels, 
while the result of Ref.~\cite{Krinner:2013cja} is used in the case of final state with two massive 
$s$-quarks, $c \to s \bar s u$.

Neglecting final state masses and approximating $|V_{ud}|^2 \approx 1$ the following expression was determined in 1991 \cite{Altarelli:1991dx}, i.e.
\begin{eqnarray}
    c_3^{\rm NLO} -  c_3^{\rm LO}& = & 8 \,\frac{\alpha_s}{4 \pi}
    \Biggl[ \,
    \underbrace{\left( \frac{25}{4} - \pi^2\right)}_{< 0}
     \underbrace{+(C_1^2 + C_2^2) \left( \frac{31}{4} - \pi^2\right)}_{< 0}
     \underbrace{- \frac 2 3 C_1 C_2 \left( \frac{7}{4} + \pi^2\right)}_{\geq 0}
    \, \Biggl] 
    \, .
    \label{eq:G3_LO_nom}
\end{eqnarray}
The first term on the r.h.s. of Eq.~(\ref{eq:G3_LO_nom}) stems from semileptonic decays and the next two terms from non-leptonic channels. For non-leptonic $b$-quark decays the NLO corrections are negative, while for charm quarks decays the third term
will dominate over the second one and the correction becomes positive. Moreover, there is a sizable enhancement of the $\alpha_s$-corrections in the non-leptonic $b$-quark decays due to finite charm quark mass effects \cite{Voloshin:1994sn,Bagan:1994zd,Bagan:1995yf,Krinner:2013cja} - the corresponding increase in charm quark decays is much less pronounced as $m_c^2/m_b^2 \approx 0.1 \gg m_s^2/m_c^2 \approx 0.005$.

The  numerical values for $\Gamma_3$ both in LO- and
NLO-QCD, for different $c$-quark mass schemes are shown in
Table~\ref{tab:Gamma_3}.
\begin{table}
\centering
\renewcommand{\arraystretch}{1.5}
   \begin{tabular}{|l||C{3cm}|C{3cm}|}
   \hline
     Mass scheme &  
     $\Gamma_3^{\rm LO}$  [ps$^{-1}$]  &  
     $\Gamma_3^{\rm NLO}$ [ps$^{-1}$] 
     \\
     \hline  \hline
     Pole ($m_c^{\rm} = 1.48$ GeV)  &
     $1.45_{-0.14}^{+0.17}$  &
     $1.52_{-0.16}^{+0.20}$ 
     \\
     \hline
     $\overline{\rm MS}$ (Eq.~\eqref{eq:c-quark-mass-Pole-to-MS}) &
     $0.69_{-0.09}^{+0.06}$  &
     $1.32_{-0.03}^{+0.06}$
     \\
     \hline
     Kinetic (Eq.~\eqref{eq:Pole-Kin-scheme})  &
     $0.97_{-0.11}^{+0.10}$  &
     $1.47_{-0.30}^{+0.27}$ 
     \\
     \hline
     $1S$ (Eq.~\eqref{eq:mc-1S}) &
     $1.25_{-0.13}^{+0.14}$  &
     $1.50_{-0.25}^{+0.31}$ 
     \\
     \hline
    \end{tabular} 
\caption{ Numerical values of 
$\Gamma_3^{\rm LO} = \Gamma_3^{(0)}$ and 
$\Gamma_3^{\rm NLO} = \Gamma_3^{(0)} + 
\alpha_s (m_c)/(4 \pi) \, \Gamma_3^{(1)}$ using different schemes for the $c$-quark mass. The uncertainties are obtained by varying the renormalisation scale $\mu_1$ between 1 GeV and 3 GeV.}
\label{tab:Gamma_3}
\end{table}
The range of NLO-QCD values from $1.3\, {\rm ps}^{-1}$ to $1.5\, {\rm ps}^{-1}$ 
for the  free charm-quark decay at NLO-QCD, is in good agreement with the experimental determinations in
Table~\ref{tab:exp-data}. Moreover we observe small ($<5 \% $) corrections
due to a non-vanishing strange quark mass. 
Interestingly the NLO-QCD result is affected by strong cancellations.
We in fact observe a suppression of the non-leptonic contribution
because of the opposite sign between the NLO corrections to the
diagrams describing QCD corrections to the upper left diagram of
Fig.~\ref{fig:HQE} and the QCD corrections intrinsic to the
$\Delta C=1$ Wilson coefficients. A further cancellation is 
then present between the semileptonic and the non-leptonic modes. This behaviour can be nicely read of the result in the Pole scheme:
\begin{eqnarray}
\Gamma_3 & = &  \Gamma_3^{\rm LO} \, \left[ 1 + \left(  \overbrace{\underbrace{1.84}_{\rm diag.} - \underbrace{0.74}_{\rm WC}}^{\rm NL} - \overbrace{0.67}^{\rm SL} \right) \, \frac{\alpha_s}{\pi} + {\cal O}\left(\frac{\alpha_s}{\pi}\right)^2 \right] \,. \label{eq:cancel1}
\end{eqnarray}
Expressing the pole mass in terms of a short distance mass like
the $\overline{\rm MS}$ scheme, an additional large NLO
correction arises from the conversion factor of $m_c^5$, which 
is the origin of the large shift between the LO and
the NLO values in the $\overline{\rm MS}$-, the kinetic and 
the $1S$-schemes, see Table \ref{tab:Gamma_3}.
We find e.g. in the $\overline{\rm MS}$ scheme
\begin{eqnarray}
\qquad \Gamma_3 & = &  \Gamma_3^{\rm LO} \, \left[ 1 + \left(\overbrace{ \underbrace{2.10}_{\rm diag.} - \underbrace{0.70}_{\rm WC}}^{\rm NL} - \overbrace{0.71}^{\rm SL} + \overbrace{6.66}^{\rm conv. fac.} \right) \, \frac{\alpha_s}{\pi}  + {\cal O}\left(\frac{\alpha_s}{\pi}\right)^2 \right] \, .
\label{eq:cancel2}
\end{eqnarray}
The corrections due to the mass conversion also make the overall semileptonic NLO term in the $\overline{\rm MS}$ scheme positive.

To get a first indication of the behaviour of the QCD series for the decay rate 
at higher orders,  we briefly discuss here the NNLO \cite{Biswas:2009rb} and NNNLO \cite{Fael:2020tow} 
corrections for the semileptonic $b$-quark decay and the preliminary NNLO-QCD corrections 
for the non-leptonic $b$-quark decay \cite{Czarnecki:2005vr}.
In the Pole mass scheme \cite{Fael:2020tow} 
\begin{equation}
\frac{\Gamma_3 (B \to X_c \ell \bar \nu_\ell)}{\Gamma_3^{\rm LO}(B \to X_c \ell \bar \nu_\ell)} 
= 1 
 - 1.72 \frac{\alpha_s(\mu)}{\pi} 
 - 13.09  \left(\frac{\alpha_s(\mu)}{\pi} \right)^2 \!
 \! - 163.3  \left(\frac{\alpha_s(\mu)}{\pi} \right)^3 \!
 =  1 - 0.12 -0.06 -0.05,
\label{eq:sl_NNNLO1}
\end{equation}
the semileptonic decay rate gets large negative corrections, 
and in the $\overline{\rm MS}$-scheme
\begin{equation}
\frac{\Gamma_3 (B \to X_c \ell \bar \nu_\ell)}{\Gamma_3^{\rm LO}(B \to X_c \ell \bar \nu_\ell)} 
 = 1 
 + 3.07  \frac{\alpha_s(\mu)}{\pi} 
 + 13.36 \left(\frac{\alpha_s(\mu)}{\pi} \right)^2 \!
 + 62.7  \left(\frac{\alpha_s(\mu)}{\pi} \right)^3 \!
 = 1 +0.21 +0.06 +0.02,
 \label{eq:sl_NNNLO2}
 \end{equation}
one finds  \cite{Fael:2020tow} sizable positive corrections - driven by the conversion of the quark mass from the Pole scheme to the $\overline{\rm MS}$-scheme and indicating again the importance of higher order perturbative corrections. For the semileptonic charm quark decay one finds even larger corrections\footnote{Results presented by Matteo Fael at CHARM 2020:
\\
https://indico.nucleares.unam.mx/event/1488/session/12/contribution/56/material/slides/0.pdf}, e.g. in the Pole mass scheme 
\begin{equation}
\frac{\Gamma_3 (D \to X \ell^+ \nu_\ell)}{\Gamma_3^{\rm LO}(D \to X \ell^+ \nu_\ell)} 
= 1 - 2.41  \frac{\alpha_s(\mu)}{\pi} 
 - 23.4  \left(\frac{\alpha_s(\mu)}{\pi} \right)^2 \!
 - 321.5  \left(\frac{\alpha_s(\mu)}{\pi} \right)^3
 \! = 1 - 0.25 - 0.26 - 0.37,
 \label{eq:sl_NNNLO3}
 \end{equation}
which clearly spoils the perturbative approach and makes the use of different quark
mass schemes mandatory.

Regarding the NNLO-QCD corrections to the non-leptonic decay rates, 
Ref.~\cite{Czarnecki:2005vr} presents a partial result (not resumming the large
logarithms, neglecting effects of the operator $Q_2^{q_1q_2}$ and assuming a 
vanishing charm quark mass) for the $b$-quark.
In the pole mass scheme the authors obtain
\begin{equation}
    \frac{\Gamma (b \to c \bar{u} d)}{3 \Gamma (b \to c e\bar{\nu} )}
    = 
     1 + 1 \, \frac{\alpha_s(\mu)}{\pi} 
 +67.1  \left(\frac{\alpha_s(\mu)}{\pi} \right)^2
 \, .
\end{equation}
It is interesting to note that from the coefficient of the $\alpha_s^2$ term,  67.1, a contribution of 54.7 stems 
from not summing large logarithms of the form $\ln (M_W / m_b)$ and $\ln^{2} (M_W / m_b)$. 
Using Eq.~(\ref{eq:sl_NNNLO1}) and the fact that, 
in the approximations of Ref.~\cite{Czarnecki:2005vr} the ratio between non-leptonic and semileptonic rate is equal to 3 at LO-QCD, yields
\begin{equation}
    \Gamma (b \to c \bar{u} d)
    = 
\Gamma^{\rm LO} (b \to c \bar{u} d) 
\left[ 1 - 0.7 \, \frac{\alpha_s(\mu)}{\pi} + 52.3  \left(\frac{\alpha_s(\mu)}{\pi} \right)^2\right] \, .
\end{equation}
For non-leptonic charm-quark decays the logarithms become even larger and we find 
that the coefficient of the $\alpha_s^2$ term increases from 52.3 to 91.2, which
clearly indicates the necessity of summing the large logarithms.
Neglecting final state masses seems to be well justified in the charm system.
In order to further estimate the effect of neglecting the operator $Q_2^{q_1q_2}$, we set in our code  $C_1 = 1$ and  
$C_2 = 0$ and we get in the Pole scheme 
\begin{eqnarray}
\frac{\Gamma_3^{\rm NL}}{\Gamma_3^{\rm NL, LO}} 
& = &
1 - 1.4 \, \frac{\alpha_s(\mu)}{\pi},
\end{eqnarray}
while the result with the full inclusion of the effective Hamiltonian yields a very different value of the QCD corrections
\begin{equation}
\frac{\Gamma_3^{\rm NL}}{\Gamma_3^{\rm NL, LO}} 
= 1 + 1.6 \, \frac{\alpha_s (\mu)}{\pi}. 
\end{equation}
All in all we conclude that, higher order corrections seem to be crucial for a reliable determination of~$\Gamma_3$. Despite being conceptually very interesting,
the result of Ref.~\cite{Czarnecki:2005vr} is not useful for phenomenological applications and a full NNLO determination of the non-leptonic decay rate
using the effective Hamiltonian would be highly desirable.
\subsection{Dimension-five Contribution}
\label{SubSec:SoftI}
\begin{figure}[t]
    \centering
    \includegraphics[scale=1.0]{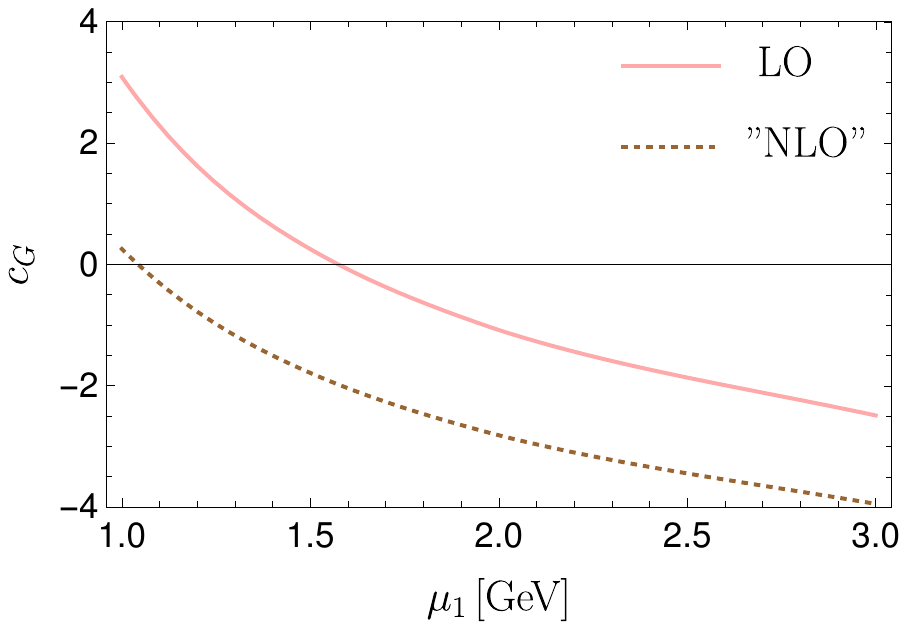}
    \caption{Scale dependence of the coefficient of the chromomagnetic operator.}
    \label{fig:cG}
\end{figure}

The first corrections to the free charm-quark decay arise at order $1/m_c^2$ and describe the effect of the kinetic and the chromomagnetic operators. Their matrix elements are parametrised by the two non-perturbative inputs $\mu_\pi^2$ and $\mu_G^2$, i.e.
\begin{eqnarray}
2 m_D \, \mu_\pi^2 (D) 
& = & 
- \langle D (p) |\bar{c}_v (i D_\mu)(i D^\mu) c_v | D(p) \rangle  \, ,
\\
2 m_D \, \mu_G^2 (D) 
& = & 
\langle D (p) | \bar{c}_v (i D_\mu)(i D_\nu) (-i \sigma^{\mu \nu}) c_v | D(p) \rangle 
\, ,
\end{eqnarray}
with $\sigma^{\mu \nu} = (i/2) [\gamma^\mu, \gamma^\nu]$.
Both the operators receive a contribution from the expansion of the dimension-three matrix element $\langle D (p) |\bar{c}_v c_v | D(p) \rangle$~\cite{Dassinger:2006md}. 
However, the chromomagnetic operator receives further contributions due to
the expansion of the short distance coefficient $c_3$ and of 
the quark-propagator in the external gluon field~\cite{Blok:1992he, Blok:1992hw, Lenz:2020oce} 
- see the second diagram on the top line of Fig.~\ref{fig:HQE}.
Hence, at order $1/m_c^2$, we can schematically write
\begin{equation}
    \Gamma_5 \frac{\langle {\cal O}_5 \rangle}{m_c^2} 
    = \Gamma_0 
    \left[
    c_{\mu_\pi} \frac{\mu_\pi^2}{m_c^2} 
    + c_G \, \frac{\mu_G^2}{m_c^2} 
    \right] .
    \label{eq:Gamma_5}
\end{equation}
The coefficient of the kinetic operator is related 
to the dimension-three contribution\footnote{ 
Since the dimension-5 contribution for non-leptonic modes is known only 
at LO in QCD, we use the dimension-three coefficient $c_3$ just at  LO-QCD for $c_{\mu_\pi}$.}, 
$c_{\mu_\pi}= - c_3^{(0)}/2$,
and the chromomagnetic coefficient $c_G$ can be  decomposed as
\begin{equation}
    c_G = 
    3 \, C_1^2    \, {\cal C}_{G,11} 
 +  2 \, C_1 C_2  \, {\cal C}_{G,12} 
 +  3 \, C_2^2    \, {\cal C}_{G,22} + {\cal C}_{G,SL} \, ,
\label{eq:CG}
\end{equation}
where again a summation over all the modes is assumed.
The individual contributions ${\cal C}_{G,nm}$ for non-leptonic modes 
can be found e.g. in the Appendix of Ref.~\cite{Lenz:2020oce}. 
In the latter reference, the coefficients of the chromomagnetic operator 
were determined for the non-leptonic $B$-meson decays, however,
since there are no IR-divergences at this order, it is straightforward to obtain the corresponding results for the charm-sector, namely by replacing 
$m_b \to m_c$, $m_c \to m_s$, etc.
For the semileptonic decay $c \to s \mu^+  \nu_\mu$, the expression for two different mass parameters $z_s \not =  0 \not =   z_\mu $ can be found in the Appendix of~Ref.~\cite{Mannel:2017jfk}.\footnote{ Since $m_s \approx m_\mu \approx 100 \, {\rm MeV} \ll m_c$, in principle one can safely set $z_s = z_\mu$ and use the non-leptonic expressions 
for the semileptonic modes, e.g. $c \to s \bar s u$ for $c \to s \mu \bar \nu_\mu$ 
by setting $N_c = 1, \, C_1 = 1, \, C_2 = 0$. }
By neglecting the strange and muon masses and by considering only the dominant CKM modes, the result for $c_G$ becomes very compact, i.e.
\begin{equation}
    c_G \approx -|V_{ud}|^2  \left[
    \frac92 \Bigl( C_1^2   + C_2^2 \Bigr)  + 19 \, C_1 C_2  \right]  - 3.
    \label{eq:cG}
\end{equation}
Because of the large coefficient in front of $C_1 C_2$ 
and of its negative value, Eq.~(\ref{eq:cG}) can be affected by cancellations. 
In Fig.~\ref{fig:cG} we plot $c_G$ in Eq.~(\ref{eq:CG}), as a function of the
renormalisation scale $\mu_1$ while in Table~\ref{tab:cG} we list the numerical result for some reference values of $\mu_1$.
For $c_G$ a change of sign occurs in the region between $1$ and $2$ GeV -- 
leading to a large uncertainty due to scale variation.
Note, that the ``NLO'' result shown in Fig.~\ref{fig:cG}
and Table~\ref{tab:cG} only includes QCD corrections due to the 
$ \Delta C= 1$ Wilson coefficients. 
A complete calculation of the NLO-QCD corrections to $c_G$ 
is still missing (these corrections are only known for the semileptonic case) and would be very desirable in order to reduce the huge scale dependence.
The numerical values of the non-perturbative parameters 
$\mu_\pi^2$ and $\mu_G^2$ will be discussed in Sections~\ref{sec:mu_pi}~and~\ref{sec:mu_G}.

\begin{table}[t]
\centering
\renewcommand{\arraystretch}{1.25}
   \begin{tabular}{|c||C{1.3cm}|C{1.3cm}|C{1.3cm}|C{1.3cm}|C{1.3cm}|C{1.3cm}|}
   \hline
   $\mu_1$ [GeV] 
   & 1 & 1.27 & 1.36 & 1.44 & 1.48 & 3 
   \\
   \hline 
   \hline
   $c_G^{\rm NL} ({\rm LO})$ &
   6.20 & 
   4.34 &
   3.91 &
   3.58 &
   3.43 &
   0.62 
   \\
   $c_G^{\rm SL} ({\rm LO})$ &
   -3.11 &
   -3.11 &
   -3.11 &
   -3.11 &
   -3.11 &
   -3.11 
   \\
   \hline
   $c_G ({\rm LO})$ &
   3.09 &
   1.23 &
   0.80 &
   0.47 &
   0.32 &
   -2.49 
   \\
   \hline \hline 
    $c_G  ({\rm "NLO"})$ &
    0.25 &
   -1.06 &
   -1.37 &
   -1.62 &
   -1.74 &
   -3.95
   \\
   \hline
  \end{tabular} 
  \caption{Comparison of the coefficients $c_G^{\rm SL}$, $c_G^{\rm NL}$,
  and $c_G = c_G^{\rm SL} + c_G^{\rm NL}$ for different values of the renormalisation scale $\mu_1$ at LO and ``NLO'', setting for reference $m_c = 1.5$ GeV.}
  \label{tab:cG}
\end{table}

\subsection{Dimension-six Two-Quark Operator Contribution
}
\label{SubSec:Darwin-coefficient}
 By determining higher order $1/m_c$ corrections in the expansion of the quark-propagator, in the expansion of the  the matrix elements of mass dimension-three and  mass dimension-five and  in the expansion of the corresponding short-distance coefficients, see e.g. Refs.~\cite{Novikov:1983gd, Blok:1992he, Blok:1992hw, Lenz:2020oce} for details, one finds the dimension-six contribution to $\Gamma(D)$, which can be compactly written as
\begin{eqnarray}
\Gamma_6 \frac{\langle {\cal O}_6 \rangle}{m_c^3}
=
\Gamma_0 \, c_{\rho_D} \frac{\rho_D^3}{m_c^3}
 \, ,
\end{eqnarray}
with the matrix element of the Darwin operator given by
\footnote{Note that with the given definition for the dimension-six two-quark operators,
in terms of full covariant derivatives, the contribution of the spin-orbit operator
to the decay width vanishes.}
\begin{eqnarray}
2 m_D \, \rho_D^3 (D) & = & 
\langle D (p) | \bar{c}_v (i D_\mu)(i v \cdot D) (i D^\mu) c_v | D (p) \rangle
\, .
\end{eqnarray}
 The coefficient $c_{\rho_D}$ can be decomposed into
\begin{equation}
    c_{\rho_D} = 
    3 \, C_1^2 \, {\cal C}_{\rho_D,11} 
 +  2 \, C_1 C_2 \, {\cal C}_{\rho_D,12} 
 +  3 \, C_2^2 \, {\cal C}_{\rho_D,22} 
 +           {\cal C}_{\rho_D,SL} 
    \, ,
\label{eq:crhoD}
\end{equation}
including both non-leptonic and semileptonic contributions.
For $B$-mesons decays, the non-leptonic coefficients were computed recently 
in Refs.~\cite{Lenz:2020oce, Mannel:2020fts, Moreno:2020rmk}.
In order to determine the corresponding expressions for the charm system, some subtleties have to be considered.
In $b$-quark decays, one assumes
$m_b \sim m_c \gg \Lambda_{\rm QCD}$, and the coefficient of the Darwin operator for the semileptonic $b \to c \ell \bar \nu_\ell$ decays is a finite function of $\rho = m_c^2/m_b^2$, which however diverges in the limit $\rho \to 0$, i.e. in correspondence of the $b \to u \ell \bar \nu_\ell$ transitions.
This is due to the fact that, the radiation of a soft gluon off a massless quark
leads to IR singularities at dimension-six. In non-leptonic $b$-quark decays,
one has to further deal with the emission of a soft gluon from the internal 
light $u$-, $d$-, and $s$-quark lines. The corresponding IR divergences are of 
the form $\log (m_q/m_b)$, for $q = u,d,s$, and are removed by taking into account the mixing between the four-quark operators with external $q$ quarks and the Darwin operator under renormalisation,
for details see e.g.~Refs.~\cite{Breidenbach:2008ua, Bigi:2009ym,Lenz:2020oce, Mannel:2020fts}.
Because of $m_c \gg m_s \sim \Lambda_{\rm QCD}$, it follows that one cannot trivially generalise the results from the $b$- to the $c$-sector, i.e.\ by only replacing $m_b \to m_c$, $m_c \to m_s$, etc., since there are further contributions due to the mixing of four-quark operators with external $s$-quarks which must be additionally included.
Specifically, this leads to a modification of the coefficients proportional to $C_1^2$ and $C_1 C_2$.
Using the same procedure as discussed in Ref.~\cite{Lenz:2020oce}, 
we have recomputed the coefficients of the Darwin operator required for the study of $D$-meson decays. 
The analytic expressions for ${\cal C}_{\rho_D,nm}$, including 
the full $s$-quark mass dependence, however finite in the limit $m_s \to 0$, are presented in Appendix~\ref{Appendix-B}
for all non-leptonic modes. 
To obtain the corresponding expression for ${\cal C}_{\rho_D,SL}$, 
it is sufficient to set in the results for the non-leptonic decays
$N_c =1$, $C_1 = 1$, $C_2 = 0$ and $z_s = z_\mu$ for the $c \to s \mu^+ \nu_\mu$ mode.  In particular, we confirm the results in Ref.~\cite{Fael:2019umf}.

\begin{figure}[th]
    \centering
    \includegraphics[scale=1.0]{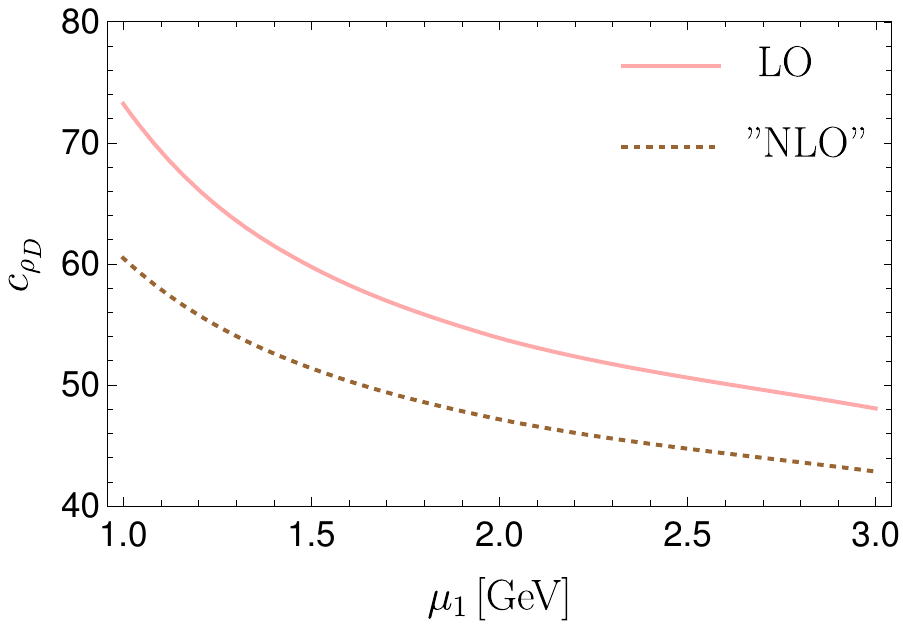}
    \caption{Scale dependence of the coefficient of the Darwin operator.}
    \label{fig:c-rho-D}
\end{figure}

Again, by neglecting the strange and muon masses and 
by considering only the dominant CKM modes, one finds
\begin{equation}
    c_{\rho_D} \approx 
     |V_{ud}|^2 \, \left(18 \, C_1^2 \, -\frac{68}{3} \, C_1 C_2 
     + 18 \, C_2^2\right) + 12        
    \, .
    \label{eq:Darwin-coeff}
\end{equation}
It is interesting to note that in this
combination all terms have the same sign and no
cancellations arise.
In Fig.~\ref{fig:c-rho-D} we show the dependence of the function $c_{\rho_D}$ in Eq.~(\ref{eq:crhoD}) on the
renormalisation scale~$\mu_1$ and in Table~\ref{tab:c-rho-D} 
we quote the numerical result for some reference values of $\mu_1$.
The determination of the matrix element of the Darwin operator will be discussed in Section~\ref{sec:rho_D}.

\begin{table}[th]
\centering
\renewcommand{\arraystretch}{1.25}
   \begin{tabular}{|c||C{1.3cm}|C{1.3cm}|C{1.3cm}|C{1.3cm}|C{1.3cm}|C{1.3cm}|}
   \hline
   $\mu_1$ [GeV] 
   & 1 & 1.27 & 1.36 & 1.44 & 1.48 & 3 
   \\
   \hline 
   \hline
   $c_{\rho_D}^{\rm NL} ({\rm LO})$ &
   60.6 & 
   51.7 &
   49.6 &
   48.1 &
   47.5 &
   35.4
   \\
   $c_{\rho_D}^{\rm SL} ({\rm LO})$ &
   12.6 &
   12.6 &
   12.6 &
   12.6 &
   12.6 &
   12.6 
   \\
   \hline
   $c_{\rho_D} ({\rm LO})$ &
   73.2 &
   64.3 &
   62.3 &
   60.8 &
   60.1 &
   48.1 
   \\
   \hline \hline 
   $c_{\rho_D}  ({\rm "NLO"})$ &
   60.5 &
   54.5 &
   53.1 &
   52.1 &
   51.6 &
   42.8
   \\
   \hline
  \end{tabular} 
  \caption{Numerical values of $c_{\rho_D}^{\rm SL}$, $c_{\rho_D}^{\rm NL}$,
  and $c_{\rho_D} = c_{\rho_D}^{\rm SL} + c_{\rho_D}^{\rm NL}$ for different values of the renormalisation scale $\mu_1$ at LO and "NLO" with 
  $\mu_0 = m_c = 1.5$~GeV.}
  \label{tab:c-rho-D}
\end{table}

\subsection{Dimension-six Four-Quark Operator Contribution }
\label{sec:dim-6-4q}
\begin{figure}[th]
\centering
\includegraphics[scale=0.5]{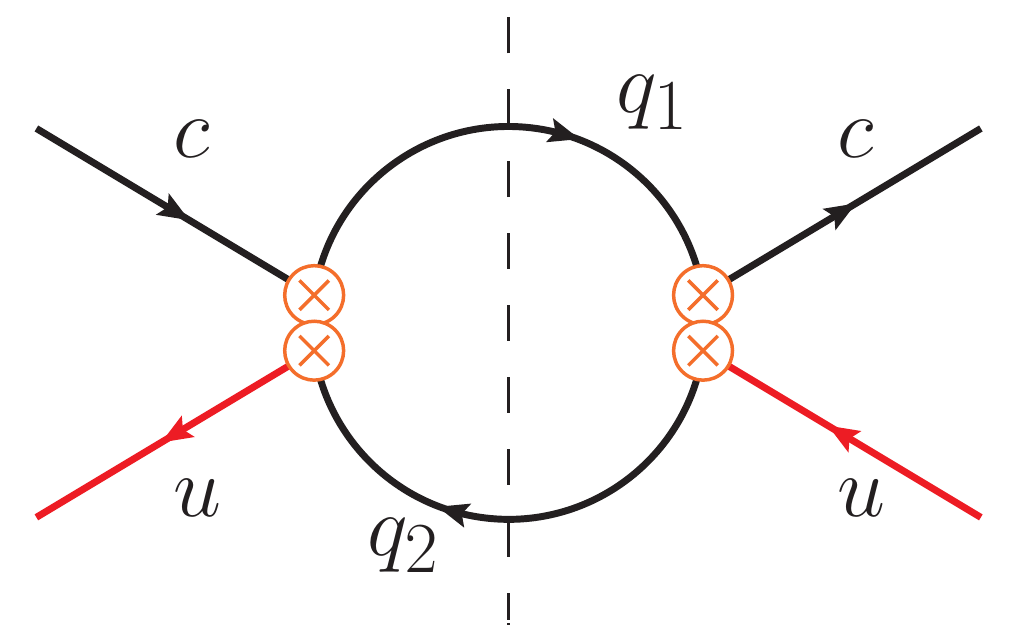}
\includegraphics[scale=0.5]{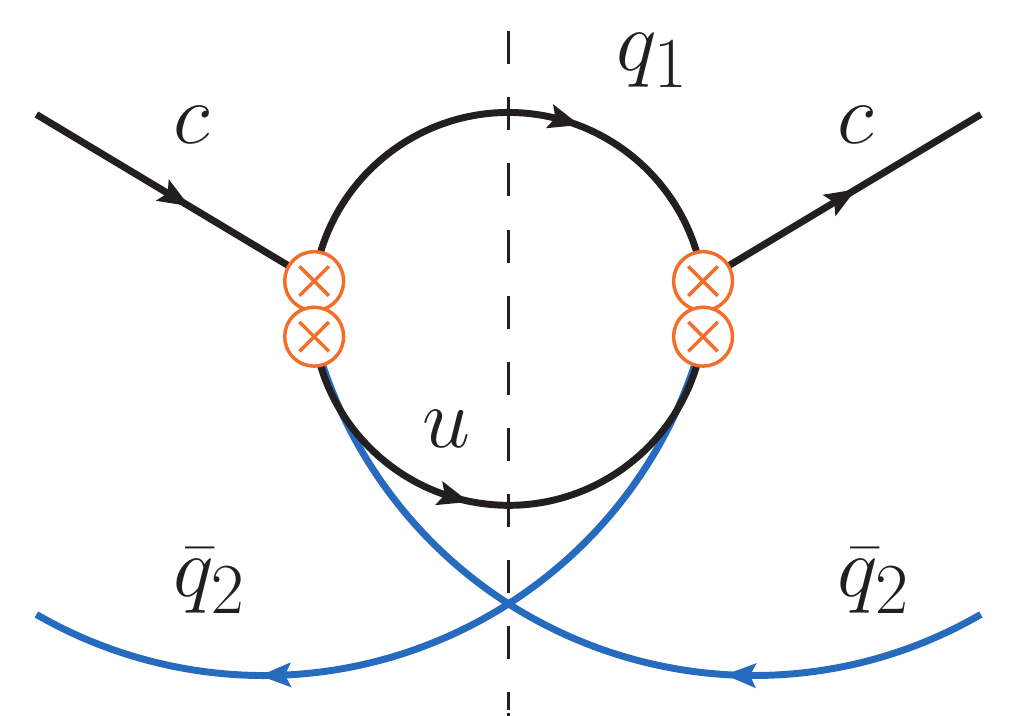}
\includegraphics[scale=0.5]{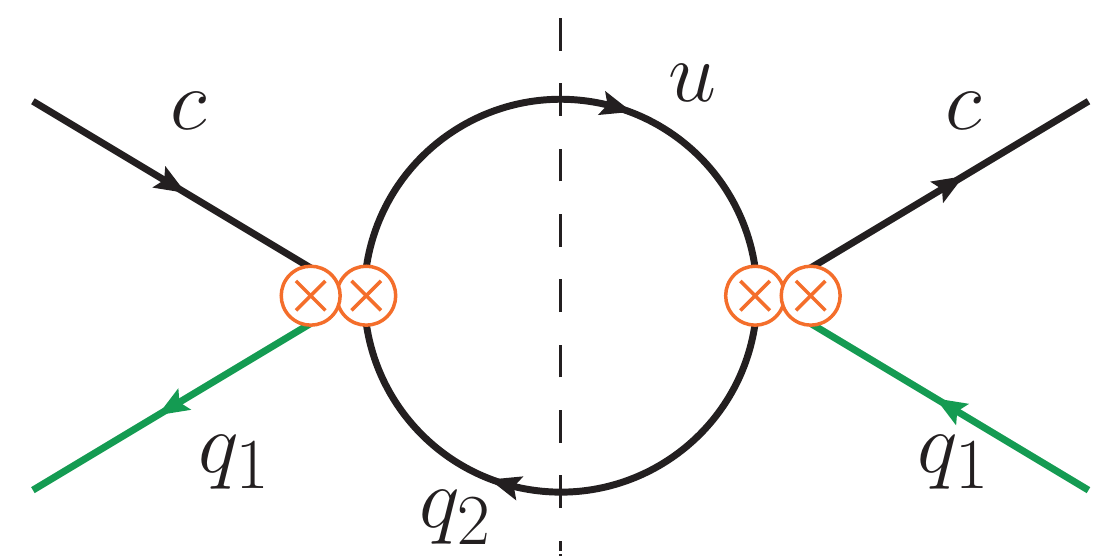}
\caption{Spectator quark effects in the HQE expansion: WE (left), PI (middle) and WA (right).}
\label{fig:WE-PI-WA}
\end{figure}

The perturbative coefficients in Eq.~(\ref{eq:HQE}) considered so far are
independent of the spectator quark in the $D$ meson, in fact its effect appears only in the corresponding matrix elements of the dimension-five and dimension-six operators.
Starting at order $1/m_c^3$, there are also one-loop contributions, c.f.~$\tilde{\Gamma}_6$ in Eq.~(\ref{eq:HQE}), in which the spectator quark is directly involved. These correspond respectively to the weak exchange (WE), Pauli interference (PI) and weak annihilation (WA) diagrams, depicted in Fig.~\ref{fig:WE-PI-WA} \footnote{In the case of semileptonic decays, only the WA topology can contribute.}. Note that compared to the terms discussed above, these contributions imply
a phase space enhancement factor of $16 \pi^2$. The corresponding $ \Delta C = 0$ four quark operators of dimension-six are
\footnote{Sometimes, we will use the short-hand notation 
$O_i^q$, $i = 1,2,3,4$ assuming 
$O_3^q \equiv T_1^q$, $O_4^q \equiv T_2^q$. }: 
\begin{eqnarray}
  O_1^q  
  & = &  
  (\bar{c}\,\gamma_\mu (1-\gamma_5) q)\,(\bar{q}\,\gamma^\mu (1-\gamma_5) c) ,
  \label{eq:O1} \\[1mm]
  O_2^q  
  & = & 
  (\bar{c} (1 - \gamma_5) q)\,(\bar{q} (1 + \gamma_5) c) ,
  \label{eq:O2} \\[1mm]
  T_1^q  
  & = & 
  (\bar{c} \, \gamma_\mu (1-\gamma_5) T^A q) 
  \, (\bar{q} \, \gamma^\mu (1-\gamma_5) T^A  c) ,
  \label{eq:T1} \\[1mm]
  T_2^q 
  & = & 
  (\bar{c} (1-\gamma_5) T^A q)\,(\bar{q}(1 + \gamma_5) T^A c) 
  \label{eq:T2},
\end{eqnarray}
where $T^A$ is a colour matrix and a summation over colour indices is implied.
The parameterisation of the matrix elements of the operators in Eqs.~(\ref{eq:O1}) - (\ref{eq:T2}) in QCD  is given in Appendix~\ref{Appendix-C}. However, by evaluating the matrix elements in the framework of the HQET, one obtains the following set of operators, i.e. \footnote{ 
Note that all quantities defined in HQET are labelled by a tilde, contrary to those in QCD.}
\begin{eqnarray}
  {\tilde O}_1^q  
  & = &  
  (\bar{h}_v\,\gamma_\mu (1-\gamma_5) q)\,(\bar{q}\,\gamma^\mu (1-\gamma_5) h_v) ,
  \label{eq:O1-HQET} \\[1mm]
  {\tilde O}_2^q  
  & = & 
  (\bar{h}_v (1 - \gamma_5) q)\,(\bar{q} (1 + \gamma_5) h_v) ,
  \label{eq:O2-HQET} \\[1mm]
  {\tilde T}_1^q  
  & = & 
  (\bar{h}_v \, \gamma_\mu (1-\gamma_5) T^A q) 
  \, (\bar{q} \, \gamma^\mu (1-\gamma_5) T^A  h_v) ,
  \label{eq:T1-HQET} \\[1mm]
  {\tilde T}_2^q 
  & = & 
  (\bar{h}_v (1-\gamma_5) T^A q)\,(\bar{q}(1 + \gamma_5) T^A h_v) 
  \label{eq:T2-HQET},
\end{eqnarray}
here $h_v$ denotes the HQET field defined by Eqs.
(\ref{eq:phase-redef}), (\ref{eq:cv-hv-relation}). 
The matrix elements of these operators in HQET are parameterised as
\begin{eqnarray}
\langle {D}_q | {\tilde O}_i^q \, | {D}_q \rangle 
& = & 
F^2(m_c) \, m_{D_q} \, \tilde B_i^q 
\, = \, 
f_{D_q}^2 m_{D_q}^2 
\left(1 + \frac{4}{3} \frac{\alpha_s (m_c)}{ \pi} \right)  \tilde B_i^q, 
\label{eq:ME-dim-6-HQET-q-q}
\\[2mm]
\langle {D}_q | {\tilde O}_i^{q^\prime} | {D}_q \rangle 
& = & 
F^2 (m_c) \, m_{D_q} \, \tilde \delta^{q^\prime q}_i
\, = \, 
f_{D_q}^2 m_{D_q}^2 
\left(1 + \frac{4}{3} \frac{\alpha_s (m_c)}{\pi} \right)
\tilde \delta^{ q^\prime q}_i, \qquad q \not = q^\prime \,,
\label{eq:ME-dim-6-HQET-q-q-prime}
\end{eqnarray}
where $q, q^\prime = u, d, s$,
${\tilde B}_i^q$ denote the Bag parameters in HQET,
with ${\tilde B}_{1,2}^q$ corresponding to the colour-singlet operators, and 
${\tilde B}_{3,4}^q \equiv \tilde \epsilon_{1,2}^q$ to the colour-octet ones, and 
$F (\mu)$ and $f_{D_q}$ are the HQET and QCD decay constants
defined, respectively, as \footnote{The subscript `QCD' or `HQET' on the states is usually omitted, however for clarity it is specified in the definition of the decay constant.}
\begin{equation}
\langle 0 | \bar q \gamma^\mu \gamma_5 c|{D_q (p)} \rangle_{\rm QCD} 
= i f_{D_q} \, p^\mu,
\label{eq:DecayConstQCD}
\end{equation}
with $p = m_D v$, and 
\begin{equation}
\langle 0 | \bar q \gamma^\mu \gamma_5 h_v| {D}_q (v) \rangle_{\rm HQET} = i \, F (\mu) \, \sqrt{m_{D_q}} \, v^\mu.
\label{eq:DecayConstHQET}
\end{equation}
The relation between $f_D$ and $F$ up to $\alpha_s$ and $1/m_c$ corrections, is given e.g. in Refs.~\cite{Neubert:1992fk, Kilian:1992cj}. At the scale $\mu = m_c$, it reads 
\begin{eqnarray}
f_D = \frac{F (m_c)}{\sqrt{ m_D}} 
\left(1 - \frac{2}{3} \frac{\alpha_s (m_c)}{\pi} 
+ \frac{G_1 (m_c)}{m_c} + 6 \, \frac{G_2 (m_c)}{m_c}
- \frac 1 2 \frac{\bar \Lambda}{m_c} \right),
\label{eq:decay_constant-conversion}
\end{eqnarray}
where $\bar \Lambda = m_D - m_c$, and the parameters $G_1$ and $G_2$ characterise matrix elements of  non-local operators.
Note that in Eqs.~\eqref{eq:ME-dim-6-HQET-q-q}, \eqref{eq:ME-dim-6-HQET-q-q-prime}, in expressing the HQET decay constant in terms of the QCD one, we have included only the $\alpha_s$ corrections -- which  become part of NLO dimension-six contribution -- but not the $1/m_c$ ones. 
The latter, as it will be explained in detail in Section~\ref{sec:dim-7-4q},
can be absorbed in the contribution of some of the dimension-seven operators in HQET. 

\begin{figure}[t]\centering
\includegraphics[scale = 0.5]{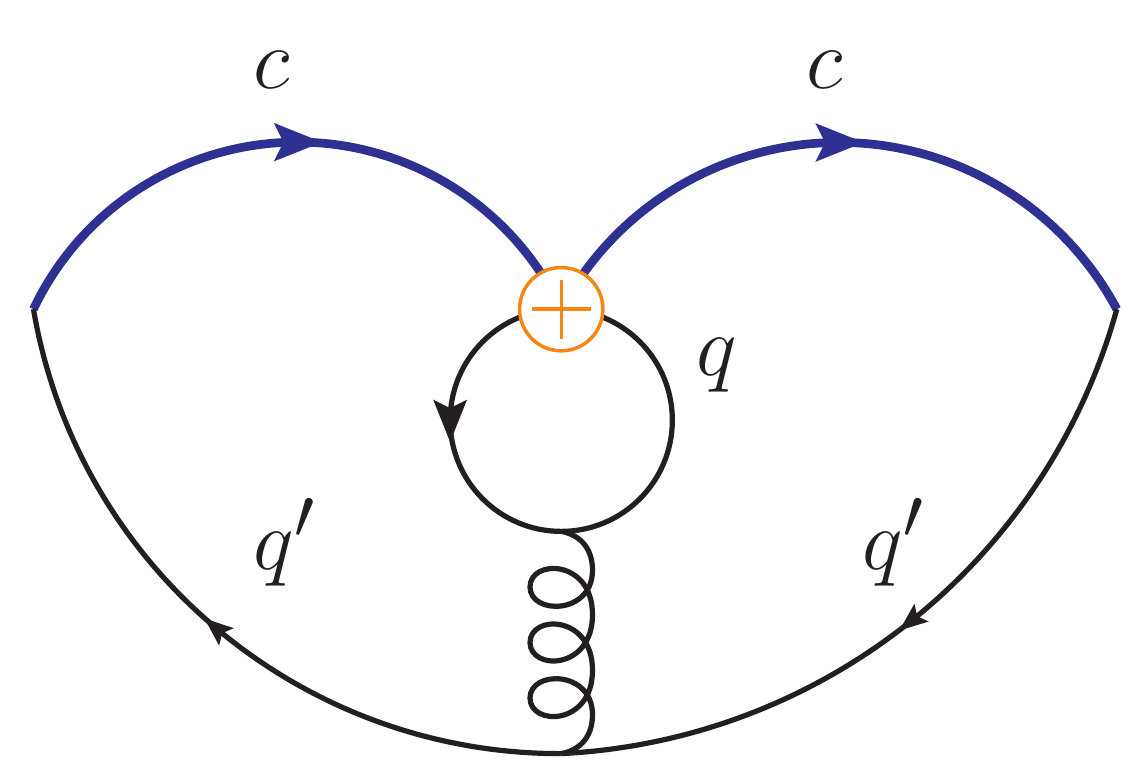}
\caption{Diagram describing the eye-contractions.}
\label{fig:eye-contractions}
\end{figure}

In vacuum insertion approximation (VIA), the Bag parameters of the colour-singlet operators are equal to one, ${\tilde B}_{1,2}^q = 1$, and the Bag parameters of the colour-octet
operators vanish, $\tilde \epsilon^q_{1,2} = 0$.
Note that throughout this work we assume  isospin symmetry, i.e. 
\begin{equation}
{\tilde B}_i^u = {\tilde B}_i^d.
\end{equation}
The quantities $\tilde \delta^{ q^\prime q}_i$ in Eq.~\eqref{eq:ME-dim-6-HQET-q-q-prime} describe the so-called eye-contractions, see Fig.~\ref{fig:eye-contractions},
and characterize "subleading" (compared to the large Bag parameters) 
effects in the non-perturbative matrix elements -- in VIA 
all eye-contractions vanish i.e.\ $\tilde \delta^{q^\prime q}_i = 0$. However, beyond VIA, the matrix elements of the four-quark operators with external  $q^\prime$ quarks differ from zero even when the spectator quark $q$ in the $D_q$ meson does not coincide with the quark $q^\prime$, as reflected by $\tilde \delta^{q^\prime q}_i$ in Eq.~\eqref{eq:ME-dim-6-HQET-q-q-prime}.
 Note that in our notation the eye-contractions with $q = q^\prime$,
are in fact included in the Bag parameters $\tilde B_i^q$.
And again, due to isospin symmetry we will use:
$$
\tilde \delta^{u q^\prime}_i = \tilde \delta^{d q^\prime}_i, 
\qquad 
\tilde \delta^{q^\prime u}_i = \tilde \delta^{q^\prime d}_i, 
\qquad q^\prime = u, d, s \, .
$$
The Bag parameters ${\tilde B}_i^q$ and $ \tilde \delta_i^{q q^\prime}$ 
have been determined using HQET sum rules, specifically
the Bag parameters $ {\tilde B}_i^q$ for the $D^{+,0}$ 
mesons were calculated in Ref.~\cite{Kirk:2017juj}, while
corrections due to the strange quark mass as well as the
contribution of the eye-contractions, see
Fig.~\ref{fig:eye-contractions},
have been computed for the first time in Ref.~\cite{King:2020}.
The numerical values of the Bag parameters will be briefly
discussed in Section~\ref{sec:Bag-par} and they are summarised in
Appendix~\ref{Appendix-A}.

By considering only the dominant CKM modes and by neglecting 
the effect of the eye-contractions, at LO-QCD and at dimension-six, 
the contributions of four-quark operators to the $D$-mesons decay rate
reads
\begin{eqnarray}
 16 \pi^2 \, \tilde{\Gamma}^{D^0}_6 \frac{\langle {\tilde O}_6\rangle^{D^0}}{m_c^3}
 & = & 
 \Gamma_0 |V_{ud}^*|^2 \, 
 16 \pi^2 \frac{M_{{D^0}} f_{D^0}^2}{m_c^3}\, (1 - z_s)^2 
 \nonumber \\
  & & 
  \left\{
  \left(\frac 1 3 C_1^2 + 2 \, C_1 C_2 + 3 \, C_2^2 \right) 
  \left[
  ({\tilde B}_2^{u} -  {\tilde B}_1^{u}) + z_s
  \left(2 {\tilde B}_2^{u} - \frac{{\tilde B}_1^{u}}{2}\right) 
  \right]
  \right.
  \nonumber
  \\
  & & \hspace{3.4cm}
  \left.
  + 2 \, C_1^2 \left[ ({\tilde \epsilon}_2^{u} -  \tilde \epsilon_1^{u}) 
  + z_s 
  \left(2 \, {\tilde \epsilon}_2^{u} - \frac{\tilde \epsilon_1^{u}}{2} \right) \right] \right\}  \, ,
  \label{eq:dim-6-4q-LO-D0} \\ 
  16 \pi^2 \, \tilde{\Gamma}^{D^+}_6 \frac{\langle {\tilde O}_6 \rangle^{D^+}}{m_c^3}
  & = & 
  \Gamma_0 |V_{ud}^*|^2 \, 16 \pi^2 \frac{M_{{D^+}} f_{D^+}^2}{m_c^3} (1-z_s)^2
  \nonumber
  \\
  & &
  \left\{
  \left( C_1^2 + 6\,  C_1 C_2 + C_2^2 \right)  {\tilde B}_1^{d}
  + 6 \left( C_1^2 +  C_2^2 \right)  \tilde \epsilon_1^{d}  
  \right\} 
  \, ,
  \label{eq:dim-6-4q-LO-Dp} \\
  16 \pi^2 \, \tilde{\Gamma}^{D_s^+}_6 \frac{\langle {\tilde O}_6 \rangle^{D_s^+}}{m_c^3}
  & = & 
  \Gamma_0 |V_{ud}^*|^2  \, 16 \pi^2
  \frac{M_{{D_s^+}}f_{D_s^+}^2}{m_c^3}
  \nonumber
  \\
  & & \left\{
  \left( 3 \, C_1^2 + 2 \, C_1 C_2 + \frac 1 3 C_2^2 + \frac{2}{|V_{ud}^*|^2 }  \right) 
  \left({\tilde B}_2^{s} -  {\tilde B}_1^{s} \right)
  + 2 \, C_2^2 \left({\tilde \epsilon}_2^{s} -  \tilde \epsilon_1^{s} \right) 
  \right\} \, ,
  \label{eq:dim-6-4q-LO-Ds}
\end{eqnarray}
respectively, for the WE, PI and WA topologies. 
Note that in the latter, the contribution due to the muon mass in the semileptonic decay $c \to s \mu^+ \nu_\mu $ has been neglected.
Here some interesting numerical effects are arising.
First, in the charm system, one expects that the contribution due to the spectator quark is of similar size compared to 
      the leading term
      $\Gamma_3$ in the HQE, unless some additional cancellations are present. Using the pole mass $m_c^{\rm Pole} = 1.48$ GeV and Lattice QCD values for the decay constants~\cite{Aoki:2019cca} we roughly obtain that
      \begin{eqnarray}
          16 \pi^2
        \frac{M_{{D^0}}f_{D^0}^2}{m_c^3} & = & {4.1} \approx 
        {\cal  O} (c_3) \, ,
        \\
        16 \pi^2
         \frac{M_{{D_s^+}}f_{D_s^+}^2}{m_c^3}
          & = & { 6.0}  \approx 
        {\cal O} (c_3) \, .
      \end{eqnarray}
This result has led the authors of Ref.~\cite{Mannel:2021uoz} to propose a different
ordering for the HQE series in the charm sector.
However, to investigate further the size of four-quark contributions, 
we consider the combinations of Wilson coefficients that appear in 
Eqs.~\eqref{eq:dim-6-4q-LO-D0} - \eqref{eq:dim-6-4q-LO-Ds}, i.e.
\begin{align}
      & C_{\rm WE}^S = \frac 1 3 C_1^2 + 2 \, C_1 C_2 + 3 \, C_2^2, 
      & & C_{\rm WE}^O =  2 \, C_1^2,
      \label{eq:C-WE} \\
      & C_{\rm PI}^S = C_1^2 + 6 \, C_1 C_2 +  C_2^2,
      & & C_{\rm PI}^O = 6 \, (C_1^2  + C_2^2)  \, ,
      \label{eq:C-PI} \\
      & C_{\rm WA}^S = 3 \, C_1^2 + 2 \, C_1 C_2 + \frac 1 3 C_2^2, 
      & & C_{\rm WA}^O = 2 \, C_2^2  \, ,
      \label{eq:C-WA}
\end{align}
where the superscript $S$ and $O$ refers to coefficient in front of 
the colour-singlet and colour-octet Bag parameters, respectively. 
A comparison of these combinations for different values of the 
renormalisation scale $\mu_1$ is
shown in Table~\ref{tab:C-WE-PI-WA}.

\begin{table}[ht]
\centering
\renewcommand{\arraystretch}{1.3}
   \begin{tabular}{|c||C{1.3cm}|C{1.3cm}|C{1.3cm}|C{1.3cm}|C{1.3cm}|C{1.3cm}|}
   \hline
   $\mu_1$ [GeV] & 1 & 1.27 & 1.36 & 1.44 & 1.48 & 3 
   \\
   \hline 
   \hline
   $C_{\rm WE}^{S} (\rm LO)$ &
   $\phantom{-}0.09 $ & 
   $\phantom{-}0.03 $ &
   $\phantom{-}0.02 $ &
   $\phantom{-}0.02 $ &
   $\phantom{-}0.01 $ &
   $\phantom{-}0.01 $ 
   \\
   $C_{\rm WE}^{S} (\rm NLO)$ &
   $-0.03 $ &
   $-0.03 $ &
   $-0.03 $ &
   $-0.02 $ &
   $-0.02 $ &
   $\phantom{-}0.04$
   \\
   \hline
   $C_{\rm WE}^{O} (\rm LO)$ &
   $\phantom{-}3.57 $ & 
   $\phantom{-}3.24 $ &
   $\phantom{-}3.16 $ &
   $\phantom{-}3.11 $ &
   $\phantom{-}3.08 $ &
   $\phantom{-}2.63 $
   \\
   $C_{\rm WE}^{O} (\rm NLO)$ &
   $\phantom{-}3.11 $ & 
   $\phantom{-}2.89 $ &
   $\phantom{-}2.83 $ &
   $\phantom{-}2.79 $ &
   $\phantom{-}2.77 $ &
   $\phantom{-}2.44 $
   \\
   \hline
   $C_{\rm PI}^{S} (\rm LO)$ &
   $-2.80 $ & 
   $-2.12 $ &
   $-1.96 $ &
   $-1.85 $ &
   $-1.79 $ &
   $-0.79 $
   \\
   $C_{\rm PI}^{S} (\rm NLO)$ &
   $-1.74$ & 
   $-1.28 $ &
   $-1.16 $ &
   $-1.08 $ &
   $-1.04 $ &
   $-0.27 $
   \\
   \hline
   $C_{\rm PI}^{O} (\rm LO)$ &
   $\phantom{-}13.0 $ & 
   $\phantom{-}11.4 $ &
   $\phantom{-}11.0 $ &
   $\phantom{-}10.7 $ &
   $\phantom{-}10.6 $ &
   $\phantom{-}8.50 $
   \\
   $C_{\rm PI}^{O} (\rm NLO)$ &
   $\phantom{-}10.6 $ & 
   $\phantom{-}9.55 $ &
   $\phantom{-}9.31 $ &
   $\phantom{-}9.13 $ &
   $\phantom{-}9.05 $ &
   $\phantom{-}7.60 $
   \\
   \hline
   $C_{\rm WA}^{S} (\rm LO)$ &
   $\phantom{-}3.82 $ & 
   $\phantom{-}3.61 $ &
   $\phantom{-}3.56 $ &
   $\phantom{-}3.53 $ &
   $\phantom{-}3.51 $ &
   $\phantom{-}3.24 $
   \\
   $C_{\rm WA}^{S} (\rm NLO)$ &
   $\phantom{-}3.57 $ & 
   $\phantom{-}3.42 $ &
   $\phantom{-}3.38 $ &
   $\phantom{-}3.36 $ &
   $\phantom{-}3.35 $ &
   $\phantom{-}3.16 $
   \\
   \hline
   $C_{\rm WA}^{O} (\rm LO)$ &
   $\phantom{-}0.77 $ & 
   $\phantom{-}0.55 $ &
   $\phantom{-}0.51 $ &
   $\phantom{-}0.47 $ &
   $\phantom{-}0.46 $ &
   $\phantom{-}0.21 $   
   \\
   $C_{\rm WA}^{O} (\rm NLO)$ &
   $\phantom{-}0.41 $ & 
   $\phantom{-}0.30 $ &
   $\phantom{-}0.27 $ &
   $\phantom{-}0.25 $ &
   $\phantom{-}0.24 $ &
   $\phantom{-}0.10 $
   \\
   \hline
  \end{tabular} 
  \caption{Comparison of the combinations $C_{\rm WE, PI, WA}^{S, O} $, respectively at LO- and NLO-QCD, for different values of the renormalisation scale $\mu_1$.}
  \label{tab:C-WE-PI-WA}
\end{table}

As one can see, the combination of Wilson coefficients multiplying the colour-singlet
Bag parameters of WE are strongly suppressed. 
Note that, depending on whether we disregard $\alpha_s^2 $ corrections in these combinations of $\Delta C =1$ Wilson coefficients -- as we do -- or  not, we can get even different signs for $C_{\rm WE}^{\rm S}$ at NLO. 
Moreover, in Eq.~(\ref{eq:dim-6-4q-LO-D0}) the Bag parameters of the colour singlet operators exactly cancel in VIA at leading order in $1/m_c$. 
The coefficient of the colour-octet operator
is on the other hand not suppressed for weak exchange, indicating that both singlet and octet operators might be equally important in this case.
For Pauli interference, the combinations of Wilson coefficients 
multiplying the colour-singlet operators are significantly
enhanced compared to those in WE, the same holds for the
colour-octet operators.
Note that $C_{\rm PI}^O$ and $C_{\rm PI}^S$ get
large modifications (and even a flip of sign) compared to 
the case $C_1 = 1$ and $C_2 = 0$ revealing the importance of 
gluon radiative corrections.
Moreover $C_{\rm PI}^O$ is enhanced compared to $C_{\rm PI}^S$,
indicating that both singlet and octet operators might be equally important for Pauli interference.
For weak annihilation, the corresponding combination in front of the
colour-singlet operators is large.
On the other hand, the Bag parameters of the colour singlet operators exactly cancel in VIA at leading order in $1/m_c$.
      
The above arguments show that, by neglecting the effect of the colour-octet operators in VIA, one might be led to misleading conclusions, and therefore an accurate determination of the deviation 
of the Bag parameters from their VIA values, using non-perturbative methods like HQET sum rules or lattice simulations, is necessary.

Finally, by including all CKM modes as well as NLO-QCD corrections,
the contribution of four-quark operators to the total decay width at order $1/m_c^3$ schematically reads
\begin{eqnarray}
\hspace*{-5mm}
16 \pi^2 \, \tilde{\Gamma}^{D_q}_6 \frac{\langle \tilde{\cal O}_6\rangle^{D_q}}{m_c^3}
& = & 
\frac{\Gamma_0}{{|V_{cs}|^2}} 
\sum_{i = 1}^4 
\Biggr\{ 
\sum_{q_1, q_2 = d, s} \!\!\! 
\left| \lambda_{q_1 q_2} \right|^2
\Biggr[
A_{i, q_1 q_2}^{\rm WE} \frac{\langle D_q | {\tilde  O}_i^u | D_q \rangle}{m_c^3}
+ A_{i, q_1 q_2}^{\rm PI} \frac{\langle D_q | {\tilde  O}_i^{q_2} | D_q \rangle}{m_c^3}
\nonumber 
\\[1mm]
& &
\qquad + \, A_{i, q_1 q_2}^{\rm WA} \frac{\langle D_q | {\tilde  O}_i^{q_1} | D_q \rangle}{m_c^3}
\Biggl]
+ \sum_{q_1 = d, s} |V_{c q_1}|^2 \sum_{\ell = e, \mu} 
\left[
A_{i, q_1 \ell}^{\rm WA} \frac{\langle D_q | {\tilde  O}_i^{q_1} | D_q \rangle}{m_c^3}
\right]
\Biggl\},
\label{eq:dim-6-4q-NLO-scheme}
\end{eqnarray}
where the matrix elements of the four-quark operators are given
in Eqs.~\eqref{eq:ME-dim-6-HQET-q-q}, \eqref{eq:ME-dim-6-HQET-q-q-prime},  
and the short-distance coefficients for the WE, PI and WA topologies,
c.f.  Fig.~\ref{fig:WE-PI-WA}
are denoted by $A_{i, q_1 q_2}^{\rm WE}$, 
$A_{i, q_1 q_2}^{\rm PI}$ and $A_{i, q_1 q_2}^{\rm WA}$, 
$A_{i, q_1 \ell}^{\rm WA}$, respectively.  NLO corrections to
$A_{i, q_1 q_2}^{\rm WE}$ and $A_{i, q_1 q_2}^{\rm PI}$ have been
computed for HQET operators in Ref.~\cite{Franco:2002fc}.
The corresponding results for $A_{i, q_1 q_2}^{\rm WA}$ can be
obtained by Fierz transforming the $\Delta C = 1$ operators 
given in Eqs.~\eqref{eq:Q1}, \eqref{eq:Q2}. 
Since the Fierz symmetry is respected also at one-loop level, the
functions $A_{i, q_1 q_2}^{\rm WA}$ are derived from 
$A_{i, q_1 q_2}^{\rm WE}$ by replacing  $C_1 \leftrightarrow
C_2$. For the semileptonic modes, the coefficients
$A_{i, q_1 \ell}^{\rm WA}$ have been determined in Ref.~\cite{Lenz:2013aua}.
Note that in our analysis,  we treat the contribution of the $\tilde \delta_i^{ q^\prime q}$ parameters as a subleading ``NLO'' effect, therefore their coefficients are included only at LO-QCD. 
To demonstrate the importance of the NLO-QCD corrections to the spectator effects, 
we show in Table~\ref{tab:dim-6-NLO-vs-LO} the dimension-six contributions to the $D$-meson decay widths (see Eq.~\eqref{eq:dim-6-4q-NLO-scheme}) 
splitting the LO and NLO parts, both in VIA and using HQET SR results for the Bag parameters. 
NLO-QCD corrections turn out to have an essential
numerical effect for the four-quark contributions. In the case of the
$D^0$ and $D_s^+$ mesons these corrections lift the helicity
suppression of weak exchange and weak annihilation being present
in LO-QCD when using VIA.
For the $D^+_s$ meson -- in addition to the CKM dominant WA contribution -- 
there is a correction due to CKM suppressed, but
nevertheless large PI topology. 
In the case of the $D^+$ meson the overall contribution from Pauli interference turns out 
to be huge, of the order of $- 2.5$ ps$^{-1}$.
In addition, the NLO correction to Pauli interference turn also out to be  very large, 
$50 \% - 100 \%$ of the LO term depending on the mass scheme. Already in the $B$ system 
this NLO-QCD corrections were found to be of the order of $30\%$ for the ratio
$\tau (B^+)/\tau (B_d)$, see e.g. Ref.~\cite{Beneke:2002rj} in the Pole scheme. 
Thus, neglecting these contributions for charm lifetime studies, 
as done in Ref.~\cite{Cheng:2018rkz}, is clearly not justified 
and a knowledge of NNLO-QCD corrections to the four-quark contributions 
would be highly desirable.
\begin{table}[ht]
    \centering
    \renewcommand{\arraystretch}{1.6}
    \begin{tabular}{|c|c|c|c|}
    \hline
     Mass scheme & $D^0$ & $D^+$ & $D^+_s$ 
    \\[2mm]
    \hline
    \multicolumn{4}{|c|}{VIA}
    \\
    \hline
    Pole 
    &
    $\underbrace{-0.014}_{\rm NLO}
    = \underbrace{0.000}_{\rm LO} \underbrace{-0.014}_{\rm \Delta NLO}$ 
    & 
    $\underbrace{-2.64}_{\rm NLO} 
    = \underbrace{-1.68}_{\rm LO} \underbrace{-0.97}_{\rm \Delta NLO}$ 
    & 
    $\underbrace{-0.20}_{\rm NLO} 
    = \underbrace{-0.12}_{\rm LO} \underbrace{-0.08}_{\rm \Delta NLO}$ 
    \\[5mm]
    \hline
    $\overline{\rm MS}$
    &
    $\underbrace{-0.010}_{\rm NLO} 
    = \underbrace{0.000}_{\rm LO} \underbrace{- 0.010}_{\rm \Delta NLO}$ 
    & 
    $\underbrace{-2.49}_{\rm NLO} 
    = \underbrace{-1.23}_{\rm LO} \underbrace{- 1.25}_{\rm \Delta NLO}$ 
    & 
    $\underbrace{-0.18}_{\rm NLO} 
    = \underbrace{-0.08}_{\rm LO} \underbrace{- 0.10}_{\rm \Delta NLO}$ 
    \\[5mm]
    \hline
    Kinetic
    &
    $\underbrace{-0.012}_{\rm NLO} 
    = \underbrace{0.000}_{\rm LO} \underbrace{- 0.012}_{\rm \Delta NLO}$ 
    & 
    $\underbrace{-2.53}_{\rm NLO} 
    = \underbrace{-1.42}_{\rm LO} \underbrace{- 1.11}_{\rm \Delta NLO}$ 
    & 
    $\underbrace{-0.19}_{\rm NLO} 
    = \underbrace{-0.10}_{\rm LO} \underbrace{- 0.09}_{\rm \Delta NLO}$ 
    \\[5mm]
    \hline
    $1S$
    &
    $\underbrace{-0.013}_{\rm NLO}
    = \underbrace{0.000}_{\rm LO} \underbrace{-0.013}_{\rm \Delta NLO}$ 
    & 
    $\underbrace{-2.60}_{\rm NLO} 
    = \underbrace{-1.58}_{\rm LO} \underbrace{-1.02}_{\rm \Delta NLO}$ 
    & 
    $\underbrace{-0.19}_{\rm NLO} 
    = \underbrace{-0.11}_{\rm LO} \underbrace{-0.08}_{\rm \Delta NLO}$ 
    \\[5mm]
    \hline
    \multicolumn{4}{|c|}{HQET SR}
    \\
    \hline
    Pole 
    &
    $\underbrace{0.007}_{\rm NLO}
    = \underbrace{0.019}_{\rm LO} \underbrace{-0.012}_{\rm \Delta NLO}$ 
    & 
    $\underbrace{-2.89}_{\rm NLO} 
    = \underbrace{-1.87}_{\rm LO} \underbrace{-1.02}_{\rm \Delta NLO}$ 
    & 
    $\underbrace{-0.21}_{\rm NLO} 
    = \underbrace{-0.16}_{\rm LO} \underbrace{-0.05}_{\rm \Delta NLO}$ 
    \\[5mm]
    \hline
    $\overline{\rm MS}$
    &
    $\underbrace{0.020}_{\rm NLO} 
    = \underbrace{0.014}_{\rm LO} \underbrace{+ 0.006}_{\rm \Delta NLO}$ 
    & 
    $\underbrace{-2.72}_{\rm NLO} 
    = \underbrace{-1.37}_{\rm LO} \underbrace{- 1.35}_{\rm \Delta NLO}$ 
    & 
    $\underbrace{-0.20}_{\rm NLO} 
    = \underbrace{-0.12}_{\rm LO} \underbrace{- 0.08}_{\rm \Delta NLO}$ 
    \\[5mm]
    \hline
    Kinetic
    &
    $\underbrace{0.014}_{\rm NLO} 
    = \underbrace{0.016}_{\rm LO} \underbrace{- 0.002}_{\rm \Delta NLO}$ 
    & 
    $\underbrace{-2.76}_{\rm NLO} 
    = \underbrace{-1.58}_{\rm LO} \underbrace{- 1.18}_{\rm \Delta NLO}$ 
    & 
    $\underbrace{-0.20}_{\rm NLO} 
    = \underbrace{-0.13}_{\rm LO} \underbrace{- 0.07}_{\rm \Delta NLO}$ 
    \\[5mm]
    \hline
    $1S$
    &
    $\underbrace{0.009}_{\rm NLO}
    = \underbrace{0.018}_{\rm LO} \underbrace{-0.008}_{\rm \Delta NLO}$ 
    & 
    $\underbrace{-2.84}_{\rm NLO} 
    = \underbrace{-1.76}_{\rm LO} \underbrace{-1.08}_{\rm \Delta NLO}$ 
    & 
    $\underbrace{-0.21}_{\rm NLO} 
    = \underbrace{-0.15}_{\rm LO} \underbrace{-0.06}_{\rm \Delta NLO}$ 
    \\[5mm]
    \hline
    \end{tabular}
    \caption{Dimension-six contributions to $D$-meson decay widths (see Eq.~\eqref{eq:dim-6-4q-NLO-scheme}) (in ps$^{-1}$) and split up into LO-QCD and NLO-QCD corrections within different mass schemes and 
    both in VIA and using the HQET SR for Bag parameters.}
    \label{tab:dim-6-NLO-vs-LO}
\end{table}

\subsection{Dimension-seven Four-Quark Operator Contribution }
\label{sec:dim-7-4q} 
The dimension-six four-quark operator contribution discussed in the previous section, is obtained by neglecting in the expression of the incoming momentum $p^\mu = p_c^\mu + p_q^\mu$ the effect due to the small momentum of the light spectator quark $p_q \sim \Lambda_{\rm QCD}$. Including also corrections linear in the quantity~$p_q/m_c$, leads to the contribution of order $1/m_c^4$ to $\Gamma(D)$, which can be described in terms of the following basis of dimension-seven operators, defined in full QCD, i.e. \footnote{
Note that e.g. in Ref.~\cite{Lenz:2013aua} it is used a redundant basis, containing also the additional operator, denoted by $P_2^q$, which however is related to $P_1^q$ by hermitean conjugation, namely
$P_2^q = m_q \, (\bar c (1 + \gamma_5) q) (\bar q (1 +  \gamma_5) c) 
= (P^q_1)^\dagger$. Since it leads to the same matrix element, we do not include this operator in our basis.
}
\begin{eqnarray}
P_1^q 
& = &  
m_q\, (\bar c (1- \gamma_5) q) (\bar q (1- \gamma_5) c) \, ,
\label{eq:P1q}
\\[2mm]
P_2^q 
& = & 
\frac{1}{m_c} (\bar c \overset{\leftarrow}{D_\nu} \gamma_\mu (1- \gamma_5)  D^\nu q) (\bar q \gamma^\mu (1 - \gamma_5) c) \, ,
\label{eq:P3q}
\\[2mm]
P_3^q
& = & 
\frac{1}{m_c} (\bar c \overset{\leftarrow}{D_\nu} (1- \gamma_5)  D^\nu q) 
(\bar q (1 + \gamma_5) c)\, , 
\label{eq:P4q}
\end{eqnarray}
together with the corresponding colour-octet operators $S_1^q, S_2^q, S_3^q$, containing the generators $T^A$, and again a summation over colour indices is implied. Due to the presence in Eqs.~(\ref{eq:P3q}), (\ref{eq:P4q}) of a covariant derivative acting on the charm quark field, 
which scales as $m_c$ at this order, there is no immediate power counting for these operators, cf. the HQET operators in Eqs.~(\ref{eq:P3q-HQET}), (\ref{eq:P4q-HQET}). 
Moreover, note that this basis differs from the one 
used in Ref.~\cite{Gabbiani:2004tp} for the computation of dimension-seven and dimension-eight contributions.

In order to evaluate the matrix element of the dimension-seven four-quark operators using the framework of the HQET, one has to further expand the charm quark momentum, according to
$p^\mu = m_c v^\mu + k^\mu + p_q^\mu$, see Eq.~\eqref{eq:c-quark-momentum}, as well as to include $1/m_c$ corrections to the effective heavy quark field and to the HQET Lagrangian, retaining only terms linear in $k/m_c$ and $p_q/m_c$.
The small residual momentum of the charm quark $k^\mu$ will result
in a covariant derivative acting on $h_v$ and the small momentum
 of the spectator quark $p_q^\mu$ will result
 in a covariant derivative acting on the light quark field $q$.
In this case, one obtains the following basis, which includes the local operators 
\begin{eqnarray}
  {\tilde P}_1^q 
 & = &  
 m_q \, (\bar h_v (1- \gamma_5) q) (\bar q (1- \gamma_5) h_v)\, , 
 \label{eq:P1q-HQET}
\\[2mm]
 {\tilde P}_2^q 
& = & 
(\bar h_v \gamma_\mu (1- \gamma_5)  (i v \cdot D) q) (\bar q \gamma^\mu (1- \gamma_5) h_v)\, ,
\label{eq:P3q-HQET}
\\[2mm]
 {\tilde P}_3^q
& = & 
 (\bar h_v(1- \gamma_5)  (i v \cdot D) q) (\bar q  (1+ \gamma_5) h_v)\, ,
\label{eq:P4q-HQET}
\end{eqnarray}
and
\begin{eqnarray}
 {\tilde R}_1^q 
& = & 
 (\bar h_v \gamma_\mu (1- \gamma_5)  q) (\bar q \gamma^\mu (1- \gamma_5) (i \slashed D) h_v) \, ,
\label{eq:P5q-HQET}
\\[2mm]
 {\tilde R}_2^q
& = & 
 (\bar h_v (1- \gamma_5) q) (\bar q  (1+ \gamma_5)  (i \slashed D)  h_v)\, ,
\label{eq:P6q-HQET}
\end{eqnarray}
supplemented by the corresponding colour-octet operators 
${\tilde  S}_{1,2,3}^q$ and ${\tilde  U}_{1,2}^q$,  and the non-local operators 
\begin{eqnarray}
\tilde M_{1, \pi}^{q}
& = &
  i \int d^4 y \,  T
\left[ 
 {\tilde O}_1^q (0), 
(\bar h_v (i D)^2 h_v) (y)
\right],
\label{eq:M1-pi} 
\\
\tilde M_{2, \pi}^{q}
& = &
 i \int d^4 y \, T
\left[ 
 {\tilde  O}_2^q (0), 
(\bar h_v (i D)^2 h_v) (y)
\right],
\label{eq:M2-pi}
\\
\tilde M_{1, G}^{q}
& = &
 i \int d^4 y \, T
\left[ 
 { \tilde O}_1^q (0), 
\frac{1}{2} g_s \left(\bar h_v \sigma_{\alpha \beta} G^{\alpha \beta} h_v \right) (y)
\right],
\label{eq:M1-G}
\\
\tilde M_{2, G}^{q}
& = &
 i \int d^4 y \,  T
\left[ 
 { \tilde O}_2^q (0), 
\frac{1}{2} g_s \left(\bar h_v \sigma_{\alpha \beta} G^{\alpha \beta} h_v \right) (y)
\right],
\label{eq:M2-G}
\end{eqnarray}
also supplemented by the corresponding colour-octet operators
\footnote{Operators which vanish due to the equation of motion $(i v \cdot D) h_v = 0$ are not shown.}.
We see that, compared to the QCD basis, there are in addition the two local operators
${\tilde  R}_1^q$ and ${\tilde  R}_2^q$
(and also the corresponding colour-octet ones),
which emerge from the expansion in Eq.~\eqref{eq:cv-hv-relation}, and the four non-local operators $\tilde M_{1, \pi}^q$,  $\tilde M_{2, \pi}^q$, 
$\tilde M_{1, G}^q$ and $\tilde M_{2, G}^q$ 
(and the corresponding colour-octet ones) which are obtained by taking the time-ordered product of the dimension-six operators with the $1/m_c$ correction to the HQET Lagrangian, see e.g. Ref.~\cite{Neubert:1993mb} for details. 

We parametrise the matrix elements of the operators in Eqs.~(\ref{eq:P1q-HQET}) - (\ref{eq:M2-G}) using VIA and account for deviations from it by including the corresponding Bag parameters, as it is explicitly shown in Appendix \ref{Appendix-C}. However, since for these matrix elements there is no non-perturbative evaluation available yet, in our analysis we have to rely only on VIA.
It follows that, at LO-QCD the matrix element of the dimension-seven operators listed above, can be expressed in terms of the HQET
non-perturbative parameters $F (\mu)$, $G_1 (\mu)$, 
$G_2 (\mu)$, and $\bar \Lambda$, so far determined only with large uncertainties.
For this reason, we prefer to use as an input the QCD decay
constant $f_D$, which is computed very precisely using Lattice QCD
\cite{Aoki:2019cca}. In doing so, we obtain that in VIA and at 
the matching scale $\mu = m_c$,  
the contribution of the local operators 
${\tilde R}_{1,2}^q$ as well as that of the non-local ones
$\tilde M_{1, \pi}^q$, 
$\tilde M_{2, \pi}^q$, 
$\tilde M_{1, G}^q$ and
$\tilde M_{2, G}^q$ 
can be entirely absorbed in the QCD decay constant $f_D$, cf. Eq.~\eqref{eq:decay_constant-conversion}
(more precisely, in the matrix element of the dimension-six QCD operators in Eqs.~(\ref{eq:O1}), (\ref{eq:O2}), which are proportional to $f_D$), so that we are left only with the $1/m_c$
contribution due to the operators $ \tilde P_{1,2,3}^q$, analogously to the QCD case \footnote{In the matrix element of  $\tilde P_{1,2,3}^q$ one can replace the HQET decay constant with the QCD one, up to higher order corrections.}.

To make this point more clear, we consider as an example the contribution due to Pauli interference at LO-QCD and up to order $1/m_c^4$, in the case of $c \to s \bar d u $ transition, which constitutes the dominant correction to $\Gamma(D^+)$,
\begin{eqnarray}
\hspace*{-2.5mm}
{\rm Im} \, {\cal T}^{\rm PI} 
& = & 
\Gamma_0 \, |V_{ud}^*|^2 \, 
\frac{32 \pi^2}{m_c^3} (1 - z_s)^2 
\Biggl[ 
C_{\rm PI}^S \left( {\tilde O}_1^d +  \frac{{\tilde R}_1^d}{m_c} + 
\frac{\tilde M_{1, \pi}^d}{m_c} + \frac{\tilde M_{1, G}^d }{m_c}
+ 2 \frac{1 + z_s}{1 - z_s}\frac{ {\tilde P}_3^q }{m_c} \right) 
\nonumber \\
& & 
\hspace*{40mm} + \, (\mbox{colour-octet part}) 
\Biggr],
\label{eq:Gamma-PI-c-to-s-d-u}
\end{eqnarray}
with $C_{\rm PI}^S$ defined in Eq.~\eqref{eq:C-PI}.
By evaluating the matrix element of ${\rm Im} {\cal T} ^{\rm PI}$ in VIA, the contribution due to the colour-octet operators vanishes. Moreover, using the parametrisation for the matrix elements of the four-quark operators given in Eq.~\eqref{eq:ME-dim-6-HQET-q-q} and in Appendix~\ref{Appendix-C}, we obtain in VIA and setting $\mu  = m_c$, that
\begin{eqnarray}
\langle  D^+|  {\tilde O}_1^d +  \frac{{\tilde R}_1^d}{m_c} + \frac{\tilde M_{1, \pi}^d}{m_c} + \frac{\tilde M_{1, G}^d}{m_c} | D^+ \rangle_{\rm HQET}
& = & 
F^2 (m_c) \, m_{D^+} 
\left[1 - \frac{\bar \Lambda}{m_c} + \frac{2 \,G_1 (m_c)}{m_c} + \frac{12 \, G_2 (m_c)}{m_c} \right]
\nonumber \\[2mm]
& = & f_D^2 \, m_{D^+}^2 = \langle D^+| O_1^d | D^+ \rangle_{\rm QCD},
\end{eqnarray}
where in the second line we have used the conversion between the QCD and HQET decay constants given in Eq.~\eqref{eq:decay_constant-conversion}, showing that the contribution of the local operators $\tilde R_i^q$ and
non-local operators $\tilde M^q_{i, \pi}$ and
$\tilde M^q_{i, G}$ is entirely absorbed in the QCD decay
constant. Note that, by neglecting the effect due to the strange quark mass and using
VIA we reproduce the approximate result of Eq.~(19) in Ref.~\cite{Mannel:2021uoz}.

The same argument applies also to the remaining topologies i.e. WE and WA. 
However, it is worth mentioning that in VIA and neglecting the strange quark mass, the contribution of WE and WA exactly vanishes at LO-QCD, due to the helicity suppression.
This suppression 
is lifted once the $s$-quark mass or perturbative
gluon corrections are included, and in this case it becomes again manifest that the contributions
of ${\tilde R}_i^q$, $\tilde {\cal M}^q_{i, \pi}$ and 
$\tilde {\cal M}^q_{i, G}$ in HQET can be completely absorbed in $f_D$ by evaluating the matrix elements in VIA
\footnote{Note, that for the operator ${ O}_2^q$ the contribution
of $ { R}_2^q$ is absorbed by the combination $(m_D \, f_D/m_c)^2
\approx (1 + 2 \, \bar \Lambda/m_c) \, f_D^2$.}.
We note that a detailed analysis of the dimension-seven
contributions within the HQET has been performed in
Ref.~\cite{Kilian:1992cj} for the case of $B - \bar B$-mixing.
Specifically, it was found that in VIA, subleading power
corrections due to non-local operators can be entirely absorbed
in the definition of the QCD decay constant,  and that the
residual $1/m_b$ corrections, due to the running of the local
dimension-seven operators from the scale $m_b$ to
$\mu \sim$~1~GeV, is numerically small ($\sim 5 \%$ for
Ref.~\cite{Kilian:1992cj}).\footnote{By neglecting the effect of
running down to a lower scale, from Ref.~\cite{Kilian:1992cj} one
can see that in VIA the QCD decay constant entirely absorbs all
the $1/m_b$ contributions.}

Finally, by summing over all the CKM
modes, at LO-QCD, the dimension-seven contribution can 
therefore be presented as (with $q = u, d, s$)
\begin{eqnarray}
16 \pi^2 \, \tilde{\Gamma}^{D_q}_7 
\frac{\langle \tilde{\cal O}_7\rangle^{D_q}}{m_c^4}
& = & 
\frac{\Gamma_0}{{|V_{cs}|^2}} 
\sum_{i = 1}^3 
\Bigg\{ \sum_{q_1, q_2 = d, s} \!\!\! 
\left|\lambda_{q_1 q_2} \right|^2 \!
\Bigg[G_{i, q_1 q_2}^{\rm WE} \frac{\langle D_q |P_i^u| D_q \rangle}{m_c^4} 
+ G_{i, q_1 q_2}^{\rm PI} \frac{\langle D_q |  P_i^{q_2} | D_q \rangle}{m_c^4}
\nonumber
\\[2mm]
& &  
\qquad 
+G_{i, q_1 q_2}^{\rm WA} \frac{\langle D_q |  P_i^{q_1} | D_q \rangle}{m_c^4}
\Bigg]
+ 
\sum_{q_1 = d, s} |V_{c q_1}|^2
\sum_{\ell = e, \mu}  
\left[
G_{i, q_1 \ell}^{\rm WA} \frac{\langle D_q | {  P}_i^{q_1} | D_q \rangle}{m_c^4}
\right] \Bigg\} 
\nonumber 
\\[2mm]
& & 
\qquad  + (\mbox{colour-octet part})\,,
\label{eq:dim-7-4q-LO-scheme}
\end{eqnarray}
where the matrix elements of the dimension-seven operators are presented in Appendix~\ref{Appendix-C}.
We confirm the results for the short-distance coefficients $G_{i, q_1 q_2}^{\rm WE}$, $G_{i, q_1 q_2}^{\rm PI}$
and $G_{i, q_1 q_2}^{\rm WA}$, $G_{i, q_1 \ell}^{\rm WA}$ presented in Ref.~\cite{Lenz:2013aua}. 
Note that, due to the current accuracy of the analysis, at dimension-seven we include only the contribution of the valence-quark, therefore e.g. 
$\langle D^0 | {  P}_i^s | D^0 \rangle = 0$.
Numerical values of the dimension-seven contributions 
to the decay rates and the ratios will be presented in Section~\ref{sec:Results}. In
Table~\ref{tab:dim-7-VIA}  we show the central values of the dimension-seven contributions
in ps$^{-1}$ in the kinetic mass scheme and we find for the $D^+$ meson a correction that is almost as large as the leading dimension three term, see
Table \ref{tab:Gamma_3}.

\begin{table}[th]
    \centering
    \renewcommand{\arraystretch}{2.0}
    \begin{tabular}{|C{5cm}||C{2.0cm}|C{2.0cm}|C{2.0cm}|}
    \hline
    & $D^0$ & $D^+$ & $D^+_s$ 
    \\
    \hline
    \hline
    $\displaystyle 16 \pi^2 \, \tilde{\Gamma}^{D_q}_7 \frac{\langle \tilde{\cal O}_7\rangle^{D_q}}{m_c^4} \, [{\rm ps}^{-1}]$
    & $-1.4 \times 10^{-6}$
    & $1.05 $
    & $0.10 $
    \\[2mm]
    \hline
    \end{tabular}
    \caption{Dimension-seven contributions to $D$-meson decay widths (see Eq.~\eqref{eq:dim-7-4q-LO-scheme}) in ps$^{-1}$
    within VIA in the kinetic mass scheme. }
    \label{tab:dim-7-VIA}
\end{table}

\section{Determination of the Non-perturbative Parameters}
\label{sec:HQE-NP-parameters}
In the present section, we discuss the numerical determination for the matrix elements
of the operators introduced in Sections \ref{SubSec:Lead} -  \ref{sec:dim-7-4q}.
We start with the operators of the lowest mass dimension.
\subsection{Parameters of the Chromomagnetic Operator}
\label{sec:mu_G}
For the $B$ system many of non-perturbative parameters have been 
determined by performing fits to the experimental data for inclusive
semileptonic decays \cite{Alberti:2014yda}. In the case of the chromomagnetic operator, one finds
\begin{eqnarray}
 \mu_G^2 (B) & = &   (0.332 \pm 0.062) \,  {\rm GeV}^2
  \, .
  \label{eq:muG_B_fit}
 \end{eqnarray}
Assuming heavy quark symmetry we expect the corresponding
parameter in the $D$ system to have a similar size.  
Another way of estimating the value of $\mu_G^2$  is to use the well-known spectroscopy relation~\cite{Uraltsev:2001ih}
\begin{equation}
\mu_G^2 (D_{(s)}) = 
\frac{3}{2} m_c \, (M_{D_{(s)}^*} - M_{D_{(s)}}) \, ,
\label{Eq:mupi_spec}
\end{equation}
which holds up to power corrections.
Using the value for the meson masses given in the PDG \cite{Zyla:2020zbs} and setting $m_c = 1.27 \, {\rm GeV}$, we obtain the following estimates:
\begin{equation}
\mu_G^2 (D) = (0.268 \pm  0.107) \,  {\rm GeV}^2, 
\qquad
\mu_G^2 (D_{s}) = (0.274 \pm  0.110) \,  {\rm GeV}^2, 
\label{eq:mu_G}
\end{equation}
where we have conservatively added an uncertainty of $40 \%$ 
due to unknown power corrections of order~$1/m_c$. 
The values in Eq.~(\ref{eq:mu_G}) are roughly 19$\%$ smaller than those obtained from experimental fits for semileptonic $B$-meson decays, 
see Eq.~(\ref{eq:muG_B_fit}).
Moreover, Eq.~(\ref{Eq:mupi_spec}) leads to a tiny amount of
$SU(3)_f$-symmetry breaking of 
$ \approx 2 \%$, which might, however,
be enhanced by the neglected power
corrections.
In the literature many times instead of Eq.~(\ref{Eq:mupi_spec})
the relation \cite{Falk:1992wt, Neubert:1993mb}
 \begin{eqnarray}
 \mu_G^2 (D_{(s)})  
 & = & \frac 3 4 \left(M_{D_{(s)}^*}^2 - M_{D_{(s)}}^2 \right)
 \label{Eq:mupi_spec2}
 \end{eqnarray}
 is adopted, which coincides with
 Eq.~(\ref{Eq:mupi_spec}) up to 
 corrections of order $1/m_c$.
 Numerically we find that
 Eq.~(\ref{Eq:mupi_spec2}) yields 
 \begin{eqnarray}
 \mu_G^2 (D)  =  0.41 \, {\rm GeV}^2
 \, , 
 & \qquad &
 \mu_G^2 (D_s^+)  =  0.44 \, {\rm GeV}^2 \, ,
 \label{eq:mu_G2}
 \end{eqnarray}
 which are roughly 23$\%$ higher than 
 that  in 
 Eq.~(\ref{eq:muG_B_fit}).
 In our numerical analysis, we will use the average value of the two determinations in  Eq.~(\ref{eq:mu_G})
and Eq.~(\ref{eq:mu_G2}). This gives
\begin{equation}
\mu_G^2 (D)  =  (0.34 \pm 0.10) \, {\rm GeV}^2,
\qquad 
\mu_G^2 (D_s^+)  =  (0.36 \pm 0.10) \, {\rm GeV}^2 \, ,
\label{eq:mu_G3}
\end{equation}
which agrees well with the one in Eq.~(\ref{eq:muG_B_fit}).

Thus, from Eq.~(\ref{eq:Gamma_5}), we expect corrections to the total
decay rate due to the chromomagnetic operator,
$c_G \, \mu_G^2/(c_3 \, m_c^2)$ ranging between $-6 \%$ and $+8 \%$
with respect to the leading free-quark decay contribution.
A large part of the sizable uncertainty derives from the
cancellations in  the coefficient~$c_G$, shown in
Table~\ref{tab:cG} and Fig.~\ref{fig:cG}, which could be
reduced with a complete determination of the NLO-QCD corrections to~$c_G$.
For semileptonic rates the contribution of the chromomagnetic operator
can be even of the order of $20\%$, see Section
\ref{sub:semileptonic}.

An experimental determination of $\mu_G^2(D)$ from inclusive
semileptonic $D$-meson decays could further reduce the uncertainties
and could in particular give some insight into the numerical size of
$SU(3)_F$ breaking.
\subsection{Parameters of the Kinetic Operator}
\label{sec:mu_pi}
For the matrix element of the kinetic operator no precise
determination is available so far in the charm sector. 
Several predictions of $\mu_\pi^2$ available in the literature 
for the  $B$-meson cover a large range of values, see Table~\ref{tab:mu-pi-sq}.
\begin{table}[ht]
\centering
\renewcommand{\arraystretch}{1.6}
\begin{tabular}{|c||c|c|c|c|c|}
  \hline
  {\rm Source}  
  & LQCD~\cite{FermilabLattice:2018est}
  & LQCD~\cite{Gambino:2017vkx}
  & Exp. fit~\cite{Alberti:2014yda} 
  & QCD SR~\cite{Neubert:1996wm} 
  & QCD SR~\cite{Ball:1993xv} 
  \\
  \hline
  $\mu_\pi^2 [{\rm GeV}^2]$   
  & 0.05(22)  
  & 0.314(15) 
  & 0.465(68)
  & 0.10(5) 
  & 0.6(1) \\
  \hline
\end{tabular}
\caption{Different determinations of $\mu_\pi^2 (B)$ available in the literature.}
\label{tab:mu-pi-sq}
\end{table}
Assuming heavy quark symmetry one can use the value obtained from the fit of the semileptonic $B$-meson decays~\cite{Alberti:2014yda}:
\begin{eqnarray}
  \mu_\pi^2 (B)  & =  &
  (0.465 \pm 0.068) \, {\rm GeV}^2 \, ,
 \end{eqnarray}
to get a following estimate for the $D$-meson
\begin{eqnarray}
  \mu_\pi^2 (D)  & = &  (0.465  \pm 0.198) \,  {\rm GeV}^2.
 \end{eqnarray}
In the above, we have again added an conservative uncertainty of $40 \% $ to account for the breaking of the heavy 
quark symmetry. This value clearly fulfills the theoretical bound $\mu_\pi^2 \geq \mu_G^2$, 
see e.g. the review \cite{Bigi:1997fj}.
Thus we expect  from Eq.~(\ref{eq:Gamma_5}) corrections 
due to the kinetic  operator of the order of $-10\%$, which is also
found  in Section \ref{sub:semileptonic} -- both for the total decay rate and the semileptonic one.

The $SU(3)_F$ breaking effects for the kinetic operator 
have been estimated in Refs.~\cite{Bigi:2011gf, Lenz:2013aua}
\begin{eqnarray}
\mu_\pi^2 (D_s^+) - \mu_\pi^2 (D^0) & \approx &  0.09 \,  {\rm GeV}^2 \, ,
\end{eqnarray}
leading to the following estimate we use for the $D_s$ meson:
\begin{equation}
\mu_\pi^2 (D_s^+) = (0.555 \pm 0.232) \, {\rm GeV}^2.
\end{equation}
Again a more precise experimental determination of $\mu_\pi^2$ from
fits to semileptonic $D^+$, $D^0$ and $D_s^+$ meson decays -- as it was
done for the $B^{+}$ and $B^{0}$ decays -- would be very desirable.

\subsection{Parameters of the Darwin Operator}
\label{sec:rho_D}
For the matrix element of the Darwin operator no theoretical 
determination for the charm sector is available. We again could 
assume heavy quark symmetry and use the corresponding value in the $B$-system,
obtained from fits of the semileptonic decays \cite{Alberti:2014yda}:
\begin{eqnarray}
  \rho_D^3 (B) & = &  (0.170 \pm 0.038) \, {\rm GeV}^3 \, ,
  \label{eq:rho_d-fit_B}
 \end{eqnarray}
and add quadratically an uncertainty of  {$40\%$ } for the transition from the $B$ to the $D$ system, leading to a first estimate of
\begin{eqnarray}
  \rho_D^3 (D)^I & = &  (0.17 \pm 0.07) \, {\rm GeV}^3 \, .
 \label{eq:Darwin_MEI}
 \end{eqnarray}
Alternatively the Darwin parameter can be
related to the matrix elements of the
dimension-six four-quark operators
through the equation of motion 
for the gluon field.
At leading order in $1/m_Q$ one obtains:
\begin{equation}
\rho_D^3 (H) = 
\frac{g_s^2}{18} f_H^2 \, m_H 
\left[
2 \, \tilde B_2^{q^\prime} -  \tilde B_1^{q^\prime} 
+ \frac{3}{4}  \tilde \epsilon_1^{q^\prime}
- \frac{3}{2}  \tilde \epsilon_2^{q^\prime} 
+ \sum_{q = u, d, s} 
\left(
2 \tilde \delta_2^{q q^\prime} -  \tilde \delta_1^{q q^\prime } 
+ \frac{3}{4}  \tilde \delta_3^{q q^\prime} - \frac{3}{2}  \tilde \delta_4^{q q^\prime }
\right)
\right]
+ {\cal O} \left(\frac{1}{m_Q}\right),
\label{eq:EoM-Darwin}
\end{equation}
where $H$ is a heavy hadron with the mass $m_H$ and the decay constant $f_H$,
$q^\prime = u, d, s$ is the light valence quark in the $H$-hadron, 
and the Bag parameters
$ \tilde B_1^q$, 
$ \tilde B_2^q$, 
$ \tilde \epsilon_1^q $, 
$ \tilde \epsilon_2^q $,
$ \tilde \delta_1^{q q^\prime } $ 
$ \tilde \delta_2^{q q^\prime } $ 
$ \tilde \delta_3^{q q^\prime } $
and
$ \tilde \delta_4^{q q^\prime }$
were introduced in Section~\ref{sec:dim-6-4q}.
Their numerical values are summarised in Table~\ref{tab:Bag-parameters} in Appendix~\ref{Appendix-C}.
The strong coupling $g_s$ has its origin in the non-perturbative
regimes -- e.g. Ref.~\cite{Bigi:1993ex} suggests to set 
$\alpha_s = 1$.

Using the input from the Appendix~\ref{Appendix-A}
and applying Eq.~(\ref{eq:EoM-Darwin}) we derive
estimates of $\rho_D^3$ for $B$- and $D$-mesons both in VIA 
and using the HQET SR results for the Bag parameters.
The values are summarised in Table~\ref{tab:rhoD}
for the three different choices, $\alpha_s(\mu = 1.5 \, {\rm GeV})$, $\alpha_s(\mu = 1 \, {\rm GeV})$ and $\alpha_s = 1$.
\begin{table}[ht]
\renewcommand{\arraystretch}{1.5}
\centering
\begin{tabular}{|c||c|c||c|c||c|c|}
\hline
&
\multicolumn{2}{|c||}{$\mu = 1.5$ GeV} &
\multicolumn{2}{|c||}{$\mu = 1.0$ GeV} &
\multicolumn{2}{|c|}{$\alpha_s = 1$} 
\\
\hline
$\rho_D^3 [{\rm GeV^3}]$
& VIA & \mbox{ HQET} 
& VIA & \mbox{ HQET}
& VIA & \mbox{ HQET}
\\
\hline \hline
$B^+, B_d$ &  
0.048  & 0.047 &  
0.066  & 0.064 & 
0.133  & 0.129
\\
\hline
$B_s$ &  
0.072  & 0.070 &  
0.098  & 0.095 & 
0.199  & 0.193
\\
\hline
$D^+, D^0 $
& 0.021 & 0.020  
& 0.027 & 0.026 
& 0.059 & 0.056 
\\
\hline
$D_s^+ $
& 0.030 & 0.029 
& 0.040 & 0.038 
& 0.086 & 0.082 
\\
\hline
\end{tabular}
\caption{Values of $\rho_D^3 (H)$ for $B$- and $D$-mesons
in VIA and using HQET SR for Bag parameters for three different choices of $\alpha_s$ in Eq.~\eqref{eq:EoM-Darwin}.}
\label{tab:rhoD}
\end{table}

Setting $\alpha_s = 1$ in Eq.~(\ref{eq:EoM-Darwin}), yields values for $\rho_D^3$ that are close to the one determined from the fit of semileptonic $B$ meson decays, Eq.~(\ref{eq:rho_d-fit_B}), indicating  $1/m_b$-corrections in Eq.~(\ref{eq:EoM-Darwin})
of the order of $+30 \%$.
Moreover, we find that VIA gives in Eq.~(\ref{eq:EoM-Darwin}) values, which are very close to the HQET sum rule ones.
We emphasise that due to the sizeable $SU(3)_F$ breaking in the decay constants, Eq.~(\ref{eq:EoM-Darwin}) leads also to a sizable $SU(3)_F$ breaking for the non-perturbative parameters $\rho_D^3(D)$, $\rho_D^3(D_s^+)$. 
Taking the values corresponding to $\alpha_s = 1$ and using HQET SR results for the bag parameters we get the second estimate (last column in Table~\ref{tab:rhoD}) 
\begin{eqnarray}
  \rho_D^3 (D)^{II}  =  (0.056 \pm 0.022) \, {\rm GeV}^3 \, ,
  & \quad &
  \rho_D^3 (D_s^+)^{II}  =  (0.82 \pm 0.033) \, {\rm GeV}^3 \, ,
 \label{eq:Darwin_MEII}
 \end{eqnarray}
where we have again added a $40\%$ uncertainty.
Finally, another possibility to extract $\rho_D^3(D)$ is to substitute in Eq.~(\ref{eq:EoM-Darwin}) the values of the Bag parameters in VIA,  which gives
\begin{equation}
\rho_D^3 (H) \approx \frac{g_s^2}{18} f_H^2 \, m_H.
\label{eq:EoM-Darwin-VIA}
\end{equation}
Assuming a similar size for the strong coupling in both the $B$- and $D$-meson matrix elements, 
from Eq.~(\ref{eq:EoM-Darwin-VIA}) one obtains:
\begin{eqnarray}
\rho_D^3 (D) 
\approx 
\frac{f_D^2 \, m_D}{f_B^2 \, m_B} \, \rho_D^3 (B) \, ,
& \quad &
\rho_D^3 (D_s) 
 \approx 
\frac{f_{D_s}^2  m_{D_s}}{f_B^2 \, m_B} \, \rho_D^3 (B) \, .
\end{eqnarray}
Using the most precise determination of the decay constants from Lattice QCD \cite{Aoki:2019cca}, and of the meson masses from PDG \cite{Zyla:2020zbs} and taking into account the value of $\rho_D^3 (B)$ in Eq.~(\ref{eq:rho_d-fit_B}), leads to the following estimates:
\begin{eqnarray}
\rho_D^3 (D)^{III} 
 = 
(0.075 \pm 0.034) \, {\rm GeV^3} \, ,
& \quad &
\rho_D^3 (D_s)^{III} 
 = 
(0.110 \pm 0.050) \, {\rm GeV^3} \, ,
\label{eq:Darwin_ME}
\end{eqnarray}
where we again assign in addition a conservative $40 \%$
uncertainty due to missing power corrections.
These values are consistent with the numbers shown in Table~\ref{tab:rhoD} for $\alpha_s = 1$.
Contrary to the case of the dimension-five non-perturbative parameters, in Eq.~(\ref{eq:Darwin_ME}) one observes a large $SU(3)_f$ -symmetry breaking of $\approx 46 \% $, similar to the $\approx 49 \%$ that one obtains for the $B_{(s)}$-mesons, mostly stemming from the ratios $f_{B_s}/f_{B_d}$ and  $f_{D_s^+}/f_{D^0}$.
In our numerical analysis we use the values shown in Eq.~(\ref{eq:Darwin_ME}), which lies between the estimates 
obtained in Eq.~(\ref{eq:Darwin_MEI}) and
Eq.~(\ref{eq:Darwin_MEII}).

Again, here a more precise experimental determination of $\rho_D^3$ from
fits to semileptonic $D^+$, $D^0$ and $D_s^+$ meson decays -- as it was
done for the $B^{+}$ and $B^{0}$ decays -- would be very desirable
and could have a significant effect on the phenomenology of inclusive charm decays.

\subsection{Bag parameters of Dimension-six and Dimension-seven}
\label{sec:Bag-par}
\noindent
 
The dimension-six Bag parameters of the $D^+$ and $D^0$ mesons
have been determined using  HQET Sum Rules \cite{Kirk:2017juj};
strange quark mass corrections, relevant for the Bag parameter of
the  $D_s^+$  meson, as well as eye-contractions have been
computed for the first time in
Ref.~\cite{King:2020}.
The results are collected in Table~\ref{tab:Bag-parameters} and 
the HQET sum rules suggest values for the Bag parameter that are very
close to~VIA.

For the dimension-seven Bag parameters (defined in HQET), we apply
 VIA. As one can see from Appendix~\ref{Appendix-C}, the matrix
 elements of dimension-seven operators in HQET depend also on the
 parameters $\bar \Lambda_{(s)} = m_{D_{(s)}} - m_c $, for which
 we use the following ranges \cite{King:2020}
\begin{align}
\bar \Lambda ={}& (0.5 \pm 0.1 ) \, {\rm GeV},
\nonumber\\
\bar \Lambda_s ={}& (0.6 \pm 0.1 ) \, {\rm GeV}.
\end{align}

\section{Numerical Results }
\label{sec:Results}
In this section, using all the ingredients described above, we present the theoretical prediction
for the total and semileptonic decay rates, and for their ratios. 
All the input used in our numerical analysis are collected in Appendix~\ref{Appendix-A}.
For each observable, we investigate several quark mass schemes
(with the kinetic scheme as default) and compare the corresponding results using both VIA and HQET SR
values for the Bag parameters. 
The uncertainties quoted below are obtained by varying all the input parameters
within their intervals.
For the renormalisation scales, we fix the central values to
$\mu_1 = \mu_0 = 1.5 \, {\rm GeV}$ and vary both of them independently between 1 and 3 GeV.
Moreover we add an estimated uncertainty due to missing 
higher power corrections. 
The results are discussed in the following subsections and they are summarised in Tables~\ref{tab:summary-diff-schemes-VIA},
\ref{tab:summary-diff-schemes-HQET-SR},
\ref{tab:summary-with-uncertainties} and in Fig.~\ref{fig:summary-comparison}. 
\begin{table}[t]\centering
\renewcommand{\arraystretch}{1.7}
\begin{tabular}{|c|C{2.3cm}|C{2.2cm}|C{2.2cm}|C{2.2cm}||C{1.8cm}|}
\hline
\multicolumn{6}{|c|}{VIA} 
\\
\hline
Observable 
& 
Pole
&
$\overline{\rm MS}$
&
Kinetic
&
$1S$
& Exp. value  \\
\hline
\hline
$\Gamma (D^0) [{\rm ps}^{-1}]$ 
& $1.68 $ & $1.47 $ & $1.56 $ & $1.64 $
& $2.44 $ \\
\hline
$\Gamma (D^+) [{\rm ps}^{-1}]$ 
& $0.19 $ & $-0.03 $ & $0.09 $ & $0.16 $
& $0.96 $ \\
\hline
$\bar \Gamma (D_s^+) [{\rm ps}^{-1}]$ 
& $1.73 $  & $1.49 $ & $1.59 $ & $1.68 $
& $1.88 $ \\
\hline
\hline
$\tau (D^+)/\tau(D^0) $ 
& $2.55 $ & $2.56 $ & $2.53 $ & $2.54 $
& $2.54 $ \\
\hline
$\bar \tau (D_s^+)/\tau(D^0) $ 
& $0.98 $ & $0.99 $ & $0.99 $ & $0.98 $
& $1.30 $ \\
\hline
\hline
$B_{sl}^{D^0} [\%]$ 
& $5.33 $ & $6.47 $ & $6.05 $ & $5.65 $
& $6.49 $ \\
\hline
$B_{sl}^{D^+} [\%]$ 
& $13.5$ & $16.4$ & $15.3$ & $14.3$
& $16.07$ \\
\hline
$B_{sl}^{D^+_s} [\%]$ 
& $6.94 $ & $8.29 $ & $7.80 $ & $7.33 $
& $6.30 $ \\
\hline
\hline
$\Gamma_{sl}^{D^+}/\Gamma_{sl}^{D^0}$
& $1.00 $ & $1.00 $ & $1.00 $ & $1.00 $
& $0.985 $ \\
\hline
$\Gamma_{sl}^{D^+_s}/\Gamma_{sl}^{D^0}$
& $1.05$ & $1.04$ & $1.05$ & $1.05$
& $0.790$ \\
\hline
\end{tabular}
\caption{Central values of the charm observables in different
quark mass schemes using VIA for the matrix elements of the
4-quark operators compared to the corresponding experimental
values (last column).}
\label{tab:summary-diff-schemes-VIA}
\end{table}

\begin{table}[th]\centering
\renewcommand{\arraystretch}{1.7}
\begin{tabular}{|c|C{2.3cm}|C{2.2cm}|C{2.2cm}|C{2.6cm}||C{1.8cm}|}
\hline
\multicolumn{6}{|c|}{HQET SR} 
\\
\hline
Observable 
& 
Pole
&
$\overline{\rm MS}$
&
Kinetic
&
$1S$
& Exp. value  \\
\hline
\hline
$\Gamma (D^0) [{\rm ps}^{-1}]$ 
& $1.71 $ & $1.50 $ & $1.58 $ & $1.66 $
& $2.44 $ \\
\hline
$\Gamma (D^+) [{\rm ps}^{-1}]$ 
& $-0.06 $ & $-0.26$ & $-0.15 $ & $-0.08$
& $0.96 $ \\
\hline
$\bar \Gamma (D_s^+) [{\rm ps}^{-1}]$ 
& $1.71 $  & $1.47 $ & $1.57 $ & $1.66 $
& $1.88 $ \\
\hline
\hline
$\tau (D^+)/\tau(D^0) $ 
& $2.83 $ & $2.83 $ & $2.80$ & $2.82$
& $2.54 $ \\
\hline
$\bar \tau (D_s^+)/\tau(D^0) $ 
& $1.00 $ & $1.02 $ & $1.01 $ & $1.00$
& $1.30 $ \\
\hline
\hline
$B_{sl}^{D^0} [\%]$ 
& $5.16 $ & $6.35 $ & $5.91 $ & $5.50$
& $6.49 $ \\
\hline
$B_{sl}^{D^+} [\%]$ 
& $13.1$ & $16.1$ & $15.0$ & $14.0$
& $16.07$ \\
\hline
$B_{sl}^{D^+_s} [\%]$ 
& $6.93 $ & $8.22 $ & $7.76 $ & $7.31$
& $6.30 $ \\
\hline
\hline
$\Gamma_{sl}^{D^+}/\Gamma_{sl}^{D^0}$
& $1.002 $ & $1.001 $ & $1.001 $ & $1.002$
& $0.985 $ \\
\hline
$\Gamma_{sl}^{D^+_s}/\Gamma_{sl}^{D^0}$
& $1.07$ & $1.05$ & $1.06$ & $1.07$
& $0.790$ \\
\hline
\end{tabular}
\caption{Central values of the charm observables in different
quark mass schemes using  HQET sum rule results
\cite{Kirk:2017juj,King:2020} for the matrix elements of the
4-quark operators compared to the corresponding experimental
values (last column).}
\label{tab:summary-diff-schemes-HQET-SR}
\end{table}

\begin{table}[ht]\centering
\renewcommand{\arraystretch}{1.7}
\begin{tabular}{|C{3.0cm}|C{5.0cm}|C{3.0cm}|}
\hline
Observable & HQE prediction &  Exp. value \\
\hline
\hline
$\Gamma (D^0) [{\rm ps}^{-1}]$ 
& $1.59 \pm 0.36^{+0.45 \, +0.01}_{-0.36 \, -0.01}$
& $2.44 \pm 0.01 $
\\
\hline
$\Gamma (D^+) [{\rm ps}^{-1}]$ 
& $-0.15 \pm 0.76^{+0.58 \, +0.25}_{-0.27 \, -0.10}$
& $0.96 \pm 0.01$
\\
\hline
$\bar \Gamma (D_s^+) [{\rm ps}^{-1}]$ & 
$1.57 \pm 0.43^{+0.51 \, +0.02}_{-0.40 \, -0.01}$ & 
$1.88 \pm 0.02$ 
\\
\hline
\hline
$\tau (D^+)/\tau(D^0) $ 
& $2.80 \pm 0.85^{+0.01 \, +0.11}_{-0.14 \, -0.26} $
& $2.54 \pm 0.02$
\\
\hline
$\bar \tau (D_s^+)/\tau(D^0) $ 
& $1.01 \pm 0.15^{+0.02 \, +0.01}_{-0.03 \, -0.01} $
& $1.30 \pm 0.01 $
\\
\hline
\hline
$B_{sl}^{D^0} [\%]$ 
& $5.91 \pm 1.57^{+0.33}_{-0.28}$
& $6.49 \pm 0.11$
\\
\hline
$B_{sl}^{D^+} [\%]$ 
& $15.0 \pm 4.04^{+0.83}_{-0.72}$
& $16.07 \pm 0.30$
\\
\hline
$B_{sl}^{D^+_s} [\%]$ 
& $7.76 \pm 2.62^{+0.43}_{-0.38}$
& $6.30 \pm 0.16$
\\
\hline
\hline
$\Gamma_{sl}^{D^+}/\Gamma_{sl}^{D^0}$
& $1.001 \pm 0.008 \pm 0.001$
& $0.985 \pm 0.028$
\\
\hline
$\Gamma_{sl}^{D^+_s}/\Gamma_{sl}^{D^0}$
& $1.06 \pm 0.23 \pm 0.01 $
& $0.790 \pm 0.026 $
\\
\hline
\end{tabular}
\caption{HQE predictions for all the ten observables in the kinetic scheme (second column), using HQET SR results for the Bag parameters.
The first uncertainty is parametric one, second and third uncertainties are due to $\mu_1$- and $\mu_0$-scales variation, respectively.
The results are compared with the corresponding experimental measurements (third column).}
\label{tab:summary-with-uncertainties}
\end{table}

\subsection{The Total Decay Rates}
\label{SubSec:ResultsTotal}

We start by investigating the theory prediction of the total decay rates, 
which are expected to have sizable uncertainties due to the dependence of the
free quark decay on the fifth power of the charm quark mass and due to large 
perturbative and power corrections. 
The central values for the HQE prediction of the decay widths in several mass
schemes are shown in the three first rows of
Table~\ref{tab:summary-diff-schemes-VIA},
using VIA for the Bag parameters and of
Table~\ref{tab:summary-diff-schemes-HQET-SR} 
using the HQET sum rules results.
In Table~\ref{tab:summary-with-uncertainties} we show the theoretical
prediction including the corresponding uncertainties within the kinetic
mass scheme and using the HQET SR values for the
dimension-six Bag parameters -- the same result is visualised  in
Fig.~\ref{fig:summary-comparison}. In each table, 
the corresponding experimental determinations are listed in the last column. 
For the $D_s^+$ meson an additional subtlety is arising due to the
large branching fraction of the leptonic decay $D_s^+ \to \tau^+ \nu_\tau$, which is not included in the HQE, as the tau lepton is more massive than the charm quark. 
Using the experimental value of the leptonic branching ratio 
\cite{Zyla:2020zbs} (online update)
\begin{eqnarray}
{\rm Br} (D_s^+ \to \tau^+ \nu_\tau) & = & (5.48 \pm 0.23)\% \, ,
\end{eqnarray}
we therefore define a reduced decay rate $\bar{\Gamma} (D_s^+)$:
\begin{eqnarray}
    \bar{\Gamma} (D_s^+)  \equiv  
    {\Gamma} (D_s^+) - \Gamma (D_s^+ \to \tau^+ \nu_\tau) 
    & = & 
    (1.88 \pm 0.02) \, \mbox{ps}^{-1} \, ,
\end{eqnarray}
leading also to a reduced lifetime ratio
\begin{eqnarray}
    \frac{\bar{\tau} (D_s^+)}{\tau (D^0)}  & = &  1.30 \pm 0.01 \, .
\end{eqnarray}
The first and main result we deduce from
Table~\ref{tab:summary-with-uncertainties} 
and Fig.~\ref{fig:summary-comparison}, is that the HQE 
gives values of $\Gamma (D^0)$, $\Gamma (D^+)$ and 
$\Gamma (D_s^+)$ which lie in the ballpark of the experimental numbers.
Looking closer we find that our prediction for
$\Gamma (D_s)$ is in good agreement with experiment (within large uncertainties), 
while the total decay rates of  the $D^0$ and $D^+$ mesons are underestimated. As a reason for that we suspect
 missing  NNLO-QCD corrections to the free charm quark decay. 
Second, using different mass schemes yields similar results,
and further higher order correction will reduce the 
differences between these schemes.
Due to the fact that the values of the HQET Bag parameters
\cite{Kirk:2017juj,King:2020} listed in 
Table~\ref{tab:Bag-parameters} are close to the corresponding ones in VIA,
the predictions shown in Table~\ref{tab:summary-diff-schemes-VIA} 
and in Table~\ref{tab:summary-diff-schemes-HQET-SR} do not differ much. 
A peculiar role is played by the $D^+$ meson, where we get a huge
theoretical uncertainty stemming from the large negative value of the Pauli
interference contribution at dimension-six. This term actually dominates the 
total decay rate. Moreover the large negative value is further enhanced by
NLO-QCD corrections, but partly compensated by the dimension-seven contribution.
Here further studies of the Bag parameters, e.g.~via an
independent confirmation of the HQET sum rule results with lattice QCD,
as well as calculation of higher order QCD corrections to dimension-six and dimension-seven might yield 
deeper insights. 

In order to further analyse the size of the individual contributions  
to the total decay rate, we show below the numerical coefficients
of each non-perturbative parameter in the HQE, 
using the central values for the input in Appendix~\ref{Appendix-A} and 
(as an example) the kinetic scheme for the charm mass with 
$\mu^{\rm cut} = 0.5$~GeV,  namely\footnote{Here and hereafter, 
in the Bag parameters we use the same label $q$ both for $u$ or $d$-quarks, 
reflecting the isospin symmetry, namely
$\tilde B_i^u = \tilde B_i^d \equiv \tilde B_i^q$ and
$\tilde \delta_i^{ud} = \tilde \delta_i^{du} \equiv \tilde \delta_i^{qq}$,
$\tilde \delta_i^{us} = \tilde \delta_i^{ds} \equiv \tilde \delta_i^{qs}$,
$\tilde \delta_i^{su} = \tilde \delta_i^{sd} \equiv \tilde \delta_i^{sq}$.}

\begin{eqnarray}
\Gamma (D^0) 
& = & 
\Gamma_0
\biggl[ 
\underbrace{6.15}_{c_3^{\rm LO}} + 
\underbrace{2.95}_{\Delta c_3^{\rm NLO}} 
- \, 1.66 \, \frac{\mu_{\pi}^2 (D)}{\rm GeV^2}
+ 0.13 \, \frac{\mu_{G}^2 (D)}{\rm GeV^2}
+ 23.6 \, \frac{\rho_{D}^3 (D)}{\rm GeV^3}
\nonumber \\
& &
\quad 
- \, 1.60 \,  {\tilde B}_1^q 
+ 1.53 \,  {\tilde B}_2^q
- 21.0 \,  \tilde \epsilon_1^q
+ 19.2 \,  \tilde \epsilon_2^q
+ \!\!\! \underbrace{0.00}_{\rm dim-7 ,VIA}
\nonumber \\
& &
\quad
- \, 10.7 \,  \tilde \delta^{qq}_{1} 
+ 1.53 \,  \tilde \delta^{qq}_{2} 
+ 54.6 \,  \tilde \delta^{qq}_{3} 
+ 0.13 \,  \tilde \delta^{qq}_{4} 
- \, 29.2 \,  \tilde \delta^{sq}_{1} 
+ 28.8 \,  \tilde \delta^{sq}_{2} 
+ 0.56 \,  \tilde \delta^{sq}_{3} 
+ 2.36 \,  \tilde \delta^{sq}_{4} 
\biggr]
\nonumber
\\[2mm]
& = & 
6.15 \, \Gamma_0
\biggl[1 + 0.48
- 0.13 \, \frac{\mu_{\pi}^2 (D)}{0.465 \, \rm GeV^2}
+ 0.01 \, \frac{\mu_{G}^2 (D)}{0.34 \, \rm GeV^2} 
+ 0.29 \, \frac{\rho_{D}^3 (D)}{0.075 \, \rm GeV^3} 
\nonumber \\[2mm] 
& &
\quad 
- \! \underbrace{0.01}_{\rm dim-6, VIA} \!\!
- \, 0.005 \, \frac{\delta \tilde{B}_1^q}{0.02}
+ 0.005 \, \frac{\delta \tilde{B}_2^q}{ 0.02}
+ 0.137 \,\frac{\tilde \epsilon_1^q}{-0.04}
- 0.125 \, \frac{ \tilde \epsilon_2^q}{-0.04}
+ \!\!\! \underbrace{0.00}_{\rm dim-7, VIA}
\nonumber \\[2mm]
& &
\quad
- \, 0.0045 \,  r^{qq}_{1}
- 0.0004 \,  r^{qq}_{2} 
- 0.0035 \,  r^{qq}_{3}
+ 0.0000 \,  r^{qq}_{4}
\nonumber \\[2mm]
& &
\quad
- \, 0.0109 \, r^{sq}_{1} 
- 0.0079 \, r^{sq}_{2}
- 0.0000 \, r^{sq}_{3}
+ 0.0001 \, r^{sq}_{4}
\biggr] \, .
\label{eq:Gamma-D0}
\end{eqnarray}
In the second equality in Eq.~\eqref{eq:Gamma-D0} we have normalised
the HQE parameters
 $\mu_{\pi}^2 (D)$, $ \mu_{G}^2 (D)$ and $ \rho_{D}^3 (D) $
to their central values. Moreover, we introduce
\begin{eqnarray}
 \tilde{B}_i^q  =  1 +  \delta   \tilde{ B}_i^q \, ,
\end{eqnarray}
to indicate deviations from VIA and we conservatively normalise
$\delta   \tilde{ B}_i^q $ to 0.02.
The matrix elements of the colour-octet operators are normalised
to $- 0.04$ -- here using the central value of the HQET 
determination for $\tilde \epsilon_2^{\,q}$ might underestimate its effect 
due to the quoted HQET uncertainties.
Furthermore, we introduce also the ratios
$ r^{qq^\prime}_{i} \equiv \tilde \delta^{qq^\prime}_{i}/\langle \tilde \delta^{qq^\prime}_{i} \rangle$, 
with $\langle \tilde \delta^{qq^\prime}_{i} \rangle$ being the central values 
listed in Table~\ref{tab:Bag-parameters}.

For the neutral $D$ meson we find a convergent  series,
with the largest correction due to the QCD corrections to the free quark decay
and  the contribution of the Darwin operator.
Here a calculation of the NNLO-QCD corrections to the free-quark decay would be very
desirable, as  well as a more profound determination of the value of the matrix
element of the Darwin operator.
Note that since we take as a central value $\mu_1 = 1.5$ GeV, 
the coefficient of the chromomagnetic operator in Eq.~\eqref{eq:Gamma-D0} 
turns out accidentally to be very small, see Fig.~\ref{fig:cG}.
In fact, varying the renormalisation scale $\mu_1$ between 1 and 3 GeV
one finds quite sizable contribution of $\sim 5 - 10 \%$ due to $\mu_G^2 (D)$.
Because of the helicity suppression, we get only small contributions from the weak
exchange diagrams. In LO-QCD and VIA these corrections actually vanish, the small
value  $\approx -0.01$ stems from NLO-QCD corrections, which break the
helicity suppression. Nevertheless, depending on the size of the $\tilde
\epsilon_i^q$, the colour-octet operator could give contributions of a similar size
as the kinetic operator.
Finally, according to the HQET SR determination, the numerical effect of 
the eye-contractions does not seem to be pronounced.

Similarly, we get for the $D^+$-meson decay width:
\begin{eqnarray}
\Gamma (D^+) 
& = & 
\Gamma_0
\biggl[ 
\underbrace{6.15}_{c_3^{\rm LO}} + 
\underbrace{2.95}_{\Delta c_3^{\rm NLO}} 
- 1.66 \, \frac{\mu_{\pi}^2 (D)}{\rm GeV^2}
+ 0.13 \, \frac{\mu_{G}^2 (D)}{\rm GeV^2}
+ 23.6 \, \frac{\rho_{D}^3 (D)}{\rm GeV^3} 
\nonumber \\[2mm] 
& &
\quad
- \, 16.9 \,  {\tilde B}_1^q 
+ 0.56 \,  {\tilde B}_2^q
+ 84.0 \,  \tilde \epsilon_1^q
- 1.34 \,  \tilde \epsilon_2^q
+ \underbrace{6.76}_{\rm dim-7}
\nonumber \\[2mm]
& &
\quad
- \, 0.06 \,  \tilde \delta^{qq}_{1} 
+ 0.06 \,  \tilde \delta^{qq}_{2} 
- 16.8 \,  \tilde \delta^{qq}_{3} 
+ 16.9 \,  \tilde \delta^{qq}_{4} 
- \, 29.3 \,  \tilde \delta^{sq}_{1} 
+ 28.8 \,  \tilde \delta^{sq}_{2} 
+ 0.56 \,  \tilde \delta^{sq}_{3} 
+ 2.36 \,  \tilde \delta^{sq}_{4} 
\biggr]
\nonumber
\\[2mm]
& = & 
6.15 \, \Gamma_0
\biggl[1 + 0.48 - 0.13 \,  \frac{\mu_{\pi}^2 (D)}{0.465 \, \rm GeV^2}
+ 0.01 \, \frac{\mu_{G}^2(D)}{0.34 \, \rm GeV^2}  
+ 0.29 \, \frac{\rho_{D}^3(D)}{0.075\,  \rm GeV^3}
\nonumber \\[2mm] 
& &
\quad - \!\!\! \underbrace{2.66}_{\rm dim-6, VIA} \!\!\!
- \, 0.055\,  \frac{\delta { \tilde B}_1^q }{0.02}
+ 0.002 \,  \frac{\delta {\tilde B}_2^q}{0.02}
- 0.546 \,  \frac{\tilde \epsilon_1^q}{-0.04}
+ 0.009 \,  \frac{\tilde \epsilon_2^q}{-0.04}
+ \!\!\! \underbrace{1.10}_{\rm dim-7, VIA}
\nonumber \\[2mm]
& &
\quad
- \, 0.0000 \,  r^{qq}_{1}
- 0.0000 \,  r^{qq}_{2} 
+ 0.0011 \,  r^{qq}_{3}
+ 0.0008 \,  r^{qq}_{4}
\nonumber \\[2mm]
& &
\quad
- \, 0.0109 \, r^{sq}_{1} 
- 0.0080 \, r^{sq}_{2}
- 0.0000 \, r^{sq}_{3}
+ 0.0001 \, r^{sq}_{4}
\biggr] \, ,
\label{eq:Gamma-Dp}
\end{eqnarray}
where we observe huge negative corrections due to Pauli interference. 
In VIA we get from dimension-six (summing LO and NLO-QCD) a $\approx -270\%$
correction to the LO-free-quark decay.
Dimension-seven yields a large positive correction of $+110\%$. 
Because of the almost perfect cancellation between the three dominant terms,
 $16 \pi ^2 
\left( \tilde \Gamma_6^{(0)} + \alpha_s / \pi  \tilde \Gamma_6^{(1)} \right) \langle \tilde{\mathcal{O}}_6 \rangle^{\rm VIA} / m_c^3$,
$\Gamma_3$ 
and
$16 \pi ^2 
\tilde \Gamma_7^{(0)} \langle \tilde{\mathcal{O}}_7 \rangle^{\rm VIA} / m_c^4 $,
the HQE series for $\Gamma (D^+)$ becomes very sensitive to sub-dominant terms, e.g.
higher order QCD corrections to $\tilde{\Gamma}_6$, $\tilde{\Gamma}_7$, $\Gamma_3$, $\Gamma_5$ and
$\Gamma_6$, and to deviations of the Bag parameter from VIA. 
In this case it might also be interesting to further study estimates of 
higher orders in the HQE, see e.g.~Refs.~\cite{Gabbiani:2003pq,Gabbiani:2004tp}.
Else, we get for the two-quark $\Delta C= 0$ contributions the same (due to isospin) size of corrections as in the $D^0$ case and we find, based on the HQET sum rules estimates, 
again that the eye-contractions give only tiny corrections.

Finally, we have for the  $D^+_s$-meson decay width:
\begin{eqnarray}
\Gamma (D_s^+) 
& = & 
\Gamma_0
\biggl[ 
\underbrace{6.15}_{c_3^{\rm LO}} \, + 
\underbrace{2.95}_{\Delta c_3^{\rm NLO}} 
\quad 
- \, 1.66 \, \frac{\mu_{\pi}^2 (D_s)}{\rm GeV^2}
+ 0.13 \, \frac{\mu_{G}^2 (D_s)}{\rm GeV^2}  
+ 23.6 \, \frac{\rho_{D}^3 (D_s)}{\rm GeV^3} 
\nonumber \\[2mm] 
& &
\quad 
- \, 49.6 \,  {\tilde B}_1^s 
+ 48.4 \,  {\tilde B}_2^s
- 13.7 \,  \tilde \epsilon_1^s
+ 18.8 \,  \tilde \epsilon_2^s
+ \underbrace{0.63}_{\rm dim-7}
\nonumber \\[2mm]
& &
\quad 
- \, 15.8 \, \tilde \delta^{qs}_{1} 
+ 2.34 \,  \tilde \delta^{qs}_{2} 
+ 55.4 \,  \tilde \delta^{qs}_{3} 
+ 25.0 \,  \tilde \delta^{qs}_{4} 
\biggr]
\nonumber
\\[2mm]
& = & 
6.15 \, \Gamma_0
\biggl[1 + 0.48 
- 0.15 \, \frac{\mu_{\pi}^2 (D_s)}{0.555 \, \rm GeV^2}
+ 0.01 \, \frac{\mu_{G}^2(D_s)}{0.36 \, \rm GeV^2}  
+ 0.42 \, \frac{\rho_{D}^3(D_s)}{0.110 \,  \rm GeV^3}
\nonumber \\[2mm] 
& &
\quad 
- \!\!\! \underbrace{0.20}_{\rm dim-6, VIA} \!\!\!
- \, 0.161 \,  \frac{\delta {\tilde B}_1^s }{0.02}
+ 0.157 \,  \frac{{\tilde B}_2^s}{0.02}
+ 0.089 \,  \frac{\tilde \epsilon_1^s}{-0.04}
- 0.122 \,  \frac{\tilde \epsilon_2^s}{-0.04}
+ \!\!\! \underbrace{0.10}_{\rm dim-7, VIA}
\nonumber \\[2mm]
& &
\quad 
- \, 0.0064 \,  r^{qs}_{1}
- 0.0007 \,  r^{qs}_{2} 
- 0.0036 \,  r^{qs}_{3}
+ 0.0012 \,  r^{qs}_{4}
\biggr] \, ,
\label{eq:Gamma-Ds}
\end{eqnarray}
where we find again a converging series with the dominant contribution coming from
the NLO-QCD corrections to the free quark decay and the Darwin term. For the latter a
more reliable determination of the corresponding non-perturbative matrix elements would be highly desirable. In VIA, the four-quark operators show again a pronounced cancellation between dimension-six and dimension-seven contributions.

\subsection{The Lifetime Ratios}
\label{sub:lifetime-ratios}
In order to eliminate the contribution of the free-quark decay, we calculate the lifetime ratios as
{\begin{equation}
\frac{\tau (D^+_{(s)})}{\tau (D^0)} = 1 + 
\left[ \Gamma^{\rm HQE} (D^0) - \Gamma^{\rm HQE} (D^+_{(s)}) \right] \tau^{\rm exp} (D^+_{(s)})\,,
\label{eq:lifetime-ratio}    
\end{equation}
where $\Gamma^{\rm HQE} (D^0)$ and $\Gamma^{\rm HQE} (D^+_{(s)})$ are 
given in Eqs.~\eqref{eq:Gamma-D0} and~\eqref{eq:Gamma-Dp}, 
\eqref{eq:Gamma-Ds}, respectively.
In these ratios, $\Gamma_3$ cancels exactly and $\Gamma_5$ and 
$\Gamma_6 $ cancel up to isospin or $SU(3)_F$ breaking corrections 
in the corresponding non-perturbative matrix elements. The lifetime ratios should then be dominated by the contribution of four-quark operators.

The central values for the HQE prediction of the lifetime ratios
in several mass schemes are shown in the fourth and fifth  rows of
Table~\ref{tab:summary-diff-schemes-VIA},
Table~\ref{tab:summary-diff-schemes-HQET-SR},
Table~\ref{tab:summary-with-uncertainties}
and in
Fig.~\ref{fig:summary-comparison}
and it turns out that the large lifetime ratio 
$\tau (D^+) / \tau (D^0)$ is well reproduced in all schemes,
while in the case of 
$\tau (D_s^+) / \tau (D^0)$ the HQE predictions lie closer to one
compared to the experimental values. The latter theory result is dominated
by $SU(3)_F$ breaking differences of the non-perturbative matrix elements
$\mu_\pi^2$, $\mu_G^2$ and~$\rho_D^3$, which are only very roughly known,
see Section \ref{sec:HQE-NP-parameters}. With future,  more precise
determinations of these parameters our conclusion might significantly change
for this lifetime ratio.

The large lifetime ratio $\tau (D^+)/\tau (D^0)$ can be expressed as
\begin{eqnarray}
\frac{\tau (D^+)}{\tau (D^0)}
& = &
1 
+  2.46 \,  \tilde B_1^q 
+ 0.16 \,  \tilde B_2^q 
- 16.9 \,  \tilde \epsilon_1^q 
+ 3.31 \,  \tilde \epsilon_2^q  
\underbrace{- \, 1.09}_{\rm dim-7, VIA} 
\nonumber \\
& &
- \,1.71 \,  \tilde \delta_1^{qq}
+ 0.24 \,  \tilde \delta_2^{qq}
+ 1.15 \,  \tilde \delta_3^{qq}
- 2.71 \,  \tilde \delta_4^{qq}
+ 0.01 \,  \tilde \delta_1^{sq}
- 0.01 \,  \tilde \delta_2^{sq}
+ 0.00 \,  \tilde \delta_3^{sq}
+ 0.00 \,  \tilde \delta_4^{sq}
\nonumber
\\[2mm]
& = &
1 
\underbrace{+ \, \, 2.62}_{\rm dim-6, VIA} 
\underbrace{- \, \, 1.09}_{\rm \, \, dim-7, VIA}  \, \,
+ \, 0.049 \,  \frac{\delta \tilde B_1^q }{0.02}
+ 0.003 \,  \frac{\delta \tilde B_2^q }{0.02}
+ 0.676  \,  \frac{\tilde \epsilon_1^q }{-0.04}
- 0.132  \,  \frac{\tilde \epsilon_2^q }{-0.04}
\nonumber
\\[2mm]
& &
 - \, 0.004 \, r_1^{qq}
 - 0.000 \, r_2^{qq}
 - 0.005 \, r_3^{qq}
 - 0.001 \, r_4^{qq} \, .
\label{eq:tauDp-to-tauD0} 
\end{eqnarray}
In VIA, we predict a lifetime ratio of 2.5, 
which is already quite close to the experimental value. 
Again, we observe here a sizable cancellation between dimension-six and dimension-seven contributions.
In order to improve the theoretical prediction, a more precise determination of 
the Bag parameters of the colour-octet operators is mandatory, as well as of the perturbative higher order QCD corrections in $\tilde{\Gamma}_6$ and $\tilde{\Gamma}_7$.

And finally we get for the lifetime ratio $\tau(D_s^+)/\tau(D^0)$: 
\begin{eqnarray}
\frac{\tau (D_s^+)}{\tau (D^0)}
& = &
1 
+ 0.14 \, \frac{\mu_\pi^2 (D_s) - \mu_\pi^2 (D)}{\rm GeV^2}
- 0.01 \, \frac{\mu_G^2 (D_s) - \mu_G^2 (D)}{\rm GeV^2}
- 1.93 \, \frac{\rho_D^3 (D_s) - \rho_D^3 (D)}{\rm GeV^3}
\! \! \! \underbrace{- \, 0.05}_{\rm dim-7, VIA} 
\nonumber 
\\
& & 
- \, 0.13 \, \tilde B_1^q 
+ 0.13 \, \tilde B_2^q
+ 4.06 \, \tilde B_1^s 
- 3.96 \,  \tilde B_2^s
- 1.72 \,  \tilde \epsilon_1^q 
+ 1.57 \,  \tilde \epsilon_2^q  
+ 1.12 \,  \tilde \epsilon_1^s 
- 1.54 \,  \tilde \epsilon_2^s  
\nonumber \\[2mm]
& &
- \, 0.88 \,  \tilde \delta_1^{qq}
+ 0.13 \,  \tilde \delta_2^{qq}
+ 4.47 \,  \tilde \delta_3^{qq}
+ 0.01 \,  \tilde \delta_4^{qq}
- 2.39 \,  \tilde \delta_1^{qs}
+ 2.36 \,  \tilde \delta_2^{qs}
+ 0.05 \,  \tilde \delta_3^{qs}
+ 0.19 \,  \tilde \delta_4^{qs}
\nonumber \\[2mm]
& & 
+ \, 1.29 \,  \tilde \delta_1^{sq}
- 0.19 \,  \tilde \delta_2^{sq}
- 4.54 \,  \tilde \delta_3^{sq}
- 2.04 \,  \tilde \delta_4^{sq}
\nonumber \\[2mm]
& = &
1 
+ 0.012 \, \frac{\mu_\pi^2 (D_s) - \mu_\pi^2 (D)}{0.09 \, \rm GeV^2}
- 0.0002 \, \frac{\mu_G^2 (D_s) - \mu_G^2 (D)}{0.02 \, \rm GeV^2}
- 0.068 \, \frac{\rho_D^3 (D_s) - \rho_D^3 (D)}{0.035 \, \rm GeV^3}
\nonumber 
\\[2mm]
& & 
\underbrace{+ \, 0.10}_{\rm dim-6, VIA} 
\underbrace{- \, 0.05}_{\rm dim-7, VIA} 
- \, 0.003 \, \frac{\delta \tilde B_1^q }{0.02}
+ 0.003 \, \frac{\delta \tilde B_2^q}{0.02}
+ 0.081 \, \frac{\delta \tilde B_1^s}{0.02} 
- 0.079 \, \frac{\delta \tilde B_2^s}{0.02}
\nonumber 
\\
& & 
+ \, 0.069 \,  \frac{\tilde \epsilon_1^q}{-0.04}
- 0.063 \,  \frac{\tilde \epsilon_2^q}{-0.04}  
- 0.045 \,  \frac{\tilde \epsilon_1^s}{-0.04} 
+ 0.062 \,  \frac{\tilde \epsilon_2^s}{-0.04} 
\nonumber \\[2mm]
& &
- \, 0.0023 \,  r_1^{qq}
- 0.0002 \,  r_2^{qq}
- 0.0018 \,  r_3^{qq}
+ 0.0000 \, r_4^{qq}
\nonumber \\[2mm]
& &
- \, 0.0055 \, r_1^{qs}
- 0.0040 \,  r_2^{qs}
- 0.0000 \,  r_3^{qs}
+ 0.0001 \,  r_4^{qs}
\nonumber \\[2mm]
& & 
+ \, 0.0032 \,  r_1^{sq}
+ 0.0003 \,  r_2^{sq}
+ 0.0018 \,  r_3^{sq}
- 0.0006 \,  r_4^{sq} \, .
\label{eq:tauDs-to-tauD0}
\end{eqnarray}
With the estimates of $\mu_\pi^2$, $\mu_G^2$ and  $\rho_D^3$ 
from Section \ref{sec:HQE-NP-parameters} we find
that the largest individual $SU(3)_F$ breaking effect 
($\approx - 7\%$) comes from the Darwin term. Using VIA we obtain a
correction of $+ 5\%$ due to the four-quark contributions of dimension-six
and dimension-seven -- finite values of the matrix elements of the colour-octet
operators as well as of $\delta \tilde{B}_{1,2}^s$ might lead to numerically similar effects.
Else we have a large number of smaller $SU(3)_F$ breaking effects, which can
be both positive and negative. 
}

\subsection{The Semileptonic Decay Widths and Their Ratios}
\label{sub:semileptonic}
\begin{figure}[t]
    \centering
    \includegraphics[scale=1.0]{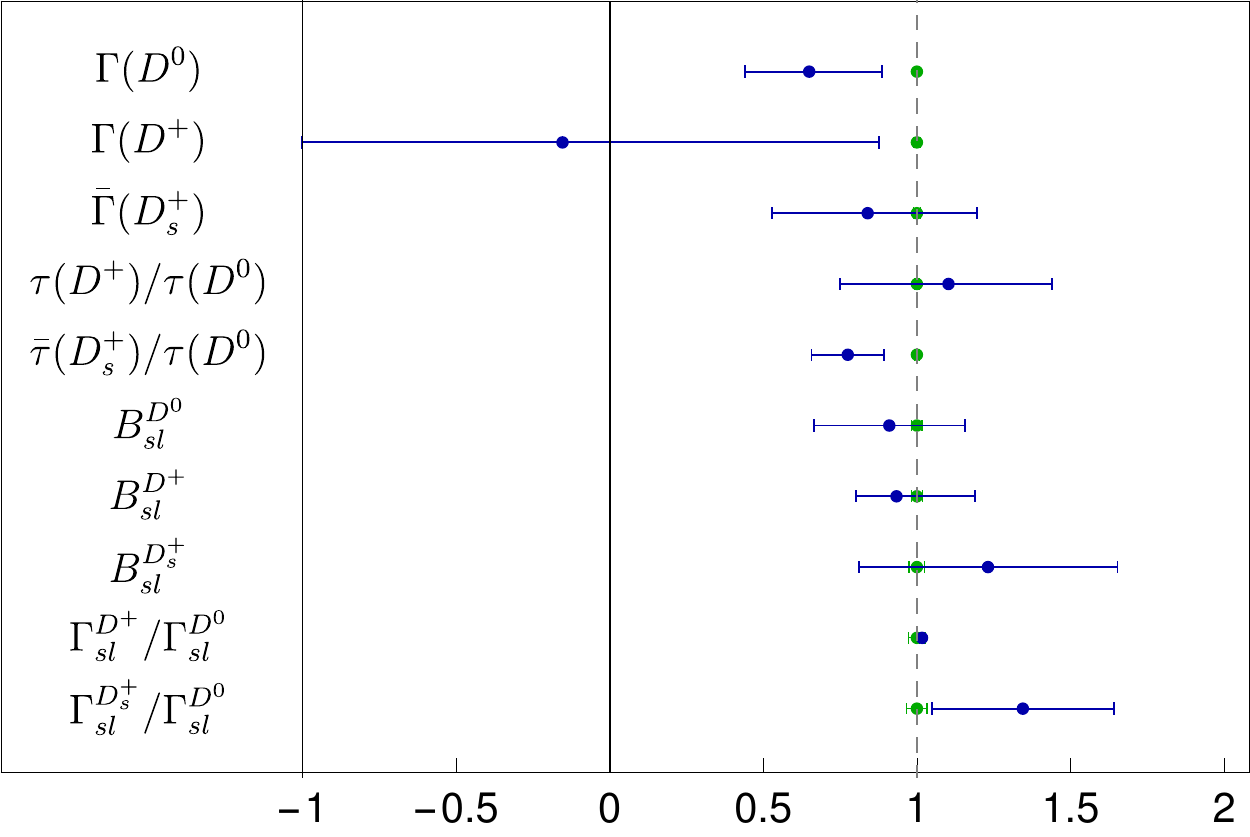}
    \caption{A comparison of the HQE prediction for the charm observables in the kinetic scheme  (blue)  with the corresponding experimental data (green).
    All the quantities are normalised to the corresponding experimental central values.}
    \label{fig:summary-comparison}
\end{figure}

For  discussing the inclusive semileptonic decays of $D$ mesons, we 
introduce the short-hand notations $\Gamma_{sl}^D \equiv \Gamma (D \to X e^+ \nu_e)$
and $B_{sl}^D \equiv {\rm Br} (D \to X e^+ \nu_e)$.
We determine the theory value of the semileptonic branching ratio as
\begin{equation}
 B_{sl}^{D, \rm HQE}    =  
 \Gamma_{sl}^{D, \rm HQE}  
 \, \cdot \, \tau (D)^{\rm exp}  \, .
\end{equation}
The central values for the HQE prediction of the lifetime ratios
in several mass schemes are shown in the sixth, seventh and eighth row of
Table~\ref{tab:summary-diff-schemes-VIA},
Table~\ref{tab:summary-diff-schemes-HQET-SR}
and
Table~\ref{tab:summary-with-uncertainties}
and in
Fig.~\ref{fig:summary-comparison}.

The semileptonic decay rate of the $D^0$ meson can be written 
(in the kinetic scheme) as
\begin{eqnarray}
\Gamma^{D^0}_{sl} 
& = & 
\Gamma_0
\biggl[ 
\underbrace{1.02}_{c_3^{\rm LO}} \, +
\underbrace{0.16}_{\Delta c_3^{\rm NLO}} 
- \, 0.27 \, \frac{\mu_{\pi}^2 (D)}{\rm GeV^2}
- 0.84 \, \frac{\mu_{G}^2 (D)}{\rm GeV^2}  
+ 2.48 \, \frac{\rho_{D}^3 (D)}{\rm GeV^3} 
\nonumber 
\\
& &
\quad 
- \, 0.28 \,  \tilde \delta^{qq}_{1} 
+ 0.28 \,  \tilde \delta^{qq}_{2} 
- 5.23 \,  \tilde \delta^{sq}_{1} 
+ 5.23 \,  \tilde \delta^{sq}_{2} 
\biggr] 
\nonumber \\
& = & 
1.02 \, \Gamma_0
\biggl[1 + 0.16
- 0.13 \, \frac{\mu_{\pi}^2 (D)}{0.465 \, \rm GeV^2}
- 0.28 \, \frac{\mu_{G}^2 (D)}{0.34 \, \rm GeV^2} 
+ 0.18 \, \frac{\rho_{D}^3 (D)}{0.075 \, \rm GeV^3} 
\nonumber \\[2mm] 
& &
\quad 
- \, 0.0007 \, r^{qq}_{1}
- 0.0005 \, r^{qq}_{2} 
- 0.0118 \, r^{sq}_{1} 
- 0.0087 \, r^{sq}_{2}
\biggr] \, ,
\label{eq:Gamma-SL-D0}
\end{eqnarray}
where as for the total $D^0$-meson decay width we find a converging series, 
with the largest correction due to the dimension-five operators, 
followed by the Darwin operator contribution and the NLO-QCD corrections 
to the free quark decay. 
Note that only the non-valence four-quark operator contributions (eye-contractions)
are present here.

For the semileptonic $D^+$-meson decay we obtain
\begin{eqnarray}
\Gamma_{sl}^{D^+} 
& = & 
\Gamma_0
\biggl[ 
\underbrace{1.02}_{c_3^{\rm LO}} \, +
\underbrace{0.16}_{\Delta c_3^{\rm NLO}} 
- \, 0.27 \, \frac{\mu_{\pi}^2 (D)}{\rm GeV^2}
- 0.84 \, \frac{\mu_{G}^2 (D)}{\rm GeV^2}  
+ 2.48 \, \frac{\rho_{D}^3 (D)}{\rm GeV^3} 
+ \! \! \! \underbrace{0.00}_{\rm dim-7, VIA}
\nonumber 
\\
& &
\quad
- \, 0.28 \,  {\tilde B}_1^q 
+ 0.28 \,  {\tilde B}_2^q
- 0.09 \,  \tilde \epsilon_1^q
+ 0.09 \,  \tilde \epsilon_2^q
- 5.24 \,  \tilde \delta^{sq}_{1} 
+ 5.24 \,  \tilde \delta^{sq}_{2} 
\biggr]
\nonumber \\
& = & 
1.02 \, \Gamma_0
\biggl[1 + 0.16
- 0.13 \, \frac{\mu_{\pi}^2 (D)}{0.465 \, \rm GeV^2}
- 0.28 \, \frac{\mu_{G}^2 (D)}{0.34 \, \rm GeV^2} 
+ 0.18 \, \frac{\rho_{D}^3 (D)}{0.075 \, \rm GeV^3} 
\nonumber \\[2mm] 
& &
\quad 
- \!\! \underbrace{0.00}_{\rm dim-6,7, VIA} \!\!\!
- \, 0.005 \, \frac{\delta {\tilde B}_1^q}{0.02}
+ 0.005 \, \frac{\delta {\tilde B}_2^q}{0.02}
+ 0.004 \,  \frac{\tilde \epsilon_1^q}{-0.04}
- 0.004 \,  \frac{\tilde \epsilon_2^q}{-0.04}
\nonumber \\
& &
\quad 
- \, 0.0118 \, r^{sq}_{1} 
- 0.0088 \, r^{sq}_{2}
\biggr] \, ,
\label{eq:Gamma-SL-Dp}
\end{eqnarray}
where we find the same series as for the neutral $D$-meson 
supplemented by contributions from CKM suppressed weak annihilation, 
which vanish in VIA both at dimension-six and dimension-seven. 
Deviations from VIA give very small corrections.

For the $D^+_s$-meson we obtain
\begin{eqnarray}
\Gamma_{sl}^{D_s^+}
& = & 
\Gamma_0
\biggl[ 
\underbrace{1.02}_{c_3^{\rm LO}} \, +
\underbrace{0.16}_{\Delta c_3^{\rm NLO}} 
- \, 0.27 \, \frac{\mu_{\pi}^2 (D_s)}{\rm GeV^2}
- 0.84 \, \frac{\mu_{G}^2 (D_s) }{\rm GeV^2}
+ 2.48 \, \frac{\rho_{D}^3 (D_s)}{\rm GeV^3}
+ \! \! \! \underbrace{0.00}_{\rm dim-7, VIA}
\nonumber 
\\
& &
\quad 
- \, 7.63 \,  {\tilde B}_1^s 
+ 7.63 \,  {\tilde B}_2^s
- 2.55 \,  \tilde \epsilon_1^s
+ 2.37 \,  \tilde \epsilon_2^s
- 0.41 \,  \tilde \delta^{qs}_{1} 
+ 0.41 \,  \tilde \delta^{qs}_{2} 
\biggr]
\nonumber \\
& = & 
1.02 \, \Gamma_0
\biggl[1 + 0.16
- 0.15 \, \frac{\mu_{\pi}^2 (D_s)}{0.555 \, \rm GeV^2}
- 0.30 \, \frac{\mu_{G}^2 (D_s)}{0.36 \, \rm GeV^2} 
+ 0.27 \, \frac{\rho_{D}^3 (D_s)}{0.110 \, \rm GeV^3} 
\nonumber \\[2mm] 
& &
\quad
- \!\! \underbrace{0.00}_{\rm dim-6, VIA} \!\!\!
- \, 0.15 \, \frac{\delta {\tilde B}_1^s}{0.02}
+ 0.15 \, \frac{\delta {\tilde B}_2^s}{0.02}
+ 0.10 \,  \frac{\tilde \epsilon_1^s}{-0.04}
- 0.09 \,  \frac{\tilde \epsilon_2^s}{-0.04}
\nonumber \\
& &
\quad 
- \, 0.0010 \, r^{qs}_{1}
- 0.0007 \, r^{qs}_{2} 
\biggr] \, ,
\label{eq:Gamma-SL-Ds}
\end{eqnarray}
where we have a larger contributions due to CKM dominant weak annihilation
as well as $SU (3)_F$ breaking corrections.
Again, in VIA the four-quark contributions vanish both at dimension-six and dimension-seven, 
but now deviations from VIA might give sizable corrections.

Using the experimental values for the $D^0$ lifetime and semileptonic branching fraction, we determine the semileptonic ratios in the following way
\begin{eqnarray}
\frac{\Gamma_{sl}^{D^+} }{\Gamma_{sl}^{D^0} }
& = & 
1 + \left[\Gamma_{sl}^{D^+}  - \Gamma_{sl}^{D^0}\right]^{\rm HQE}
\left[\frac{\tau(D^0)}{B_{sl}^{D^0}}\right]^{\rm exp} 
 \, ,
\\[2mm]
\frac{\Gamma_{sl}^{D_s^+}}{\Gamma_{sl}^{D^0}}
& = & 
1 + \left[\Gamma_{sl}^{D^+_s}  - \Gamma_{sl}^{D^0}\right]^{\rm HQE}
\left[\frac{\tau(D^0)}{B_{sl}^{D^0}}\right]^{\rm exp} 
\, ,
\end{eqnarray}
where $\left[\Gamma_{sl}^{D^0}\right]^{\rm HQE}$, $\left[\Gamma_{sl}^{D^+}\right]^{\rm HQE}$ 
and $\left[\Gamma_{sl}^{D^+_s}\right]^{\rm HQE}$ are given in Eqs.~\eqref{eq:Gamma-SL-D0}, \eqref{eq:Gamma-SL-Dp} and \eqref{eq:Gamma-SL-Ds}, respectively. \\[2mm]
The HQE values of these ratios are shown in the ninth and tenth rows of
Tables~\ref{tab:summary-diff-schemes-VIA}, \ref{tab:summary-diff-schemes-HQET-SR}
and~\ref{tab:summary-with-uncertainties}
and in Fig.~\ref{fig:summary-comparison}. In agreement with experiment
HQE predicts values for $\Gamma_{sl}^{D^+}/\Gamma_{sl}^{D^0}$
very close to one. Using the inputs from 
Appendix~\ref{Appendix-A} the HQE prefers also 
for $\Gamma_{sl}^{D_s^+}/\Gamma_{sl}^{D^0}$ values close to one,
while experiment find a value as low as 0.79 -- again a more profound
determination of $\mu_G^2$, $\mu_\pi^2$ and $\rho_D^3$ 
as well as an inclusion of dimension-seven contributions with two-quarks operators
for $D$ mesons might change this conclusion.

We expand $\Gamma_{sl}^{D^+}/\Gamma_{sl}^{D^0}$ as
\begin{eqnarray}
\frac{\Gamma_{sl}^{D^+}}{\Gamma_{sl}^{D^0}}
& = &
1
- \, 0.27 \,  \tilde B_1^q 
+ 0.27 \,  \tilde B_2^q 
- 0.09 \,  \tilde \epsilon_1^q 
+ 0.08 \,  \tilde \epsilon_2^q 
\underbrace{+ \, 0.00}_{\rm dim-7, VIA} \! \! 
\nonumber \\
& &
+ \, 0.27 \,  \tilde \delta_1^{qq}
- 0.27 \,  \tilde \delta_2^{qq}
- 0.01 \,  \tilde \delta_1^{sq}
+ 0.01 \,  \tilde \delta_2^{sq}
\nonumber
\\
& = &
1
+ \!\!\! \underbrace{0.00}_{\rm dim-6,7, VIA} \!\!\! 
- \, 0.005 \,  \frac{\delta \tilde B_1^q}{0.02}
+ 0.005 \,  \frac{\delta \tilde B_2^q}{0.02}
+ 0.004 \,  \frac{\tilde \epsilon_1^q}{-0.04} 
- 0.003 \,  \frac{\tilde \epsilon_2^q }{-0.04} \,.
\label{eq:GammaDp-to-GammaD0-SL}
\end{eqnarray}
Due to isospin symmetry, in Eq.~(\ref{eq:GammaDp-to-GammaD0-SL}) the contributions of the kinetic, chromomagnetic and the Darwin operators vanish. Moreover, in VIA there is also no correction due to the spectator quark effects. Thus this ratio, within the framework of the HQE, is predicted to be very close to one.

Finally, we obtain for the ratio 
$\Gamma_{sl}^{D_s^+} / \Gamma_{sl}^{D^0}$ \footnote{We note here a typo in the corresponding expression of this ratio in Ref.~\cite{Lenz:2013aua}. In Eq.~(40) of Ref.~\cite{Lenz:2013aua}, the sign in front of the contribution of the kinetic operator has to be changed.}
\begin{eqnarray}
\frac{\Gamma_{sl}^{D_s^+}}{\Gamma_{sl}^{D^0}}
& = &
1 
- 0.27 \, \frac{\mu_\pi^2 (D_s) - \mu_\pi^2 (D)}{\rm GeV^2}
- 0.82 \, \frac{\mu_G^2 (D_s) - \mu_G^2 (D)}{\rm GeV^2}
+ 2.42 \, \frac{\rho_D^3 (D_s) - \rho_D^3 (D)}{\rm GeV^3}
\nonumber 
\\[2mm]
& & 
- \, 7.47 \,  \tilde B_1^s 
+ 7.47 \,  \tilde B_2^s
- 2.50 \,  \tilde \epsilon_1^s 
+ 2.32 \,  \tilde \epsilon_2^s  
\underbrace{+ \, 0.00}_{\rm dim-7, VIA} \! \! 
\nonumber \\
& &
+ \, 0.27 \,  \tilde \delta_1^{qq}
- 0.27 \,  \tilde \delta_2^{qq}
+ 5.11 \,  \tilde \delta_1^{sq}
- 5.11 \,  \tilde \delta_2^{sq}
- 0.40 \,  \tilde \delta_1^{qs}
+ 0.40 \,  \tilde \delta_2^{qs}
\nonumber 
\\[2mm]
& = &
1 
- 0.024 \, \frac{\mu_\pi^2 (D_s) - \mu_\pi^2 (D)}{0.09 \, \rm GeV^2}
- 0.016 \, \frac{\mu_G^2 (D_s) - \mu_G^2 (D)}{0.02 \, \rm GeV^2}
+ 0.085 \,  \frac{\rho_D^3 (D_s) - \rho_D^3 (D)}{0.035 \, \rm GeV^3}
\nonumber 
\\
& & 
+ \!\!\! \underbrace{0.00}_{\rm dim-6,7, VIA}
- \, 0.15 \,  \frac{\delta \tilde B_1^s}{0.02}
+ 0.15 \,  \frac{\delta \tilde B_2^s}{0.02}
+ 0.10 \,  \frac{\tilde \epsilon_1^s}{-0.04} 
- 0.09 \,  \frac{\tilde \epsilon_2^s}{-0.04}  
\nonumber \\[2mm]
& &
+ \, 0.0007  \, r_1^{qq}
+ 0.0005  \, r_2^{qq}
+ 0.0118  \, r_1^{sq}
+ 0.0087  \, r_2^{sq}
- 0.0001  \, r_1^{qs}
- 0.0007  \, r_2^{qs} \, ,
\label{eq:GammaDs-to-GammaD0-SL}
\end{eqnarray}
which is dominated by $SU(3)_F$-symmetry breaking corrections. 
The Darwin operator gives a sizable positive contribution to the ratio, which is partly compensated by the kinetic and the chromomagnetic terms.
The spectator effects give in VIA a vanishing contribution, but deviations from VIA could sizably affect the ratio and also eye-contractions could yield a visible effect -- here again a more precise determination of the non-perturbative parameters is necessary in order to make more profound statements.

\section{Conclusions and Outlook}
\label{sec:conclusion}
We have performed a comprehensive study of charmed mesons
lifetimes, of their ratios and of the inclusive semileptonic
decay rates. Compared to previous studies we have included for the first
time the sizeable contribution due to the Darwin term in the charm sector (with new expressions shown in Appendix~\ref{Appendix-B}),
non-perturbative estimates of the eye-contractions 
\cite{King:2020} and strange quark mass corrections to the Bag parameters of the $D_s^+$ meson \cite{King:2020}. 
Moreover we have studied different mass schemes for the charm quark.

In particular our new study supersedes the one done by some of us in Ref.~\cite{Kirk:2017juj} and we could clarify in the present work that the dimension-seven operators $\tilde{R}_{1,2}^q$ (introduced in
Ref.~\cite{Kirk:2017juj} as $P_{5,6}^q$) can be absorbed in the definition of the QCD decay constant.
In contrast to the present work, Ref.~\cite{Kirk:2017juj}
could describe the experimental number for $\tau (D_s^+) / \tau (D^0) $ by fitting the Bag parameters in order to accommodate the experimental value
of $\Gamma_{sl}^{D_s^+} / \Gamma_{sl}^{D^0}$ -- this can be achieved by 
demanding e.g. for the difference $\tilde{B}^s_1- \tilde{B}^s_2 \approx 0.032$, 
which is in slight tension with the HQET sum rule result 
$\tilde{B}^s_1- \tilde{B}^s_2 = 0.004^{+0.019}_{-0.012}$ we are using here.

Ref.~\cite{Cheng:2018rkz} also studies charm mesons,
albeit restricting exclusively to LO-QCD expressions. 
Different quark mass schemes can only be differentiated starting
from NLO-QCD onwards -- working at leading order in QCD only, the
different quark mass schemes used in our work would induce a
relative uncertainty of the free-quark decay of $(1.48/1.27)^5
\approx 2.15$, which is clearly not acceptable. Moreover, as can
be nicely read from Table~\ref{tab:dim-6-NLO-vs-LO}, NLO-QCD
corrections to the four-quark operators can dominate over the LO
contribution. Using exclusively LO-QCD expressions is thus a far too crude 
and unnecessary assumption in the charm sector. 
Finally, Ref.~\cite{Cheng:2018rkz}  considers only the $\overline{\rm MS}$ 
scheme for the charm quark mass
and obviously the recently determined Darwin term and the eye-contractions could not have been included, since they were not known at that point of time.

Finally there is also some overlap with two recent studies of the
$B_c$ lifetime \cite{Aebischer:2021ilm, Aebischer:2021eio}. The
first paper~\cite{Aebischer:2021ilm} considers also the free
charm quark decay $\Gamma_3$ and the second one \cite{Aebischer:2021eio}
the total $D$ meson decay rate without the free charm quark
decay, i.e. $\Gamma(D) - \Gamma_3$. For $\Gamma_3$ the authors 
of  Ref.~\cite{Aebischer:2021ilm}
consider three quark mass schemes: $\overline{\rm MS}$, $1S$ and the meson
mass scheme.  They find in Table 3 and 4 of their paper values in the
$\overline{\rm MS}$/$1S$ scheme which are slightly smaller/slightly larger than our
values in Table \ref{tab:Gamma_3}: $1.0 \, {\rm ps}^{-1}$ vs 
$1.3 \, {\rm ps}^{-1}$ and $1.7 \, {\rm ps}^{-1}$ vs $1.5 \,{\rm ps}^{-1}$. 
Since they in principle use the same NLO-QCD expressions as we do, 
we expect the slight difference to root in a different treatment of higher orders 
in $\alpha_s$ and some differences in the values of the input parameters.
As in our study, they also find a relatively small effect 
due to a non-vanishing strange quark mass.
In Ref.~\cite{Aebischer:2021eio} the authors
determine the $D$ decay rate without the free charm quark decay. 
In that respect they consider all the corrections we also take into account, 
except contributions of dimension-seven and eye-contractions. 
In the end, when considering the $D^+$ meson they obtain values
for the $B_c$-meson decay rate of around $3.3 \, {\rm ps}^{-1}$ 
(see Table III of Ref.~\cite{Aebischer:2021eio}), compared to the
experimental value of $1.961(35) \, {\rm ps}^{-1}$. 
We naively estimate that an inclusion of the dimension-seven contribution to the $D^+$
meson decay rate would decrease their result by about $1.1 \, {\rm ps}^{-1}$, 
see Table \ref{tab:dim-7-VIA}, to bring it in nice agreement with the measurement.
On the other hand, these missing dimension-seven contributions might be partially compensated by the corresponding contributions to the $B_c$-meson decay rate. 
Here a further investigation might be necessary to clarify this point.

Our main numerical results are presented in Tables~\ref{tab:summary-diff-schemes-VIA}, \ref{tab:summary-diff-schemes-HQET-SR} and \ref{tab:summary-with-uncertainties} 
and in Fig.~\ref{fig:summary-comparison}. 
At a first glance all considered observables lie in the ballpark of the experimental results.
In particular, we find good agreement with experiment for the ratio $\tau(D^+)/\tau(D^0)$, for the total 
$D_s^+$-meson decay rate, for the semileptonic rates of all three mesons  $D^0$, $D^+$ and $D_s^+$, and for 
the semileptonic ratio $\Gamma_{sl}^{D^+}/\Gamma_{sl}^{D^0}$.  
The values obtained with different mass schemes for the charm quark overlap
and the exclusive use of only one scheme might  underestimate the
uncertainties. Including higher orders in the perturbative QCD expansion 
will further alleviate the differences among the mass schemes.
Looking at the structure of the contributions to the total decay rates and neglecting spectator effects for a start,
we find that the NLO-QCD corrections to the free quark decay
give the dominant correction  (of the order of $50\%$ 
of LO-QCD free quark decay), 
followed by the Darwin term  (of the order of $30\%$ of LO-QCD free quark decay). 
In the case of semileptonic decay rates the chromomagnetic term provides 
the dominant contribution (of the order of $30\% $),
followed by the Darwin term and NLO-QCD corrections to the free quark decay.
Turning now to the spectator effects, we find them to be  tiny for 
$\Gamma (D^0)$, $\Gamma_{sl}^{D^0}$ and
$\Gamma_{sl}^{D^+}$, but they provide visible corrections
to $\Gamma_{sl}^{D_s^+}$ and $\Gamma (D_s^+)$ 
-- in the latter case we find also sizable cancellations
between dimension-six and dimension-seven contributions.
For the  $D^+$ meson we find, however, a huge negative Pauli interference contribution
-- with a substantial part stemming from the NLO-QCD corrections. Moreover, one observes here a significant 
cancellation between dimension-six and dimension-seven terms related to Pauli interference.
The values of the HQET Bag parameters entering the spectator effects are close to
the VIA values, deviations from the latter can, however, lead to sizable effects in $\Gamma (D^+)$
and to visible effects in  $\Gamma (D^0)$,  $\Gamma (D_s^+)$ and  $\Gamma_{sl}^{D_s^+}$.
Based on the HQET sum rule results \cite{King:2020}
we find that eye-contractions constitute only subleading  corrections, 
 they might, however, turn out to be relevant for 
$\Gamma_{sl}^{D_s^+} / \Gamma_{sl}^{D^0}$ and $\tau (D_s^+)/\tau(D^0)$, 
 when more precise non-perturbative estimates will become available.
In the end, the total decay rates of the $D^0$ and $D^+$ mesons stay underestimated in our HQE  approach and we suspect
that this is  due to missing  higher-order QCD corrections to the free charm quark decay and the Pauli interference
contribution. For the $SU(3)_F$ breaking ratios $\tau (D_s^+) / \tau (D^0) $ and $\Gamma_{sl}^{D_s^+}/\Gamma_{sl}^{D^0} $
our predictions lie closer to one than experiment. This might originate from  the poor knowledge of the 
non-perturbative parameters $\mu_G^2$, $\mu_\pi^2$ and $\rho_D^3$ in the $D^0$ and $D_s^+$ systems, as 
discussed in Section \ref{sec:HQE-NP-parameters}.

Our numerical analysis shows that there are many possibilities for future improvements of the HQE predictions in the charm sector:
\begin{itemize}
\item $\Gamma_3^{(2)}$: 
NNLO-QCD \cite{Czarnecki:1997hc,Czarnecki:1998kt,vanRitbergen:1999gs,Melnikov:2008qs,Pak:2008cp,Pak:2008qt,Dowling:2008ap,Bonciani:2008wf,Biswas:2009rb,Brucherseifer:2013cu,Fael:2020tow,Czakon:2021ybq} contributions to the semileptonic decays have been found to be large and NLO-QCD corrections to the non-leptonic decay rates represent one of the dominant corrections.
Moreover we observe that at NLO-QCD there is pronounced cancellation - see Eq.~(\ref{eq:cancel1}) and
 Eq.~(\ref{eq:cancel2}) - which might not be necessarily present at NNLO-QCD.
Thus a first determination of the NNLO-QCD corrections to the non-leptonic decays might have some sizable impact on the numerical studies of the total decay rates. 

\item  $\Gamma_5^{(1)}$: 
Cancellations in the coefficient $c_G$ for the total decay rate, shown in Eq.~(\ref{tab:cG}) and Fig.~\ref{fig:cG} lead to large uncertainties, even the sign of these corrections 
      is ambiguous. Here a determination of the QCD-corrections to the coefficient 
      $c_G$ for the non-leptonic case might considerably improve the situation.
\item $\Gamma_6^{(1)}$: The Wilson oefficients of the Darwin operator are large, therefore QCD corrections for the non-leptonic case  might be important.

\item $\Gamma_{7,8}^{(0)}$: Since the dimension-six  contribution is sizable, the LO-QCD determination of the dimension-seven and dimension-eight contributions with two-quark operators for the non-leptonic case might bring some additional insights on the convergence of the HQE in the charm sector.

\item $\tilde{\Gamma}_{6}^{(2)}$, $\tilde{\Gamma}_{7}^{(1)}$: Pauli interference dominates the total decay rate of the $D^+$ meson. Currently
 $\tilde{\Gamma}_{6}^{(0)}$,  $\tilde{\Gamma}_{6}^{(1)}$ and $\tilde{\Gamma}_{7}^{(0)}$ are known and their numerical values were found to be huge, see e.g.~Table~\ref{tab:dim-6-NLO-vs-LO}. Thus further QCD corrections will turn out to be very important. 

\item $\tilde{\Gamma}_{8}^{(0)}$: 
since the four-quark dimension-six contribution can dominate the total decay rate and $\tilde{\Gamma}_{7}^{(0)}$ is also very sizable, a further study of the dimension-eight contributions might bring further insights on the convergence of the HQE in the charm sector, see Refs.~\cite{Gabbiani:2003pq,Gabbiani:2004tp}.

\item 
More precise determinations for the parameters $\mu_G^2$, $\mu_\pi^2$ and $\rho_D^3$ -- both for the $D^0$ and the $D_s^+$ mesons: the Darwin term and the chromomagnetic term provide large corrections to the decay rates and they are poorly known 
-- in particular the size of $SU(3)_F$ breaking effects is largely unknown.
An experimental determination of $\mu_G^2$, $\mu_\pi^2$ and $\rho_D^3$ from fits
            to semileptonic
            $D^+$-, $D^0$- and $D_s^+$-meson decays - as done in the $B$ system, see e.g.~Ref.~\cite{Alberti:2014yda} -- would be very desirable. This might be doable at BESIII,  Belle II and a future tau-charm factory. 
     Moreover, new theoretical determinations, e.g. via lattice simulations or sum rules could be undertaken.
\item Independent lattice determination of the matrix elements of the four-quark operators of dimension-six: here we have currently  only HQET sum rule determinations \cite{Kirk:2017juj,King:2020} or outdated lattice results \cite{DiPierro:1998ty,Becirevic:2001fy}.
\item A first non-perturbative determination of the matrix elements of the dimension-seven four-quark operators in order to test
the validity of VIA. A similar endeavour has already been performed for $B_s$ mixing~\cite{Davies:2019gnp}.
\end{itemize}

 Overall, we find that the HQE can describe inclusive charm observables, 
in which no pronounced GIM cancellation arises\footnote{See e.g. Ref.~\cite{Lenz:2020efu} 
for a recent discussion of the extreme GIM cancellations in mixing of neutral $D$
mesons.}, albeit with very large uncertainties. We therefore do not observe a clear signal for a breakdown of the
HQE in the charm sector or of violations of quark hadron duality, see e.g.
Ref.~\cite{Jubb:2016mvq} and we presented a long list of potential theoretical improvements,
which might shed further light into the convergence properties of the HQE in the charm sector.

\section*{Acknowledgments}
The work of M.L.P. is supported by Deutsche Forschungsgemeinschaft
(DFG, German Research Foundation) through grant  396021762 -TRR
257 “Particle Physics Phenomenology after the Higgs Discovery”,
the work of D.K. and C.V. was supported by the STFC grant of the
IPPP.
We would like to thank T. Mannel, B. Melic, D. Moreno, I. Nisandzic, 
and A. Pivovarov for insightful discussions.
\appendix

\section{Numerical Input}
\label{Appendix-A}
We use five-loop running for $\alpha_s (\mu)$ \cite{Herren:2017osy} 
with four active flavours at the scale $\mu \sim m_c$,
and the most recent value \cite{Zyla:2020zbs}
\begin{eqnarray}
    \alpha_s (M_Z) & = & 0.1179 \pm 0.0010.
\end{eqnarray}
For the CKM matrix elements we apply the standard parametrisation 
in terms of $\theta_{12}, \theta_{13}, \theta_{22}, \delta$
and use as an input \cite{Charles:2004jd} (online update)
\begin{eqnarray}
|V_{us}| & = & 
0.224834^{+0.000252}_{-0.000059} \, , \\
\frac{|V_{ub}|}{|V_{cb}|} & = & 
0.088496^{+0.001885}_{-0.002244} \, ,
\\
|V_{cb}|& = &
0.04162^{+0.00026}_{-0.00080} \, ,
\\
\delta & = & 
\left(65.80^{+0.94}_{-1.29}\right)^\circ .
\end{eqnarray}
For the $c$-quark mass, we use different values 
depending of the scheme. In the $\overline{\rm MS}$-scheme we take~\cite{Zyla:2020zbs}:
\begin{equation}
\overline{m}_c (\overline{m}_c) =  
(1.27 \pm 0.02) \, {\rm GeV}, 
\end{equation}
in the kinetic scheme we employ (at NLO)~\cite{Fael:2020njb}:
\begin{equation}
m_c^{\rm kin} (0.5 \, {\rm GeV})
= (1.306 \pm 0.02) \, {\rm GeV},
\end{equation}
and in the $1S$-scheme (see Eq.~\eqref{eq:mc-1S}) 
we use $m_{c}^{1S} = 1.44$~GeV~\cite{Herren:2017osy}.
\\
For the $s$-quark mass we take the value~\cite{Zyla:2020zbs}
\begin{equation}
m_s  = \left(93^{+11}_{-5}\right) \, {\rm MeV}. 
\end{equation}
The masses of $D$-mesons are known very precisely \cite{Zyla:2020zbs}:
$$
M_{D^0} = 1.86493 \, {\rm GeV}, \qquad  
M_{D^+} = 1.86965 \, {\rm GeV}, \qquad 
M_{D_s^+} = 1.96834 \, {\rm GeV}.
$$
The values of the non-perturbative parameters used in the analysis are shown in Tables~\ref{tab:num-input} and \ref{tab:Bag-parameters}.
\begin{table}[ht]\centering
\renewcommand{\arraystretch}{1.5}
    \begin{tabular}{|c||C{2.7cm}|C{3.5cm}||C{2.7cm}|C{3.5cm}|}
    \hline 
    Parameter 
    & $D^{+,0}$ 
    & Source
    & $D_s^+$
    & Source \\
    \hline
    \hline
     $f_D$ [GeV]
    & $0.2120 \pm 0.0007$
    & Lattice QCD \cite{Aoki:2019cca}
    & $0.2499 \pm 0.0005$ 
    & Lattice QCD \cite{Aoki:2019cca} 
    \\
    \hline 
    $\mu_\pi^2 (D)$ [GeV$^2$]
    & $0.465 \pm  0.198 $
    & Exp. fit \cite{Alberti:2014yda} 
      and HQ symmetry
    & $0.555 \pm  0.232 $
    & Exp. fit \cite{Alberti:2013kxa}, 
      $SU(3)_f$-breaking \cite{Bigi:2011gf} 
      and HQ symmetry  
    \\
    \hline
    $\mu_G^2 (D)$ [GeV$^2$]
    & $0.34 \pm  0.10 $ 
    & Spectroscopy relations~\cite{Uraltsev:2001ih}
    & $0.36 \pm  0.10 $ 
    & Spectroscopy relations~\cite{Uraltsev:2001ih} 
    \\
    \hline
    $\rho_D^3 (D)$ [GeV$^3$]
    & $0.075 \pm  0.034 $
    & Exp. fit \cite{Alberti:2014yda} and 
      E.O.M relation 
    & $0.110 \pm  0.050 $ 
    & Exp. fit \cite{Alberti:2014yda} 
      and E.O.M relation \\
    \hline 
    \end{tabular}
    \caption{Numerical values of the non-perturbative parameters used in our analysis.}
    \label{tab:num-input}
\end{table}

\begin{table}\centering
\renewcommand{\arraystretch}{1.6}
\begin{tabular}{|c||c|c|c|c|}
\hline
${\rm HQET}, \, \mu_0 = 1.5 \, {\rm GeV}$    
&  $ \tilde B_1$ 
&  $ \tilde B_2$ 
& $ \tilde \epsilon_1$ 
& $ \tilde \epsilon_2$ 
\\
\hline
\hline
    $D^{+,0}$ 
     & $\phantom{-}1.0026^{+0.0198}_{-0.0106}$ 
     & $\phantom{-}0.9982^{+0.0052}_{-0.0066}$ 
     & $-0.0165^{+0.0209}_{-0.0346}$ 
     & $-0.0004^{+0.0200}_{-0.0326}$
\\
\hline
     $D_s^+$  
     & $\phantom{-}1.0022^{+0.0185}_{-0.0099}$ 
     & $\phantom{-}0.9983^{+0.0052}_{-0.0067}$ 
     & $-0.0104^{+0.0202}_{-0.0330}$ 
     & $-0.0001^{+0.0199}_{-0.0324}$
\\
\hline
\end{tabular}
\begin{tabular}{|c||c|c|c|c|}
\hline
${\rm HQET}, \, \mu_0 = 1.5 \, {\rm GeV}$    
& $ \tilde \delta_1$
& $ \tilde \delta_2$ 
& $ \tilde \delta_3$ 
& $ \tilde \delta_4$ 
\\
\hline
\hline
$\langle D_q | \tilde O^q | D_q \rangle $
& $\phantom{-}0.0026^{+0.0142}_{-0.0092}$ 
& $-0.0018^{+0.0047}_{-0.0072}$ 
& $-0.0004^{+0.0015}_{-0.0024}$ 
& $\phantom{-}0.0003^{+0.0012}_{-0.0008}$
\\
\hline
$\langle D_s |  \tilde O^q | D_s \rangle$ 
& $\phantom{-}0.0025^{+0.0144}_{-0.0093}$ 
& $-0.0018^{+0.0047}_{-0.0072}$ 
& $-0.0004^{+0.0015}_{-0.0024}$ 
& $\phantom{-}0.0003^{+0.0012}_{-0.0008}$
\\
\hline
$\langle D_q | \tilde O^s | D_q \rangle$ 
& $\phantom{-}0.0023^{+0.0140}_{-0.0091}$ 
& $-0.0017^{+0.0046}_{-0.0070}$ 
& $-0.0004^{+0.0015}_{-0.0023}$ 
& $\phantom{-}0.0003^{+0.0012}_{-0.0008}$
\\
\hline
\end{tabular}
\caption{Numerical values of the HQET Bag parameters \cite{Kirk:2017juj,King:2020} evaluated through a traditional HQET sum rule.
}
\label{tab:Bag-parameters}
\end{table}

\clearpage

\section{Expressions for the Darwin Coefficients}
\label{Appendix-B}

The coefficients $C_{\rho_D, mn}^{(q_1 q_2)} (\rho, \mu_0)$ including 
full $\rho = m_s^2/m_c^2$ dependence are given by the expressions: 
%
%
\begin{eqnarray}
{\cal C}_{\rho_D, 11}^{(d \bar d)} 
& = &
6 + 8 \, \log \left(\frac{\mu_0^2}{m_c^2}\right),
\label{eq:CrhoDdd11}
\\ 
{\cal C}_{\rho_D, 12}^{(d \bar d)} 
& = & 
-\frac{34}{3}, 
\label{eq:CrhoDdd12}
\\
{\cal C}_{\rho_D, 22}^{(d \bar d)} 
& = &
6 + 8 \, \log \left(\frac{\mu_0^2}{m_c^2}\right),
\label{eq:CrhoDdd22}
\end{eqnarray}
%
%
\begin{eqnarray}
{\cal C}_{\rho_D, 11}^{(d \bar s)} 
& = & 
\frac{2}{3} (1 - \rho) \biggl[ 9  + 11 \rho - 12 \rho ^2 \log (\rho ) 
 - \, 24 \left(1 - \rho^2 \right) \log (1-\rho )- 25 \rho ^2  + 5 \rho^3 \biggl]
\nonumber \\
& & 
 + \,  8 \, (1 - \rho) (1 - \rho^2) \log \left(\frac{\mu_0^2}{m_c^2}\right) , 
\label{eq:CrhoDds11} 
\\
{\cal C}_{\rho_D, 12}^{(d \bar s)} 
& = & 
- \frac{2}{3} \biggl[17 + 12 \rho \left(5 + 2 \rho - 2 \rho ^2 \right) \log(\rho) 
+ \, 48 (1 - \rho) (1 - \rho^2) \log (1-\rho)
\nonumber \\
& &
\quad \quad  - 26  \rho + 18 \rho^2 - 38 \rho^3 + 5 \rho ^4 
 +  24 \, \rho \, (1 + \rho - \rho^2) \log \left(\frac{\mu_0^2}{m_c^2}\right) 
\biggl],
\label{eq:CrhoDds12} 
\\
{\cal C}_{\rho_D, 22}^{(d \bar s)} 
& = &
\frac{2}{3} (1 - \rho) \biggl[ 9  + 11 \rho - 12 \rho ^2 \log (\rho )  - \, 24 \left(1 - \rho^2 \right) \log (1-\rho )- 25 \rho ^2  + 5 \rho^3 \biggr],
\nonumber \\ 
& & 
+\, 8 \, (1 - \rho) (1 - \rho^2) \log \left(\frac{\mu_0^2}{m_c^2}\right),
\label{eq:CrhoDds22}
\end{eqnarray}
%
%
\begin{eqnarray}
{\cal C}_{\rho_D, 11}^{(s \bar d)} 
& = & 
\frac{2}{3} \biggl[9 - 16 \rho - 12 \rho ^2 + 16 \rho ^3 - 5 \rho ^4 
+ 12 \log \left(\frac{\mu_0^2}{m_c^2}\right) \biggr], 
\label{eq:CrhoDsd11} 
\\
{\cal C}_{\rho_D, 12}^{(s \bar d)} 
& = & 
- \frac{2}{3} \biggl[17 + 12 \, \rho^2 \left(3 - \rho \right) \log(\rho) 
- \, 24 (1 - \rho)^3 \log (1-\rho) 
\nonumber \\
& & 
\quad \quad  -  50 \rho + 90 \rho ^2 - 54 \rho ^3 + 5 \rho ^4
- 12  \rho \, (3 - 3 \rho +\rho^2) \log \left(\frac{\mu_0^2}{m_c^2}\right)
\biggr],  
\label{eq:CrhoDsd12} 
\\
{\cal C}_{\rho_D, 22}^{(s \bar d)} 
& = &
\frac{2}{3} (1 - \rho) \biggl[ 9  + 11 \rho - 12 \rho ^2 \log (\rho ) 
- \, 24 \left(1 - \rho^2 \right) \log (1-\rho)
\nonumber \\
& & 
\quad \quad - 25 \rho^2  + 5 \rho^3 
+ 12 \, (1 - \rho^2) \log \left(\frac{\mu_0^2}{m_c^2}\right) \biggr],
\label{eq:CrhoDsd22}
\end{eqnarray}
%
%
\begin{eqnarray}
{\cal C}_{\rho_D, 11}^{(s \bar s)} & = & 
\frac{2}{3} \Biggl[
\sqrt{1 - 4 \rho} \left(17 + 8 \rho - 22 \rho^2 - 60 \rho^3 \right)
-4 \left(2 - 3 \rho + \rho^3 \right) +
\nonumber \\
& & 
\quad \quad - \, 12 \left(1 - \rho - 2 \rho^2 + 2 \rho^3 + 10 \rho^4 \right) 
\log \left(\frac{1 + \sqrt{1 - 4 \rho^{\phantom{\! 1}}}}
{1 - \sqrt{1 - 4 \rho^{\phantom{\! 1}}}} \right) 
\nonumber \\
& & 
\quad \quad - \, 12 \, (1 - \rho)(1 - \rho^2) 
\left(\log (\rho) - \log \left(\frac{\mu_0^2}{m_c^2}\right)\right)  
\Biggr],
\label{eq:CrhoDss11} \\ 
{\cal C}_{\rho_D, 12}^{(s \bar s)} & = & 
\frac{2}{3} \Biggl[ \sqrt{1 - 4 \rho} \left(- 33 
+ 24 \, \log(\rho) - 24 \, \log (1 - 4 \rho)  
+ 46 \rho - 106 \rho^2 - 60 \rho^3 \right)
\nonumber \\
& & 
\quad \quad + \, 12 \left(3 - 2 \rho + 4 \rho^2 - 16 \rho^3 - 10 \rho^4 \right) 
\log \left(\frac{1 + \sqrt{1 - 4 \rho^{\phantom{\! 1}}}}
{1 - \sqrt{1 - 4 \rho^{\phantom{\! 1}}}} \right)  
\nonumber \\
& & 
\quad \quad + \, 4 \left(1 -\rho \right)^2 
\left(4 + 3 (1-\rho) \log(\rho) - \rho \right) 
\nonumber \\
& &
\quad \quad - 12 \, \left(1 - \sqrt{1 - 4 \rho} - 3 \rho + 3 \rho^2 - \rho^3 \right)
\, \log \left(\frac{\mu_0^2}{m_c^2}\right) 
\Biggr],  
\label{eq:CrhoDss12} 
\\ 
{\cal C}_{\rho_D, 22}^{(s \bar s)} 
& = &
\frac{2}{3} \Biggl[ \sqrt{1 - 4 \rho} \left(9 + 24 \, \log(\rho) - 24 \, \log (1 - 4 \rho)
+ 22 \rho - 34 \rho^2 - 60 \rho^3 \right)
\nonumber \\
& & 
\quad \quad  + \, 24 \left(1 - 2 \rho - \rho^2 - 2 \rho^3 - 5 \rho^4 \right) 
\log \left(\frac{1 + \sqrt{1 - 4 \rho^{\phantom{\! 1}}}}
{1 - \sqrt{1 - 4 \rho^{\phantom{\! 1}}}} \right) 
\nonumber \\
& & 
\quad \quad \, + \, 12 \, \sqrt{1 - 4 \rho} \, \log \left(\frac{\mu_0^2}{m_c^2}\right)
\Biggr].
\label{eq:CrhoDss22}
\end{eqnarray}
The numerical values of the above coefficients for $\rho = 0.006$ are shown 
in Table~\ref{tab:CrhoD-values}.
\begin{table}[th]\centering
\renewcommand{\arraystretch}{1.25}
\begin{tabular}{|c|C{2cm}|C{2cm}|C{2cm}|}
\hline
  & $3 \, C_1^2$ & $2 \, C_1 C_2$ & $3 \, C_2^2$ \\
\hline
\hline
$ c \to d \bar d u $ 
& 6 & -11.33 & 6  
\\
\hline
$ c \to d \bar s u $ 
& 6.10 & -9.81 & 6.10  
\\
\hline
$ c \to s \bar d u $ 
& 5.94 & -11.23 & 6.10  
\\
\hline
$ c \to s \bar s u $ 
& 6.04 & -9.70 & 6.21  
\\
\hline
\end{tabular}
\caption{Numerical values of $C_{\rho_D, nm}^{(q_1 q_2)}$ for $\rho = 0.006$ and $\mu_0 = m_c$.}
\label{tab:CrhoD-values}
\end{table}

\section{Parametrisation of the Matrix Element of Four-Quark Operators}
\label{Appendix-C}

The matrix elements of the dimension-six operators in QCD are parametrised in the following way
\begin{eqnarray}
\langle {D}_q | O_i^q \, | {D}_q \rangle 
& = & 
A_i \, f_{D_q}^2 m_{D_q}^2 \, B_i^q, 
\label{eq:ME-dim-6-QCD-q-q}
\\[2mm]
\langle {D}_q | O_i^{q^\prime} | {D}_q \rangle 
& = & 
A_i \, f_{D_q}^2 m_{D_q}^2 
\, \delta^{q q^\prime}_i, \qquad q \not = q^\prime \,,
\label{eq:ME-dim-6-QCD-q-q-prime}
\end{eqnarray}
where 
$$
A_1^q = A_3^q = 1, \qquad A_2^q = A_4^q = \frac{m_D^2}{(m_c + m_q)^2}.
$$
In  VIA the Bag parameters reduce to $B_1^q = B_2^q = 1$
and $B_3^q = \epsilon_1^q =0$, $B_4^q = \epsilon_2^q = 0$ and
all $\delta^{q q'}_i = 0$.

The matrix elements of the dimension-seven four-quark operators in Eqs.~\eqref{eq:P1q-HQET} - \eqref{eq:M2-G} in HQET 
are parametrised in the following way:
\begin{eqnarray}
\langle D_q |  {\tilde P}_1^q | D_q \rangle 
& = & 
- m_q F^2 (\mu_0) \, m_D \, \tilde B_{P, 1}^q \, , 
\label{eq:ME-dim-7-P1}
\\[2mm]
\langle D_q |  {\tilde P}_2^q |D_q \rangle 
& = & 
- F^2 (\mu_0) \, m_D \, \bar \Lambda \, \tilde B_{P, 2}^q\,, 
\label{eq:ME-dim-7-P3}
\\[2mm]
\langle D_q |  {\tilde P}_3^q |D_q \rangle 
& = & 
- F^2 (\mu_0) \, m_D \, \bar \Lambda \, \tilde B_{P, 3}^q\,, 
\label{eq:ME-dim-7-P4}
\\[2mm]
\langle D_q |  {\tilde R}_1^q |D_q \rangle 
& = & 
- F^2 (\mu_0) \, m_D \, (\bar \Lambda - m_q) \, \tilde B_{R, 1}^q\,, 
\label{eq:ME-dim-7-P5}
\\[2mm]
\langle D_q |  {\tilde R}_2^q |D_q \rangle 
& = & 
F^2 (\mu_0) \, m_D  \, (\bar \Lambda - m_q) \, \tilde B_{R, 1}^q\,,
\label{eq:ME-dim-7-P6}
\end{eqnarray}
with $ \bar \Lambda = m_D - m_c$, and
\begin{eqnarray}
\langle D_q |  {\tilde M}_{1, \pi}^q | D_q \rangle 
& = &
2 \, F^2 (\mu_0) \, m_D \, G_1 (\mu_0) \, \tilde L_{1, \pi}^q \, ,
\label{eq:ME-M1-pi}
\\[2mm]
\langle D_q |  {\tilde M}_{2, \pi}^q | D_q \rangle 
& = &
2 \, F^2 (\mu_0) \, m_D \, G_1 (\mu_0) \, \tilde L_{2, \pi}^q \, ,
\label{eq:ME-M2-pi}
\\[2mm]
\langle D_q |  {\tilde M}_{1, G}^q | D_q \rangle 
& = &
12 \, F^2 (\mu_0) \, m_D \, G_2 (\mu_0) \, \tilde L_{1, G}^q \, ,
\label{eq:ME-M1-G}
\\[2mm]
\langle D_q |  {\tilde M}_{2, G}^q | D_q \rangle 
& = &
12 \, F^2 (\mu_0) \, m_D \, G_2 (\mu_0) \, \tilde L_{2, G}^q \, ,
\label{eq:ME-M2-G}
\end{eqnarray}
and similar expressions for the colour-octet operators.
Again, in VIA, the dimension-seven Bag parameters are 
$\tilde B_{P, i}^q = 1$, $\tilde B_{R, i}^q = 1$, and
$\tilde L_{1, \pi}^q = 1$, $\tilde L_{1, G}^q = 1$
and the corresponding colour-octet Bag parameters vanish.

The expressions in Eqs.~\eqref{eq:ME-dim-7-P1} - \eqref{eq:ME-dim-7-P6} can be obtained using a general parametrisation of matrix elements of the HQET quark currents with a heavy pseudo-scalar meson ${\cal M}$ 
(see e.q. Ref.~\cite{Neubert:1993mb}):
\begin{eqnarray}
\langle 0 | \bar q \,  \Gamma \, h_v| M(v)\rangle 
& = & 
\frac{i}{2} F(\mu) \, {\rm Tr} [\Gamma {\cal M}(v)]\, ,
\\
\langle 0 | \bar q \,  \Gamma \,i D_\alpha h_v| M(v) \rangle 
& = & 
-\frac{i}{6} (\bar \Lambda -m_q) F(\mu) \, {\rm Tr} [(v_\alpha + \gamma_\alpha) \Gamma {\cal M}(v)]\, ,
\\
\langle 0 | \bar q (-i \overset{\leftarrow}{D}_\alpha) \,  \Gamma \, h_v| M(v)\rangle 
& = & 
- \frac{i}{6} F(\mu) \, {\rm Tr}[((4 \bar \Lambda - m_q) v_\alpha +  (\bar \Lambda - m_q) \gamma_\alpha) \Gamma {\cal M}(v)],
\end{eqnarray}
and for the non-local operators:
\begin{eqnarray}
\langle 0 |
i \int d^4 y \, T
\left[ 
(\bar q \, \Gamma \, h_v) (0), 
(\bar h_v (i D)^2 h_v) (y)
\right]
| {\cal M}(v) \rangle
& = & 
F(\mu) \, G_1 (\mu) \, {\rm Tr} [\Gamma {\cal M}(v)],
\\
\langle 0 |
i \int d^4 y \, T
\left[ 
(\bar q \, \Gamma \, h_v) (0), 
\frac{1}{2} g_s \left(\bar h_v \sigma_{\alpha \beta} G^{\alpha \beta} h_v \right) (y)
\right]
| {\cal M}(v) \rangle
& = &
6 \, F(\mu) \, G_2 (\mu) \,  {\rm Tr} [\Gamma {\cal M}(v)],
\end{eqnarray}
where $\Gamma$ is a generic Dirac structure, and
\begin{eqnarray}
{\cal M}(v) & = & - \sqrt{m_D} \, \frac{(1+\slashed v)}{2} \gamma_5\,.
\end{eqnarray}
Since we are limited to LO-QCD  for the dimension-seven contribution, we can just replace the HQET decay constant $F (\mu)$ by the full QCD one $f_D$, using $F(\mu) = f_D \, \sqrt{m_D}$.

\bibliographystyle{JHEP}
\bibliography{References}

\providecommand{\href}[2]{#2}\begingroup\raggedright\begin{thebibliography}{10}

\bibitem{Zyla:2020zbs}
{\bf Particle Data Group} Collaboration, P.~Zyla et~al., {\it {Review of
  Particle Physics}},  {\em PTEP} {\bf 2020} (2020), no.~8 083C01.

\bibitem{Belle-II:2021cxx}
{\bf Belle-II} Collaboration, F.~Abudin\'en et~al., {\it {Precise measurement
  of the $D^0$ and $D^+$ lifetimes at Belle II}},
  \href{http://arxiv.org/abs/2108.03216}{{\tt arXiv:2108.03216}}.

\bibitem{Ablikim:2021qvs}
{\bf BESIII} Collaboration, M.~Ablikim et~al., {\it {Measurement of the
  absolute branching fraction of inclusive semielectronic $D_s^+$ decays}},
  \href{http://arxiv.org/abs/2104.07311}{{\tt arXiv:2104.07311}}.

\bibitem{Khoze:1983yp}
V.~A. Khoze and M.~A. Shifman, {\it {HEAVY QUARKS}},  {\em Sov. Phys. Usp.}
  {\bf 26} (1983) 387.

\bibitem{Shifman:1986mx}
M.~A. Shifman and M.~B. Voloshin, {\it {Hierarchy of Lifetimes of Charmed and
  Beautiful Hadrons}},  {\em Sov. Phys. JETP} {\bf 64} (1986) 698. [Zh. Eksp.
  Teor. Fiz.91,1180(1986)].

\bibitem{Lenz:2015dra}
A.~Lenz, {\it {Lifetimes and heavy quark expansion}},  {\em Int. J. Mod. Phys.}
  {\bf A30} (2015), no.~10 1543005, [\href{http://arxiv.org/abs/1405.3601}{{\tt
  arXiv:1405.3601}}].

\bibitem{Herren:2017osy}
F.~Herren and M.~Steinhauser, {\it {Version 3 of RunDec and CRunDec}},  {\em
  Comput. Phys. Commun.} {\bf 224} (2018) 333--345,
  [\href{http://arxiv.org/abs/1703.03751}{{\tt arXiv:1703.03751}}].

\bibitem{Chetyrkin:1999qi}
K.~Chetyrkin and M.~Steinhauser, {\it {The Relation between the MS-bar and the
  on-shell quark mass at order $\alpha_s^3$}},  {\em Nucl. Phys. B} {\bf 573}
  (2000) 617--651, [\href{http://arxiv.org/abs/hep-ph/9911434}{{\tt
  hep-ph/9911434}}].

\bibitem{Chetyrkin:1999ys}
K.~Chetyrkin and M.~Steinhauser, {\it {Short distance mass of a heavy quark at
  order $\alpha_s^3$}},  {\em Phys. Rev. Lett.} {\bf 83} (1999) 4001--4004,
  [\href{http://arxiv.org/abs/hep-ph/9907509}{{\tt hep-ph/9907509}}].

\bibitem{Melnikov:2000qh}
K.~Melnikov and T.~v. Ritbergen, {\it {The Three loop relation between the
  MS-bar and the pole quark masses}},  {\em Phys. Lett. B} {\bf 482} (2000)
  99--108, [\href{http://arxiv.org/abs/hep-ph/9912391}{{\tt hep-ph/9912391}}].

\bibitem{Bardeen:1978yd}
W.~A. Bardeen, A.~J. Buras, D.~W. Duke, and T.~Muta, {\it {Deep Inelastic
  Scattering Beyond the Leading Order in Asymptotically Free Gauge Theories}},
  {\em Phys. Rev. D} {\bf 18} (1978) 3998.

\bibitem{Bigi:1994ga}
I.~I.~Y. Bigi, M.~A. Shifman, N.~G. Uraltsev, and A.~I. Vainshtein, {\it {Sum
  rules for heavy flavor transitions in the SV limit}},  {\em Phys. Rev.} {\bf
  D52} (1995) 196--235, [\href{http://arxiv.org/abs/hep-ph/9405410}{{\tt
  hep-ph/9405410}}].

\bibitem{Bigi:1996si}
I.~I.~Y. Bigi, M.~A. Shifman, N.~Uraltsev, and A.~I. Vainshtein, {\it {High
  power n of $m_b$ in beauty widths and $n=5 \to \infty$ limit}},  {\em Phys.
  Rev. D} {\bf 56} (1997) 4017--4030,
  [\href{http://arxiv.org/abs/hep-ph/9704245}{{\tt hep-ph/9704245}}].

\bibitem{Fael:2020njb}
M.~Fael, K.~Sch\"onwald, and M.~Steinhauser, {\it {Relation between the
  $\overline{\mathrm{MS}}$ and the kinetic mass of heavy quarks}},  {\em Phys.
  Rev. D} {\bf 103} (2021), no.~1 014005,
  [\href{http://arxiv.org/abs/2011.11655}{{\tt arXiv:2011.11655}}].

\bibitem{Hoang:1998ng}
A.~H. Hoang, Z.~Ligeti, and A.~V. Manohar, {\it {B decay and the Upsilon
  mass}},  {\em Phys. Rev. Lett.} {\bf 82} (1999) 277--280,
  [\href{http://arxiv.org/abs/hep-ph/9809423}{{\tt hep-ph/9809423}}].

\bibitem{Hoang:1998hm}
A.~H. Hoang, Z.~Ligeti, and A.~V. Manohar, {\it {B decays in the upsilon
  expansion}},  {\em Phys. Rev. D} {\bf 59} (1999) 074017,
  [\href{http://arxiv.org/abs/hep-ph/9811239}{{\tt hep-ph/9811239}}].

\bibitem{Hoang:1999zc}
A.~H. Hoang and T.~Teubner, {\it {Top quark pair production close to threshold:
  Top mass, width and momentum distribution}},  {\em Phys. Rev. D} {\bf 60}
  (1999) 114027, [\href{http://arxiv.org/abs/hep-ph/9904468}{{\tt
  hep-ph/9904468}}].

\bibitem{Beneke:1998rk}
M.~Beneke, {\it {A Quark mass definition adequate for threshold problems}},
  {\em Phys. Lett. B} {\bf 434} (1998) 115--125,
  [\href{http://arxiv.org/abs/hep-ph/9804241}{{\tt hep-ph/9804241}}].

\bibitem{Hokim:1983yt}
Q.~Ho-kim and X.-Y. Pham, {\it {Exact One Gluon Corrections for Inclusive Weak
  Processes}},  {\em Annals Phys.} {\bf 155} (1984) 202.

\bibitem{Altarelli:1991dx}
G.~Altarelli and S.~Petrarca, {\it {Inclusive beauty decays and the spectator
  model}},  {\em Phys. Lett.} {\bf B261} (1991) 303--310.

\bibitem{Voloshin:1994sn}
M.~B. Voloshin, {\it {QCD radiative enhancement of the decay $b \to c \bar{c}
  s$}},  {\em Phys. Rev.} {\bf D51} (1995) 3948--3951,
  [\href{http://arxiv.org/abs/hep-ph/9409391}{{\tt hep-ph/9409391}}].

\bibitem{Bagan:1994zd}
E.~Bagan, P.~Ball, V.~M. Braun, and P.~Gosdzinsky, {\it {Charm quark mass
  dependence of QCD corrections to nonleptonic inclusive B decays}},  {\em
  Nucl. Phys.} {\bf B432} (1994) 3--38,
  [\href{http://arxiv.org/abs/hep-ph/9408306}{{\tt hep-ph/9408306}}].

\bibitem{Bagan:1995yf}
E.~Bagan, P.~Ball, B.~Fiol, and P.~Gosdzinsky, {\it {Next-to-leading order
  radiative corrections to the decay $b \to c \bar{c} s$}},  {\em Phys. Lett.}
  {\bf B351} (1995) 546--554, [\href{http://arxiv.org/abs/hep-ph/9502338}{{\tt
  hep-ph/9502338}}].

\bibitem{Lenz:1997aa}
A.~Lenz, U.~Nierste, and G.~Ostermaier, {\it {Penguin diagrams, charmless B
  decays and the missing charm puzzle}},  {\em Phys. Rev.} {\bf D56} (1997)
  7228--7239, [\href{http://arxiv.org/abs/hep-ph/9706501}{{\tt
  hep-ph/9706501}}].

\bibitem{Lenz:1998qp}
A.~Lenz, U.~Nierste, and G.~Ostermaier, {\it {Determination of the CKM angle
  gamma and $|V_{ub} / V_{cb}|$ from inclusive direct CP asymmetries and
  branching ratios in charmless B decays}},  {\em Phys. Rev.} {\bf D59} (1999)
  034008, [\href{http://arxiv.org/abs/hep-ph/9802202}{{\tt hep-ph/9802202}}].

\bibitem{Krinner:2013cja}
F.~Krinner, A.~Lenz, and T.~Rauh, {\it {The inclusive decay $b \to c\bar{c}s$
  revisited}},  {\em Nucl. Phys.} {\bf B876} (2013) 31--54,
  [\href{http://arxiv.org/abs/1305.5390}{{\tt arXiv:1305.5390}}].

\bibitem{Czarnecki:1997hc}
A.~Czarnecki and K.~Melnikov, {\it {Two loop QCD corrections to semileptonic b
  decays at maximal recoil}},  {\em Phys. Rev. Lett.} {\bf 78} (1997)
  3630--3633, [\href{http://arxiv.org/abs/hep-ph/9703291}{{\tt
  hep-ph/9703291}}].

\bibitem{Czarnecki:1998kt}
A.~Czarnecki and K.~Melnikov, {\it {Two - loop QCD corrections to semileptonic
  b decays at an intermediate recoil}},  {\em Phys. Rev.} {\bf D59} (1999)
  014036, [\href{http://arxiv.org/abs/hep-ph/9804215}{{\tt hep-ph/9804215}}].

\bibitem{vanRitbergen:1999gs}
T.~van Ritbergen, {\it {The Second order QCD contribution to the semileptonic
  $b \to u$ decay rate}},  {\em Phys. Lett.} {\bf B454} (1999) 353--358,
  [\href{http://arxiv.org/abs/hep-ph/9903226}{{\tt hep-ph/9903226}}].

\bibitem{Melnikov:2008qs}
K.~Melnikov, {\it {$ {\cal O}(\alpha_s^2)$ corrections to semileptonic decay $
  b \to cl \bar{\nu}_l$}},  {\em Phys. Lett.} {\bf B666} (2008) 336--339,
  [\href{http://arxiv.org/abs/0803.0951}{{\tt arXiv:0803.0951}}].

\bibitem{Pak:2008cp}
A.~Pak and A.~Czarnecki, {\it {Heavy-to-heavy quark decays at NNLO}},  {\em
  Phys. Rev.} {\bf D78} (2008) 114015,
  [\href{http://arxiv.org/abs/0808.3509}{{\tt arXiv:0808.3509}}].

\bibitem{Pak:2008qt}
A.~Pak and A.~Czarnecki, {\it {Mass effects in muon and semileptonic $b \to c$
  decays}},  {\em Phys. Rev. Lett.} {\bf 100} (2008) 241807,
  [\href{http://arxiv.org/abs/0803.0960}{{\tt arXiv:0803.0960}}].

\bibitem{Dowling:2008ap}
M.~Dowling, A.~Pak, and A.~Czarnecki, {\it {Semi-Leptonic b-decay at
  Intermediate Recoil}},  {\em Phys. Rev.} {\bf D78} (2008) 074029,
  [\href{http://arxiv.org/abs/0809.0491}{{\tt arXiv:0809.0491}}].

\bibitem{Bonciani:2008wf}
R.~Bonciani and A.~Ferroglia, {\it {Two-Loop QCD Corrections to the
  Heavy-to-Light Quark Decay}},  {\em JHEP} {\bf 11} (2008) 065,
  [\href{http://arxiv.org/abs/0809.4687}{{\tt arXiv:0809.4687}}].

\bibitem{Biswas:2009rb}
S.~Biswas and K.~Melnikov, {\it {Second order QCD corrections to inclusive
  semileptonic $b \to X_c \ell \bar{\nu}_l$decays with massless and massive
  lepton}},  {\em JHEP} {\bf 02} (2010) 089,
  [\href{http://arxiv.org/abs/0911.4142}{{\tt arXiv:0911.4142}}].

\bibitem{Brucherseifer:2013cu}
M.~Brucherseifer, F.~Caola, and K.~Melnikov, {\it {On the $O(\alpha_s^2)$
  corrections to $b \to X_u e \bar \nu$ inclusive decays}},  {\em Phys. Lett.}
  {\bf B721} (2013) 107--110, [\href{http://arxiv.org/abs/1302.0444}{{\tt
  arXiv:1302.0444}}].

\bibitem{Fael:2020tow}
M.~Fael, K.~Sch\"onwald, and M.~Steinhauser, {\it {Third order corrections to
  the semi-leptonic $b\to c$ and the muon decays}},
  \href{http://arxiv.org/abs/2011.13654}{{\tt arXiv:2011.13654}}.

\bibitem{Czakon:2021ybq}
M.~Czakon, A.~Czarnecki, and M.~Dowling, {\it {Three-loop corrections to the
  muon and heavy quark decay rates}},  {\em Phys. Rev. D} {\bf 103} (2021)
  L111301, [\href{http://arxiv.org/abs/2104.05804}{{\tt arXiv:2104.05804}}].

\bibitem{Czarnecki:2005vr}
A.~Czarnecki, M.~Slusarczyk, and F.~V. Tkachov, {\it {Enhancement of the
  hadronic b quark decays}},  {\em Phys. Rev. Lett.} {\bf 96} (2006) 171803,
  [\href{http://arxiv.org/abs/hep-ph/0511004}{{\tt hep-ph/0511004}}].

\bibitem{Bigi:1992su}
I.~I.~Y. Bigi, N.~G. Uraltsev, and A.~I. Vainshtein, {\it {Nonperturbative
  corrections to inclusive beauty and charm decays: QCD versus phenomenological
  models}},  {\em Phys. Lett.} {\bf B293} (1992) 430--436,
  [\href{http://arxiv.org/abs/hep-ph/9207214}{{\tt hep-ph/9207214}}]. [Erratum:
  Phys. Lett.B297,477(1992)].

\bibitem{Blok:1992hw}
B.~Blok and M.~A. Shifman, {\it {The Rule of discarding $1/N_c$ in inclusive
  weak decays. 1.}},  {\em Nucl. Phys.} {\bf B399} (1993) 441--458,
  [\href{http://arxiv.org/abs/hep-ph/9207236}{{\tt hep-ph/9207236}}].

\bibitem{Blok:1992he}
B.~Blok and M.~A. Shifman, {\it {The Rule of discarding $1/N_c$ in inclusive
  weak decays. 2.}},  {\em Nucl. Phys.} {\bf B399} (1993) 459--476,
  [\href{http://arxiv.org/abs/hep-ph/9209289}{{\tt hep-ph/9209289}}].

\bibitem{Bigi:1992ne}
I.~I.~Y. Bigi, B.~Blok, M.~A. Shifman, N.~G. Uraltsev, and A.~I. Vainshtein,
  {\it {A QCD 'manifesto' on inclusive decays of beauty and charm}},  in {\em
  {The Fermilab Meeting DPF 92. Proceedings, 7th Meeting of the American
  Physical Society, Division of Particles and Fields, Batavia, USA, November
  10-14, 1992. Vol. 1, 2}}, pp.~610--613, 1992.
\newblock \href{http://arxiv.org/abs/hep-ph/9212227}{{\tt hep-ph/9212227}}.

\bibitem{Alberti:2013kxa}
A.~Alberti, P.~Gambino, and S.~Nandi, {\it {Perturbative corrections to power
  suppressed effects in semileptonic B decays}},  {\em JHEP} {\bf 01} (2014)
  147, [\href{http://arxiv.org/abs/1311.7381}{{\tt arXiv:1311.7381}}].

\bibitem{Mannel:2014xza}
T.~Mannel, A.~A. Pivovarov, and D.~Rosenthal, {\it {Inclusive semileptonic B
  decays from QCD with NLO accuracy for power suppressed terms}},  {\em Phys.
  Lett.} {\bf B741} (2015) 290--294,
  [\href{http://arxiv.org/abs/1405.5072}{{\tt arXiv:1405.5072}}].

\bibitem{Mannel:2015jka}
T.~Mannel, A.~A. Pivovarov, and D.~Rosenthal, {\it {Inclusive weak decays of
  heavy hadrons with power suppressed terms at NLO}},  {\em Phys. Rev.} {\bf
  D92} (2015), no.~5 054025, [\href{http://arxiv.org/abs/1506.08167}{{\tt
  arXiv:1506.08167}}].

\bibitem{Gremm:1996df}
M.~Gremm and A.~Kapustin, {\it {Order $1/m_b^3$ corrections to $B \to X(c)$
  lepton anti-neutrino decay and their implication for the measurement of
  $\bar{\Lambda}$ and $\lambda_1$}},  {\em Phys. Rev.} {\bf D55} (1997)
  6924--6932, [\href{http://arxiv.org/abs/hep-ph/9603448}{{\tt
  hep-ph/9603448}}].

\bibitem{Mannel:2019qel}
T.~Mannel and A.~A. Pivovarov, {\it {QCD corrections to inclusive heavy hadron
  weak decays at $\Lambda_{\rm QCD}^3 /m_Q^3$}},  {\em Phys. Rev.} {\bf D100}
  (2019), no.~9 093001, [\href{http://arxiv.org/abs/1907.09187}{{\tt
  arXiv:1907.09187}}].

\bibitem{Lenz:2020oce}
A.~Lenz, M.~L. Piscopo, and A.~V. Rusov, {\it {Contribution of the Darwin
  operator to non-leptonic decays of heavy quarks}},  {\em JHEP} {\bf 12}
  (2020) 199, [\href{http://arxiv.org/abs/2004.09527}{{\tt arXiv:2004.09527}}].

\bibitem{Mannel:2020fts}
T.~Mannel, D.~Moreno, and A.~Pivovarov, {\it {Heavy quark expansion for heavy
  hadron lifetimes: completing the $ 1/{m}_b^3 $ corrections}},  {\em JHEP}
  {\bf 08} (2020) 089, [\href{http://arxiv.org/abs/2004.09485}{{\tt
  arXiv:2004.09485}}].

\bibitem{Moreno:2020rmk}
D.~Moreno, {\it {Completing $1/m_b^3$ corrections to non-leptonic
  bottom-to-up-quark decays}},  {\em JHEP} {\bf 01} (2021) 051,
  [\href{http://arxiv.org/abs/2009.08756}{{\tt arXiv:2009.08756}}].

\bibitem{Gambino:2010jz}
P.~Gambino and J.~F. Kamenik, {\it {Lepton energy moments in semileptonic charm
  decays}},  {\em Nucl. Phys. B} {\bf 840} (2010) 424--437,
  [\href{http://arxiv.org/abs/1004.0114}{{\tt arXiv:1004.0114}}].

\bibitem{Fael:2019umf}
M.~Fael, T.~Mannel, and K.~K. Vos, {\it {The Heavy Quark Expansion for
  Inclusive Semileptonic Charm Decays Revisited}},
  \href{http://arxiv.org/abs/1910.05234}{{\tt arXiv:1910.05234}}.

\bibitem{Beneke:2002rj}
M.~Beneke, G.~Buchalla, C.~Greub, A.~Lenz, and U.~Nierste, {\it {The $B^+
  -B^0_d$ Lifetime Difference Beyond Leading Logarithms}},  {\em Nucl. Phys.}
  {\bf B639} (2002) 389--407, [\href{http://arxiv.org/abs/hep-ph/0202106}{{\tt
  hep-ph/0202106}}].

\bibitem{Franco:2002fc}
E.~Franco, V.~Lubicz, F.~Mescia, and C.~Tarantino, {\it {Lifetime ratios of
  beauty hadrons at the next-to-leading order in QCD}},  {\em Nucl. Phys.} {\bf
  B633} (2002) 212--236, [\href{http://arxiv.org/abs/hep-ph/0203089}{{\tt
  hep-ph/0203089}}].

\bibitem{Lenz:2013aua}
A.~Lenz and T.~Rauh, {\it {D-meson lifetimes within the heavy quark
  expansion}},  {\em Phys. Rev.} {\bf D88} (2013) 034004,
  [\href{http://arxiv.org/abs/1305.3588}{{\tt arXiv:1305.3588}}].

\bibitem{Gabbiani:2003pq}
F.~Gabbiani, A.~I. Onishchenko, and A.~A. Petrov, {\it {$\Lambda_b$ lifetime
  puzzle in heavy quark expansion}},  {\em Phys. Rev. D} {\bf 68} (2003)
  114006, [\href{http://arxiv.org/abs/hep-ph/0303235}{{\tt hep-ph/0303235}}].

\bibitem{Gabbiani:2004tp}
F.~Gabbiani, A.~I. Onishchenko, and A.~A. Petrov, {\it {Spectator effects and
  lifetimes of heavy hadrons}},  {\em Phys. Rev.} {\bf D70} (2004) 094031,
  [\href{http://arxiv.org/abs/hep-ph/0407004}{{\tt hep-ph/0407004}}].

\bibitem{Bazavov:2018omf}
{\bf Fermilab Lattice, MILC, TUMQCD} Collaboration, A.~Bazavov et~al., {\it
  {Up-, down-, strange-, charm-, and bottom-quark masses from four-flavor
  lattice QCD}},  {\em Phys. Rev.} {\bf D98} (2018), no.~5 054517,
  [\href{http://arxiv.org/abs/1802.04248}{{\tt arXiv:1802.04248}}].

\bibitem{Gambino:2017vkx}
P.~Gambino, A.~Melis, and S.~Simula, {\it {Extraction of heavy-quark-expansion
  parameters from unquenched lattice data on pseudoscalar and vector
  heavy-light meson masses}},  {\em Phys. Rev.} {\bf D96} (2017), no.~1 014511,
  [\href{http://arxiv.org/abs/1704.06105}{{\tt arXiv:1704.06105}}].

\bibitem{Aoki:2003jf}
{\bf JLQCD} Collaboration, S.~Aoki et~al., {\it {Heavy quark expansion
  parameters from lattice NRQCD}},  {\em Phys. Rev.} {\bf D69} (2004) 094512,
  [\href{http://arxiv.org/abs/hep-lat/0305024}{{\tt hep-lat/0305024}}].

\bibitem{Kronfeld:2000gk}
A.~S. Kronfeld and J.~N. Simone, {\it {Computation of Lambda-bar and lambda(1)
  with lattice QCD}},  {\em Phys. Lett.} {\bf B490} (2000) 228--235,
  [\href{http://arxiv.org/abs/hep-ph/0006345}{{\tt hep-ph/0006345}}]. [Erratum:
  Phys. Lett.B495,441(2000)].

\bibitem{Gimenez:1996av}
V.~Gimenez, G.~Martinelli, and C.~T. Sachrajda, {\it {A High statistics lattice
  calculation of lambda(1) and lambda(2) in the B meson}},  {\em Nucl. Phys.}
  {\bf B486} (1997) 227--244, [\href{http://arxiv.org/abs/hep-lat/9607055}{{\tt
  hep-lat/9607055}}].

\bibitem{Ball:1993xv}
P.~Ball and V.~M. Braun, {\it {Next-to-leading order corrections to meson
  masses in the heavy quark effective theory}},  {\em Phys. Rev.} {\bf D49}
  (1994) 2472--2489, [\href{http://arxiv.org/abs/hep-ph/9307291}{{\tt
  hep-ph/9307291}}].

\bibitem{Neubert:1996wm}
M.~Neubert, {\it {QCD sum rule calculation of the kinetic energy and chromo
  interaction of heavy quarks inside mesons}},  {\em Phys. Lett.} {\bf B389}
  (1996) 727--736, [\href{http://arxiv.org/abs/hep-ph/9608211}{{\tt
  hep-ph/9608211}}].

\bibitem{Kirk:2017juj}
M.~Kirk, A.~Lenz, and T.~Rauh, {\it {Dimension-six matrix elements for meson
  mixing and lifetimes from sum rules}},  {\em JHEP} {\bf 12} (2017) 068,
  [\href{http://arxiv.org/abs/1711.02100}{{\tt arXiv:1711.02100}}].

\bibitem{King:2019lal}
D.~King, A.~Lenz, and T.~Rauh, {\it {$B_{s}$ mixing observables and
  $|V_{td}/V_{ts}|$ from sum rules}},  {\em JHEP} {\bf 05} (2019) 034,
  [\href{http://arxiv.org/abs/1904.00940}{{\tt arXiv:1904.00940}}].

\bibitem{King:2020}
D.~King, A.~Lenz, and T.~Rauh, {\it {to appear}}, .

\bibitem{Buchalla:1995vs}
G.~Buchalla, A.~J. Buras, and M.~E. Lautenbacher, {\it {Weak decays beyond
  leading logarithms}},  {\em Rev. Mod. Phys.} {\bf 68} (1996) 1125--1144,
  [\href{http://arxiv.org/abs/hep-ph/9512380}{{\tt hep-ph/9512380}}].

\bibitem{Dassinger:2006md}
B.~M. Dassinger, T.~Mannel, and S.~Turczyk, {\it {Inclusive semi-leptonic B
  decays to order $1/m_b^4$}},  {\em JHEP} {\bf 03} (2007) 087,
  [\href{http://arxiv.org/abs/hep-ph/0611168}{{\tt hep-ph/0611168}}].

\bibitem{Neubert:1993mb}
M.~Neubert, {\it {Heavy quark symmetry}},  {\em Phys. Rept.} {\bf 245} (1994)
  259--396, [\href{http://arxiv.org/abs/hep-ph/9306320}{{\tt hep-ph/9306320}}].

\bibitem{Mannel:2017jfk}
T.~Mannel, A.~V. Rusov, and F.~Shahriaran, {\it {Inclusive semitauonic $B$
  decays to order ${\cal O} (\Lambda_{QCD}^3/m_b^3)$}},  {\em Nucl. Phys.} {\bf
  B921} (2017) 211--224, [\href{http://arxiv.org/abs/1702.01089}{{\tt
  arXiv:1702.01089}}].

\bibitem{Novikov:1983gd}
V.~A. Novikov, M.~A. Shifman, A.~I. Vainshtein, and V.~I. Zakharov, {\it
  {Calculations in External Fields in Quantum Chromodynamics. Technical
  Review}},  {\em Fortsch. Phys.} {\bf 32} (1984) 585.

\bibitem{Breidenbach:2008ua}
C.~Breidenbach, T.~Feldmann, T.~Mannel, and S.~Turczyk, {\it {On the Role of
  'Intrinsic Charm' in Semi-Leptonic B-Meson Decays}},  {\em Phys. Rev.} {\bf
  D78} (2008) 014022, [\href{http://arxiv.org/abs/0805.0971}{{\tt
  arXiv:0805.0971}}].

\bibitem{Bigi:2009ym}
I.~Bigi, T.~Mannel, S.~Turczyk, and N.~Uraltsev, {\it {The Two Roads to
  'Intrinsic Charm' in B Decays}},  {\em JHEP} {\bf 04} (2010) 073,
  [\href{http://arxiv.org/abs/0911.3322}{{\tt arXiv:0911.3322}}].

\bibitem{Neubert:1992fk}
M.~Neubert, {\it {Symmetry breaking corrections to meson decay constants in the
  heavy quark effective theory}},  {\em Phys. Rev. D} {\bf 46} (1992)
  1076--1087.

\bibitem{Kilian:1992cj}
W.~Kilian and T.~Mannel, {\it {QCD corrected $1/m_b$ contributions to $B -
  \bar{B}$ mixing}},  {\em Phys. Lett. B} {\bf 301} (1993) 382--392,
  [\href{http://arxiv.org/abs/hep-ph/9211333}{{\tt hep-ph/9211333}}].

\bibitem{Aoki:2019cca}
{\bf Flavour Lattice Averaging Group} Collaboration, S.~Aoki et~al., {\it {FLAG
  Review 2019: Flavour Lattice Averaging Group (FLAG)}},  {\em Eur. Phys. J. C}
  {\bf 80} (2020), no.~2 113, [\href{http://arxiv.org/abs/1902.08191}{{\tt
  arXiv:1902.08191}}].

\bibitem{Mannel:2021uoz}
T.~Mannel, D.~Moreno, and A.~A. Pivovarov, {\it {The Heavy Quark Expansion for
  the Charm Quark}},  \href{http://arxiv.org/abs/2103.02058}{{\tt
  arXiv:2103.02058}}.

\bibitem{Cheng:2018rkz}
H.-Y. Cheng, {\it {Phenomenological Study of Heavy Hadron Lifetimes}},  {\em
  JHEP} {\bf 11} (2018) 014, [\href{http://arxiv.org/abs/1807.00916}{{\tt
  arXiv:1807.00916}}].

\bibitem{Alberti:2014yda}
A.~Alberti, P.~Gambino, K.~J. Healey, and S.~Nandi, {\it {Precision
  Determination of the Cabibbo-Kobayashi-Maskawa Element $V_{cb}$}},  {\em
  Phys. Rev. Lett.} {\bf 114} (2015), no.~6 061802,
  [\href{http://arxiv.org/abs/1411.6560}{{\tt arXiv:1411.6560}}].

\bibitem{Uraltsev:2001ih}
N.~Uraltsev, {\it {On the chromomagnetic expectation value $\mu^2_G$ and higher
  power corrections in heavy flavor mesons}},  {\em Phys. Lett. B} {\bf 545}
  (2002) 337--344, [\href{http://arxiv.org/abs/hep-ph/0111166}{{\tt
  hep-ph/0111166}}].

\bibitem{Falk:1992wt}
A.~F. Falk and M.~Neubert, {\it {Second order power corrections in the heavy
  quark effective theory. 1. Formalism and meson form-factors}},  {\em Phys.
  Rev. D} {\bf 47} (1993) 2965--2981,
  [\href{http://arxiv.org/abs/hep-ph/9209268}{{\tt hep-ph/9209268}}].

\bibitem{FermilabLattice:2018est}
{\bf Fermilab Lattice, MILC, TUMQCD} Collaboration, A.~Bazavov et~al., {\it
  {Up-, down-, strange-, charm-, and bottom-quark masses from four-flavor
  lattice QCD}},  {\em Phys. Rev. D} {\bf 98} (2018), no.~5 054517,
  [\href{http://arxiv.org/abs/1802.04248}{{\tt arXiv:1802.04248}}].

\bibitem{Bigi:1997fj}
I.~I.~Y. Bigi, M.~A. Shifman, and N.~Uraltsev, {\it {Aspects of heavy quark
  theory}},  {\em Ann. Rev. Nucl. Part. Sci.} {\bf 47} (1997) 591--661,
  [\href{http://arxiv.org/abs/hep-ph/9703290}{{\tt hep-ph/9703290}}].

\bibitem{Bigi:2011gf}
I.~I. Bigi, T.~Mannel, and N.~Uraltsev, {\it {Semileptonic width ratios among
  beauty hadrons}},  {\em JHEP} {\bf 09} (2011) 012,
  [\href{http://arxiv.org/abs/1105.4574}{{\tt arXiv:1105.4574}}].

\bibitem{Bigi:1993ex}
I.~I.~Y. Bigi, M.~A. Shifman, N.~G. Uraltsev, and A.~I. Vainshtein, {\it {On
  the motion of heavy quarks inside hadrons: Universal distributions and
  inclusive decays}},  {\em Int. J. Mod. Phys. A} {\bf 9} (1994) 2467--2504,
  [\href{http://arxiv.org/abs/hep-ph/9312359}{{\tt hep-ph/9312359}}].

\bibitem{Aebischer:2021ilm}
J.~Aebischer and B.~Grinstein, {\it {Standard Model prediction of the $B_c$
  lifetime}},  \href{http://arxiv.org/abs/2105.02988}{{\tt arXiv:2105.02988}}.

\bibitem{Aebischer:2021eio}
J.~Aebischer and B.~Grinstein, {\it {A novel determination of the $B_c$
  lifetime}},  \href{http://arxiv.org/abs/2108.10285}{{\tt arXiv:2108.10285}}.

\bibitem{DiPierro:1998ty}
{\bf UKQCD} Collaboration, M.~Di~Pierro and C.~T. Sachrajda, {\it {A Lattice
  study of spectator effects in inclusive decays of B mesons}},  {\em Nucl.
  Phys. B} {\bf 534} (1998) 373--391,
  [\href{http://arxiv.org/abs/hep-lat/9805028}{{\tt hep-lat/9805028}}].

\bibitem{Becirevic:2001fy}
D.~Becirevic, {\it {Theoretical progress in describing the B meson lifetimes}},
   {\em PoS} {\bf HEP2001} (2001) 098,
  [\href{http://arxiv.org/abs/hep-ph/0110124}{{\tt hep-ph/0110124}}].

\bibitem{Davies:2019gnp}
{\bf HPQCD} Collaboration, C.~T.~H. Davies, J.~Harrison, G.~P. Lepage, C.~J.
  Monahan, J.~Shigemitsu, and M.~Wingate, {\it {Lattice QCD matrix elements for
  the ${B_s^0-\bar{B}_s^0}$ width difference beyond leading order}},  {\em
  Phys. Rev. Lett.} {\bf 124} (2020), no.~8 082001,
  [\href{http://arxiv.org/abs/1910.00970}{{\tt arXiv:1910.00970}}].

\bibitem{Lenz:2020efu}
A.~Lenz, M.~L. Piscopo, and C.~Vlahos, {\it {Renormalization scale setting for
  D-meson mixing}},  {\em Phys. Rev. D} {\bf 102} (2020), no.~9 093002,
  [\href{http://arxiv.org/abs/2007.03022}{{\tt arXiv:2007.03022}}].

\bibitem{Jubb:2016mvq}
T.~Jubb, M.~Kirk, A.~Lenz, and G.~Tetlalmatzi-Xolocotzi, {\it {On the ultimate
  precision of meson mixing observables}},  {\em Nucl. Phys.} {\bf B915} (2017)
  431--453, [\href{http://arxiv.org/abs/1603.07770}{{\tt arXiv:1603.07770}}].

\bibitem{Charles:2004jd}
{\bf CKMfitter Group} Collaboration, J.~Charles, A.~Hocker, H.~Lacker,
  S.~Laplace, F.~R. Le~Diberder, J.~Malcles, J.~Ocariz, M.~Pivk, and L.~Roos,
  {\it {CP violation and the CKM matrix: Assessing the impact of the asymmetric
  $B$ factories}},  {\em Eur. Phys. J. C} {\bf 41} (2005), no.~1 1--131,
  [\href{http://arxiv.org/abs/hep-ph/0406184}{{\tt hep-ph/0406184}}].

\end{thebibliography}\endgroup

\end{document}